\documentclass[twocolumn]{article}
\pdfoutput=1
\usepackage{graphicx}

\begin{document}


 \newcommand{\fig}[3]{
 \begin{figure*}
  $$#2$$\relax
  \caption{#3}
 \label{fig.#1}
 \end{figure*}                 }

 \newcommand{\fign}[3]{
 \begin{figure}
  $$#2$$\relax
  \caption{#3}
 \label{fig.#1}
 \end{figure}                 }

\newcommand{\A}[0]{{\mbox{$\mathbf{A}$}}}
\newcommand{\B}[0]{{\mbox{$\mathbf{B}$}}}
\newcommand{\C}[0]{{\mbox{$\mathbf{C}$}}}
\newcommand{\Abar}[0]{{\mbox{$\mathbf{\bar{A}}$}}}
\newcommand{\Bbar}[0]{{\mbox{$\mathbf{\bar{B}}$}}}
\newcommand{\AB}[0]{{\mbox{$\mathbf{AB}$}}}
\newcommand{\aB}[0]{{\mbox{$\mathbf{\bar{A}B}$}}}
\newcommand{\Ab}[0]{{\mbox{$\mathbf{A\bar{B}}$}}}
\newcommand{\ab}[0]{{\mbox{$\mathbf{\bar{A}\bar{B}}$}}}
\newcommand{\Ap}[0]{{\mbox{$\mathbf{A^\prime}$}}}
\newcommand{\Bp}[0]{{\mbox{$\mathbf{B^\prime}$}}}

\title{Universal Cellular Automata Based on the\\ Collisions of Soft
  Spheres\footnote{This is a posting to arXiv.org of a paper that was
    originally published in 2002 as a chapter of a book
    \cite{soft-spheres}.}}

\author{Norman Margolus} 
\date{}  
\maketitle

\begin{abstract}
{\em Fredkin's Billiard Ball Model (BBM) is a
continuous classical mechanical model of computation based on the
elastic collisions of identical finite-diameter hard spheres.  When
the BBM is initialized appropriately, the sequence of states that
appear at successive integer time-steps is equivalent to a discrete
digital dynamics.

Here we discuss some models of computation that are based on
the elastic collisions of identical finite-diameter {\em soft}
spheres: spheres which are very compressible and hence take an
appreciable amount of time to bounce off each other.  Because of this
extended impact period, these Soft Sphere Models (SSM's) correspond
directly to simple lattice gas automata---unlike the fast-impact BBM.
Successive time-steps of an SSM lattice gas dynamics can be viewed as
integer-time snapshots of a continuous physical dynamics with a
finite-range soft-potential interaction.  We present both 2D and 3D
models of universal CA's of this type, and then discuss
spatially-efficient computation using momentum conserving versions of
these models (i.e., without fixed mirrors).  Finally, we discuss the
interpretation of these models as relativistic and as semi-classical
systems, and extensions of these models motivated by these
interpretations.} 
\end{abstract}

\section{Introduction}

Cellular Automata (CA) are spatial computations.  They imitate the
locality and uniformity of physical law in a stylized digital format.
The finiteness of the information density and processing rate in a CA
dynamics is also physically realistic.  These connections with physics
have been exploited to construct CA models of spatial processes in
Nature and to explore artificial ``toy'' universes.  The discrete and
uniform spatial structure of CA computations also makes it possible to
``crystallize'' them into efficient hardware\cite{cc,isca}.

Here we will focus on CA's as realistic spatial models of ordinary
(non-quantum-coherent) computation.  As Fredkin and Banks pointed
out\cite{banks}, we can demonstrate the computing capability of a CA
dynamics by showing that certain patterns of bits act like logic
gates, like signals, and like wires, and that we can put these pieces
together into an initial state that, under the dynamics, exactly
simulates the logic circuitry of an ordinary computer.  Such a CA
dynamics is said to be {\em computation universal}.  A CA may also be
universal by being able to simulate the operation of a computer in a
less efficient manner---never reusing any logic gates for example.  A
universal CA that can perform long iterative computations within a
fixed volume of space is said to be a {\em spatially efficient} model
of computation.

We would like our CA models of computation to be as realistic as
possible.  They should accurately reflect important constraints on
physical information processing.  For this reason, one of the basic
properties that we incorporate into our models is the microscopic
reversibility of physical dynamics: there is always enough information
in the microscopic state of a physical system to determine not only
what it will do next, but also exactly what state it was in a moment
ago.  This means, in particular, that in reversible CA's (as in
physics) we can never truly erase any information.  This constraint,
combined with energy conservation, allows reversible CA systems to
accurately model thermodynamic limits on
computation\cite{bennett-thermo,mpf}.  Conversely, reversible CA's are
particularly useful for modeling thermodynamic processes in
physics\cite{raissa}.  Reversible CA ``toy universes'' also tend to
have long and interesting evolutions\cite{cc,raissa-thesis}.

All of the CA's discussed in this paper fall into a class of CA's
called Lattice Gas Automata (LGA), or simply lattice gases.  These
CA's are particularly well suited to physical modeling.  It is very
easy to incorporate constraints such as reversibility, energy
conservation and momentum conservation into a lattice gas.  Lattice
gases are known which, in their large-scale average behavior,
reproduce the normal continuum differential equations of
hydrodynamics\cite{hpp,fhp}.  In a lattice gas, particles hop around
from lattice site to lattice site.  These models are of particular
interest here because one can imagine that the particles move
continuously between lattice sites in between the discrete CA
time-steps.  Using LGA's allows us to add energy and momentum
conservation to our computational models, and also to make a direct
connection with continuous classical mechanics.

Our discussion begins with the most realistic classical mechanical
model of digital computation, Fredkin's Billiard Ball
Model\cite{conservative-logic}.  We then describe related classical
mechanical models which, unlike the BBM, are isomorphic to simple
lattice gases at integer times.  In the BBM, computations are
constructed out of the elastic collisions of very incompressible
spheres.  Our new 2D and 3D models are based on elastically colliding
spheres that are instead very compressible, and hence take an
appreciable amount of time to bounce off each other.  The universality
of these Soft Sphere Models (SSM's) depends on the finite extent in
time of the interaction, rather than its finite extent in space (as in
the BBM).  This difference allows us to interpret these models as
simple LGA's.  Using the SSM's, we discuss computation in perfectly
momentum conserving physical systems (cf. \cite{moore}), and show that
we can compute just as efficiently in the face of this added
constraint.  The main difficulty here turns out to be reusing
signal-routing resources.  We then provide an alternative physical
interpretation of the SSM's (and of all mass and momentum conserving
LGA's) as relativistic systems, and discuss some alternative
relativistic SSM models.  Finally, we discuss the use of these kinds
of models as semi-classical systems which embody realistic quantum
limits on classical computation.

\section{Fredkin's Billiard Ball Model}

\fig{bbm}{%
\begin{array}{c@{\hspace{.4in}}c@{\hspace{.4in}}c@{\hspace{.4in}}c}
\includegraphics[height=1.5in]{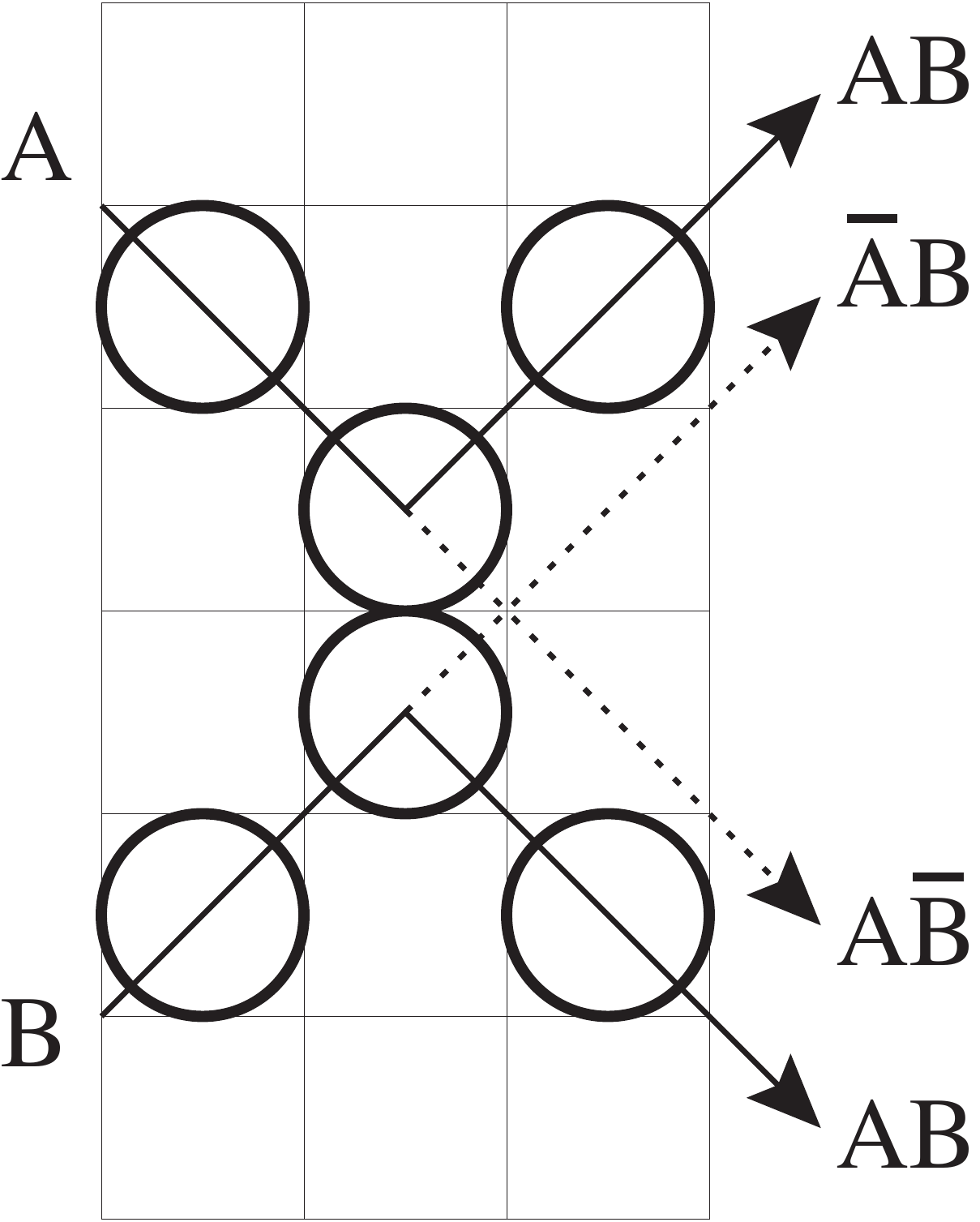} &
\includegraphics[height=1.5in]{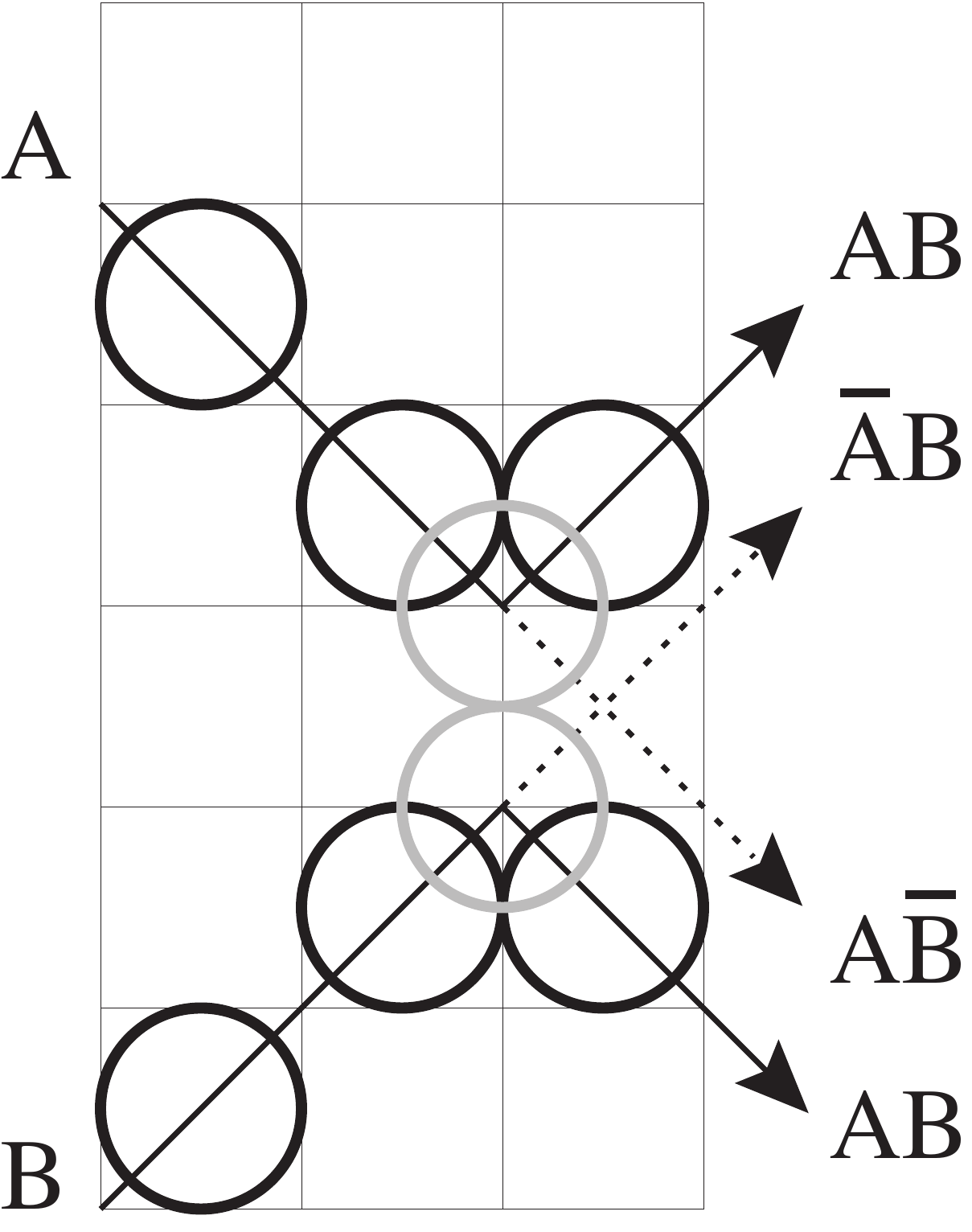} &
\includegraphics[height=1.5in]{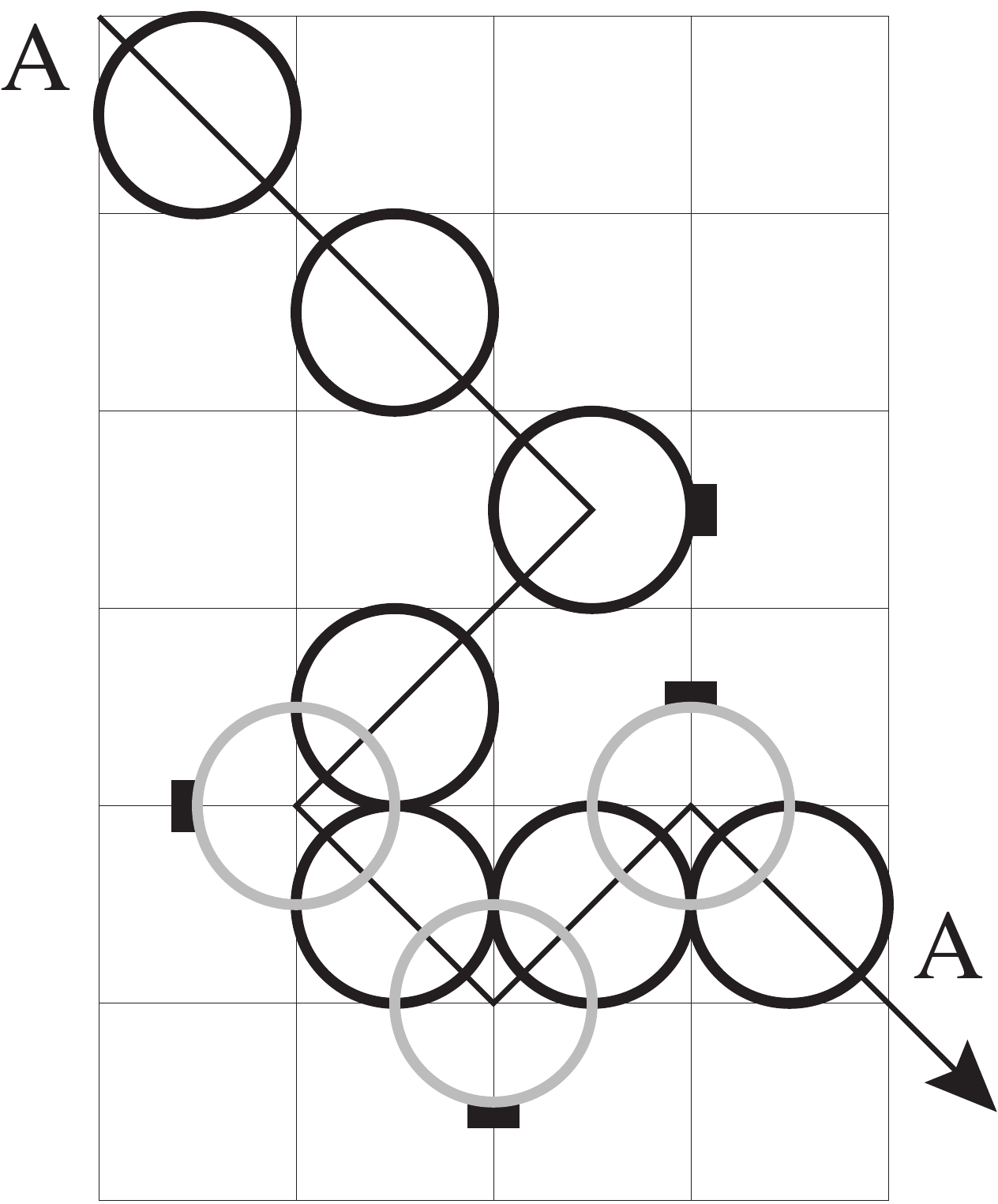} &
\includegraphics[height=1.5in]{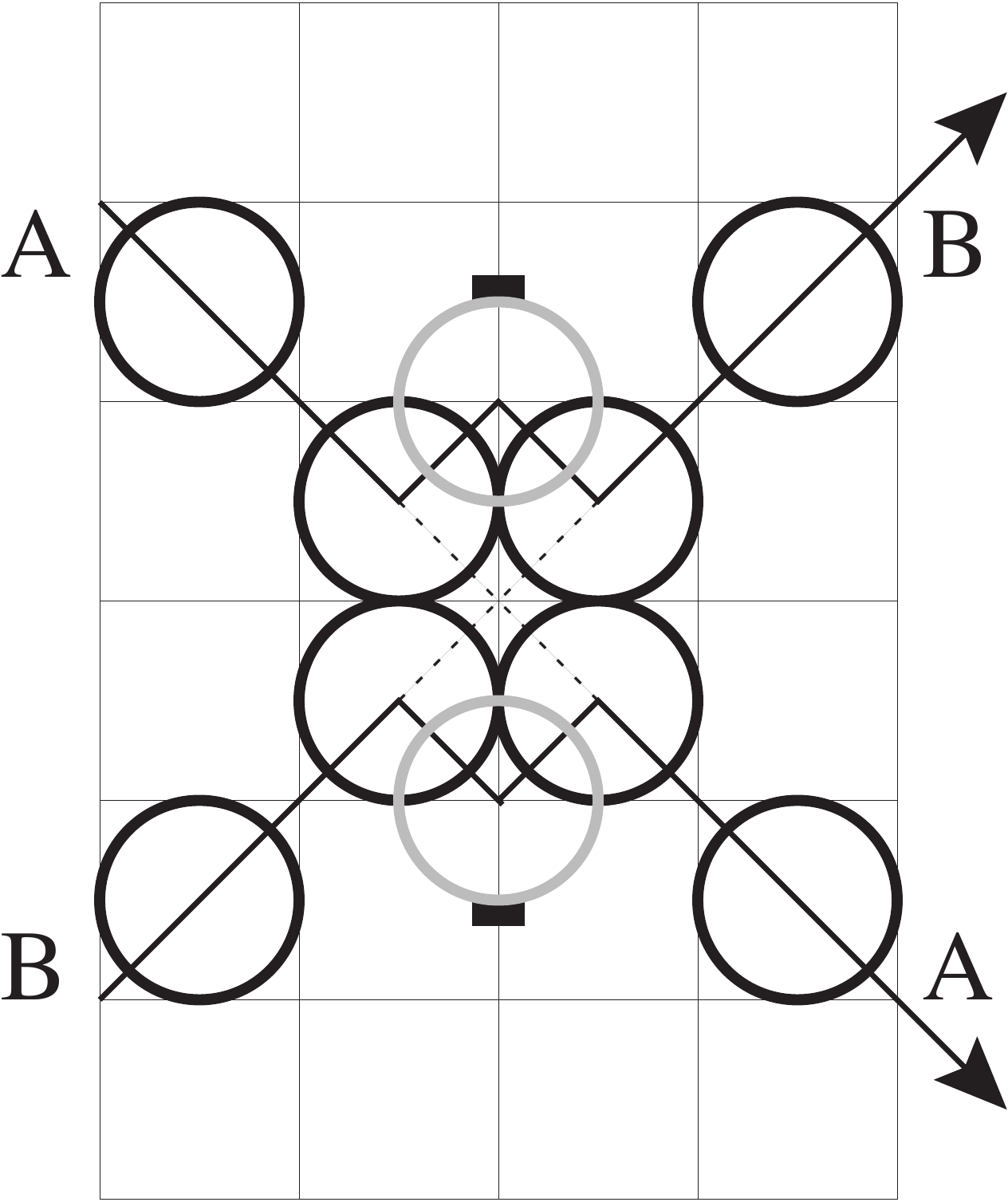} \\
\mbox{\bf (a)} & \mbox{\bf (b)} & \mbox{\bf (c)} & \mbox{\bf (d)} \\
\end{array}}
{The Billiard Ball Model.  Balls are always found at integer
coordinates at integer times.  (a)~A collision that does logic.  Two
balls are initially moving towards each other to the right.
Successive columns catch the balls at successive integer times.  The
dotted lines indicate paths the balls would have taken if only one or
the other had come in (i.e., no collision).  (b)~Balls can collide at
half-integer times (gray).  (c)~Billiard balls are routed and delayed
by carefully placed mirrors as needed to connect logic-gate collisions
together.  Collisions with mirrors can occur at either integer or
half-integer times.  (d)~Using mirrors, we can make two signal paths
cross as if the signals pass right through each other.}

In Figure~\ref{fig.bbm}, we summarize Edward Fredkin's classical
mechanical model of computation, the Billiard Ball Model.  His basic
insight is that a location where balls may or may not collide acts
like a logic gate: we get a ball coming out at certain places only if
another ball didn't knock it away!  If the balls are used as signals,
with the presence of a ball representing a logical ``1'' and the
absence a logical ``0'', then a place where signals intersect acts as
a logic gate, with different logic functions of the inputs coming out
at different places.  Figure~\ref{fig.bbm}a illustrates the idea in
more detail.  For this to work right, we need synchronized streams of
data, with evenly spaced time-slots in which a 1 (ball) or 0 (no ball)
may appear.  When two 1's impinge on the collision ``gate'', they
behave as shown in the Figure, and they come out along the paths
labeled \AB{}.  If a 1 comes in at \A{} but the corresponding slot at
\B{} is empty, then that 1 makes it through to the path labeled \Ab{}
(\A{} and not \B{}).  If sequences of such gates can be connected
together with appropriate delays, the set of logic functions that
appear at the outputs in Figure~\ref{fig.bbm}a is sufficient to build
any computer.

In order to guarantee composability of these logic gates, we constrain
the initial state of the system.  All balls are identical and are
started at integer coordinates, with the unit of distance taken to be
the diameter of the balls.  This spacing is indicated in the Figure by
showing balls centered in the squares of a grid.  All balls move at
the same speed in one of four directions: up-right, up-left,
down-right, or down-left.  The unit of time is chosen so that at
integer times, all freely moving balls are again found at integer
coordinates.  We arrange things so that balls always collide at
right angles, as in Figure~\ref{fig.bbm}a.  Such a collision leaves the
colliding balls on the grid at the next integer time.
Figure~\ref{fig.bbm}b shows another allowed collision, in which the
balls collide at half-integer times (shown in gray) but are still
found on the grid at integer times.  The signals leaving one
collision-gate are routed to other gates using fixed mirrors, as shown
in Figure~\ref{fig.bbm}c.  The mirrors are strategically placed so
that balls are always found on the grid at integer times.  Since zeros
are represented by no balls (i.e., gaps in streams of balls), zeros
are routed just as effectively by mirrors as the balls themselves are.
Finally, in Figure~\ref{fig.bbm}d, we show how two signal streams are
made to cross without interacting---this is needed to allow wires to
cross in our logic diagrams.  In the collision shown, if two balls
come in, one each at \A{} and \B{}, then two balls come out on the
same paths and with the same timing as they would have if they had
simply passed straight through.  Needless to say, if one of the input
paths has no ball, a ball on the other path just goes straight
through.  And if both inputs have no ball, we will certainly not get
any balls at the outputs, so the zeros go straight through as well.

Clearly any computation that is done using the BBM is reversible,
since if we were to simultaneously and exactly reverse the velocities
of all balls, they would exactly retrace their paths, and either meet
and collide or not at each intersection, exactly as they did going
forward.  Even if we don't actually reverse the velocities, we
know that there is enough information in the present state to recover
any earlier state, simply because we {\em could} reverse the dynamics.
Thus we have a classical mechanical system which, viewed at integer
time steps, performs a discrete reversible digital process.

The digital character of this model depends on more than just starting
all balls at integer coordinates.  We need to be careful, for example,
not to wire two outputs together.  This would result in head-on
collisions which would not leave the balls on the grid at integer
times!  Miswired logic circuits, in which we use a collision gate
backward with the four inputs improperly correlated, would also spoil
the digital character of the model.  Rather than depending on correct
logic design to assure the applicability of the digital
interpretation, we can imagine that our balls have an interaction
potential that causes them to pass through each other without
interacting in all cases that would cause problems.  This is a
bit strange, but it does conserve energy and momentum and is
reversible.  Up to four balls, one traveling in each direction, can
then occupy the same grid cell as they pass through each other.  We
can also associate the mirror information with the grid cells, thus
completing the BBM as a CA model.  Unfortunately this is a rather
complicated CA with a rather large neighborhood.

\fig{ebm}{%
\begin{array}{c@{\hspace{.4in}}c@{\hspace{.4in}}c@{\hspace{.4in}}c}
\includegraphics[height=1.3in]{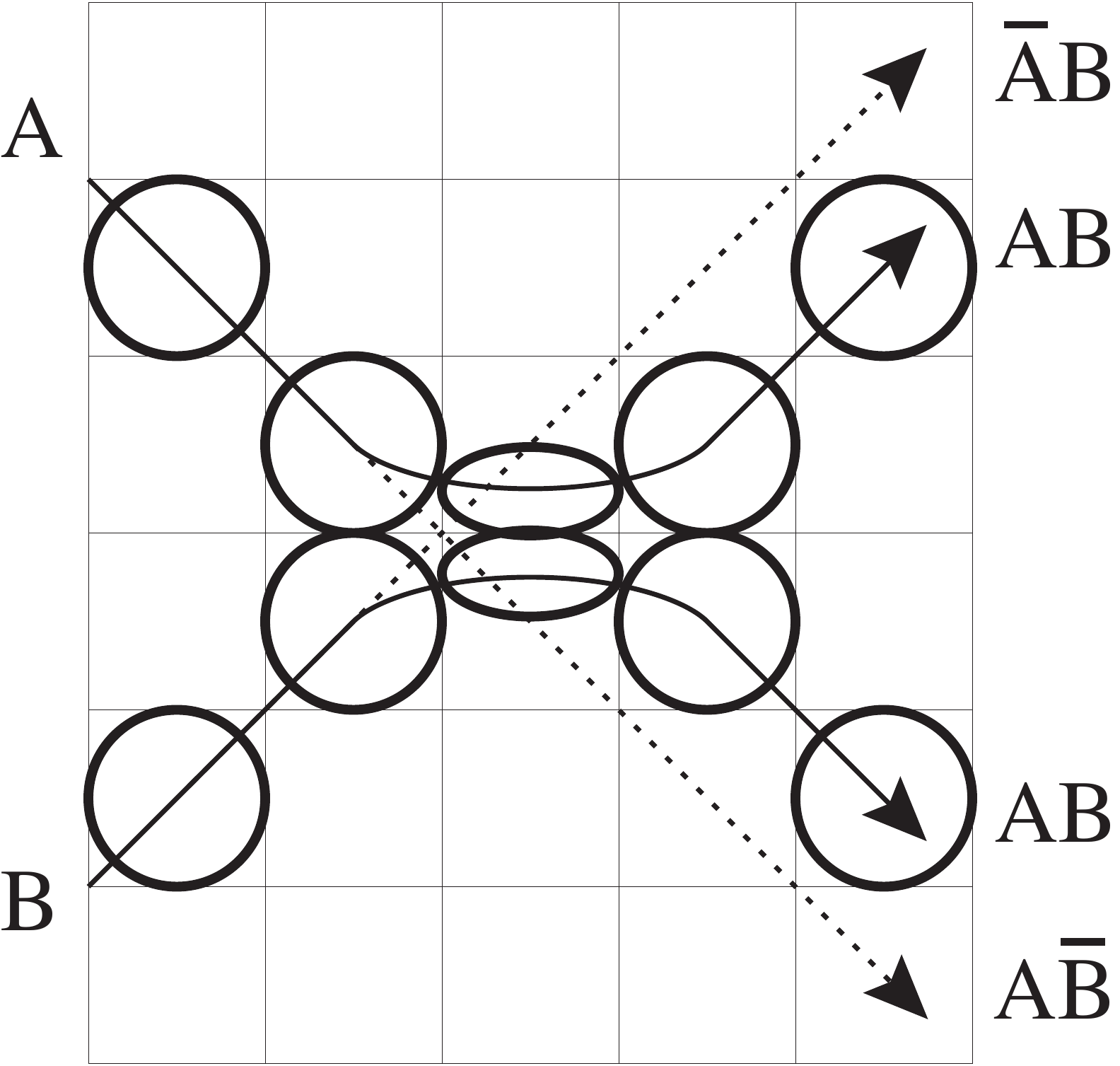} &
\includegraphics[height=1.3in]{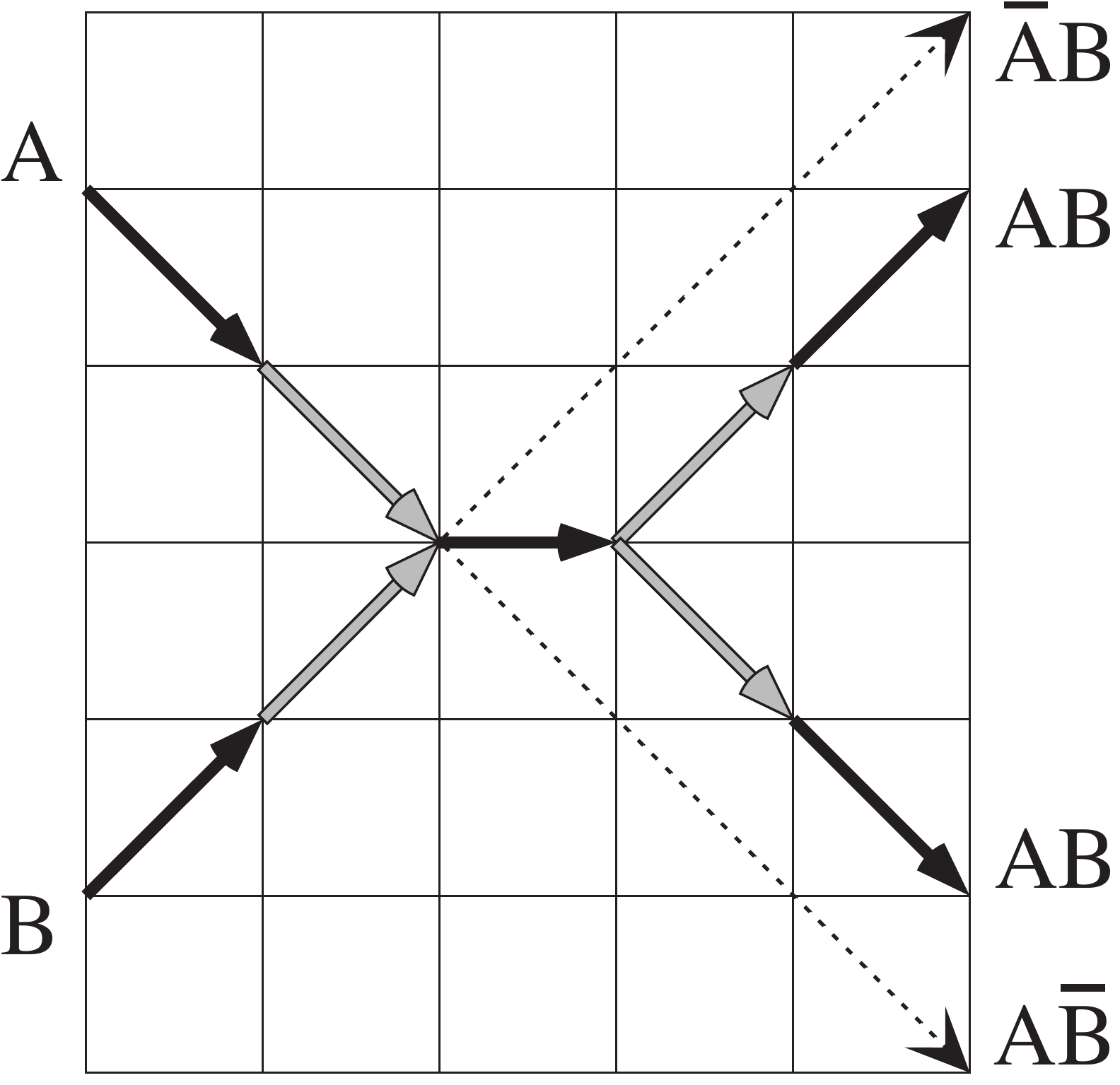} &
\includegraphics[height=1.3in]{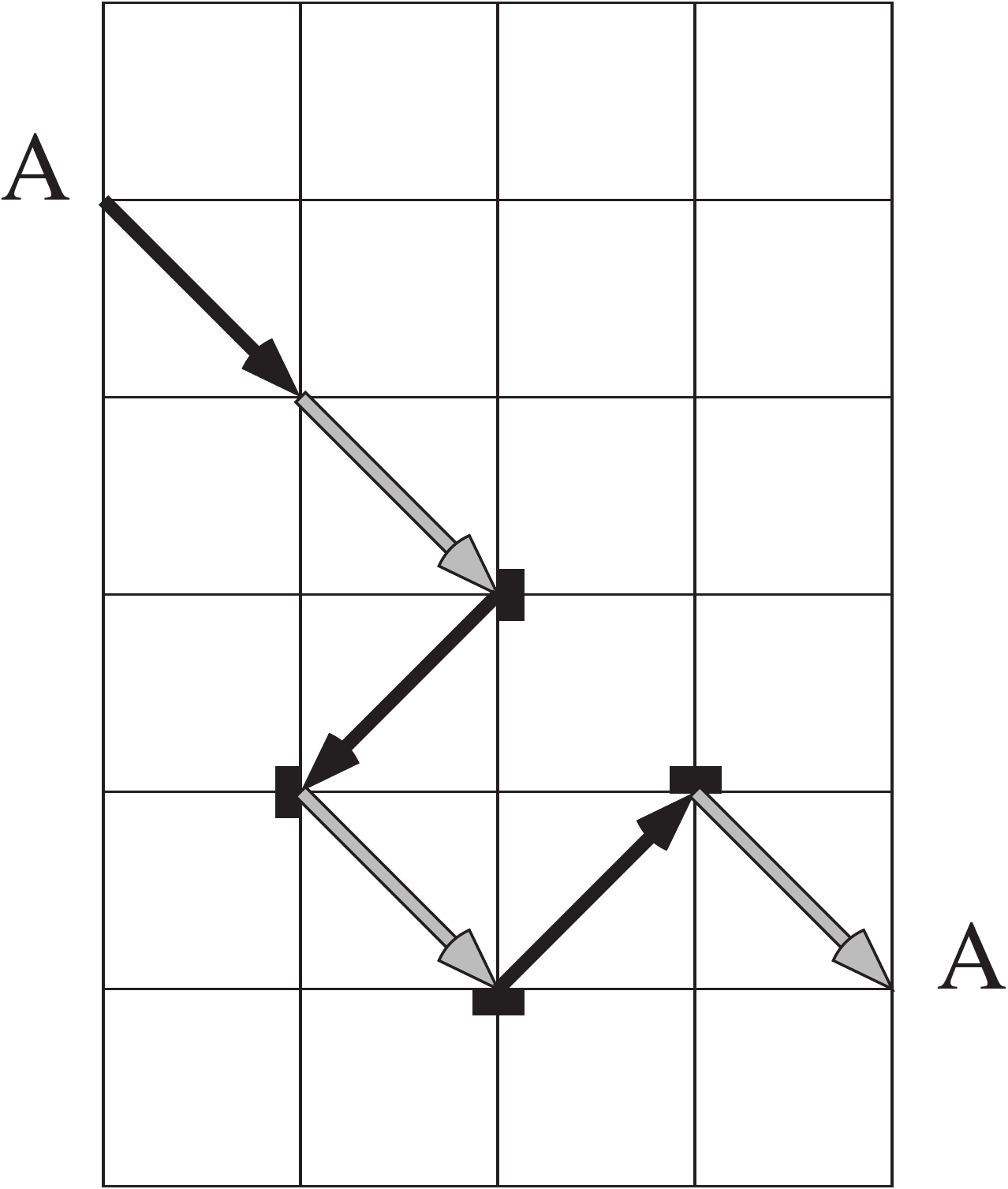} &
\includegraphics[height=1.3in]{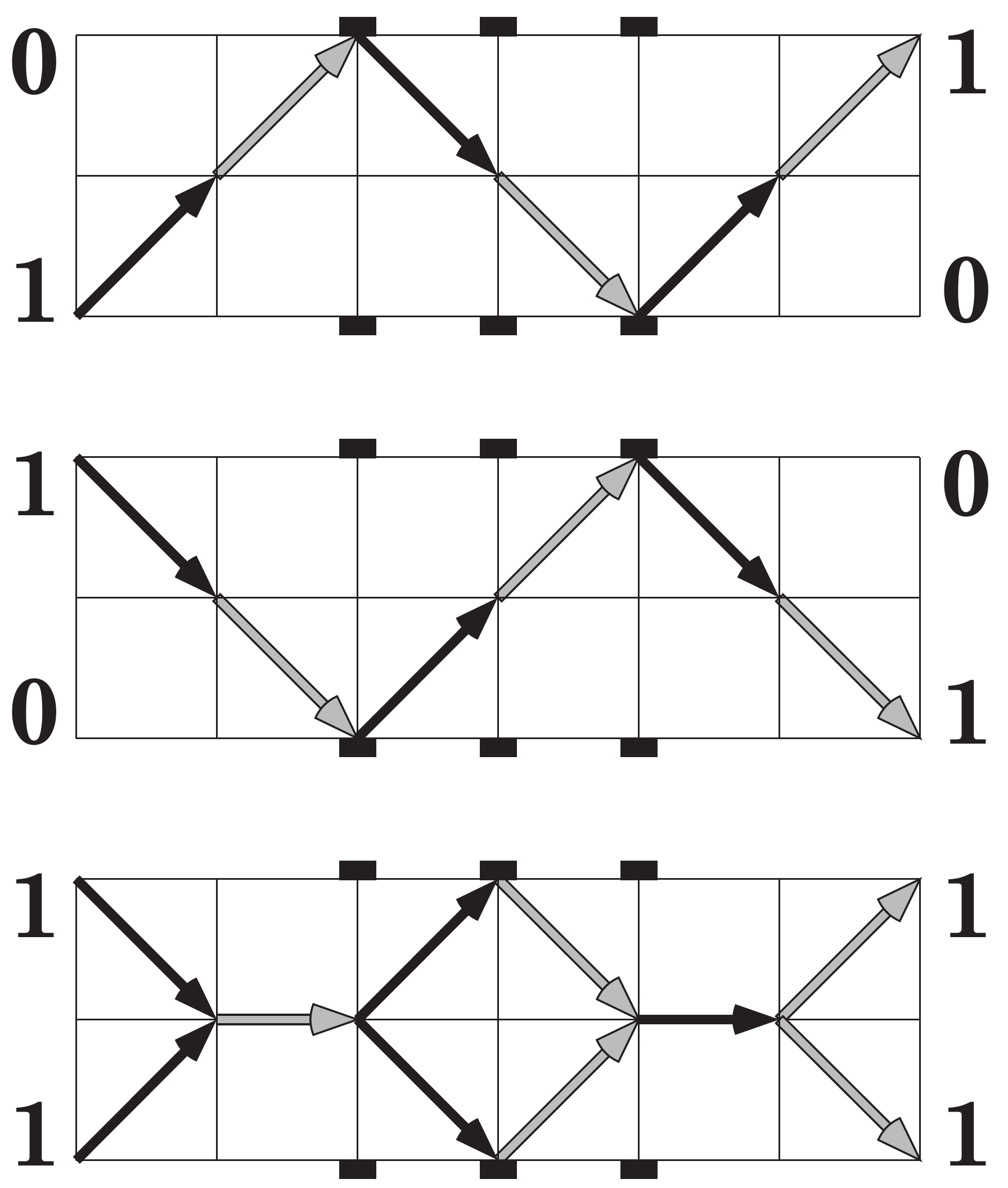} \\
\mbox{\bf (a)} & \mbox{\bf (b)} & \mbox{\bf (c)} & \mbox{\bf (d)} \\
\end{array}}
{A soft sphere model of computation.  (a)~A BBM-like collision using
very compressible balls.  The springiness of the balls is chosen so
that after the collision, the balls are again at integer sites at
integer times.  The logic is just like the BBM, but the paths are
deflected inwards, rather than outwards.  (b)~Arrows show the
velocities of balls at integer times.  During the collision, we
consider the pair to be a single mass, and draw a single arrow.
(c)~We can route and delay signals using mirrors.  (d)~We can make
signals cross.}

The complexity of the BBM as a CA rule can be attributed to the
non-locality of the hard-sphere interaction.  Although the BBM
interaction can be softened---with the grid correspondingly
adjusted---this model depends fundamentally upon information
interacting at a finite distance.  A very simple CA model based on the
BBM, the BBMCA\cite{bbmca,cc} avoids this non-locality by modeling the
front and back edges of each ball, and using a sequence of
interactions between edge-particles to simulate a billiard ball
collision.  This results in a reversible CA with just a 4-bit
neighborhood (including all mirror information!), but this model gives
up exact momentum conservation, even in simulating the collision of
two billiard balls.

In addition to making the BBMCA less physical, this loss of
conservation makes BBMCA logic circuits harder to synchronize than the
original BBM.  In the BBM, if we start a column of signals out, all
moving up-right or down-right, then they all have the same horizontal
component of momentum.  If all the mirrors they encounter are
horizontal mirrors, this component remains invariant as we pass the
signals through any desired sequence of collision ``gates.''  We don't
have to worry about synchronizing signals---they all remain in a
single column moving uniformly to the right.  In the BBMCA, in
contrast, simulated balls are delayed whenever they collide with
anything.  In a BBMCA circuit with only horizontal mirrors (or even
without any mirrors), the horizontal component of momentum is not
conserved, the center of mass does not move with constant horizontal
velocity, and appropriate delays must be inserted in order to bring
together signals that have gotten out of step.  The BBMCA has energy
conservation, but not momentum conservation.

It turns out that it is easy to make a model which is very similar to
the BBM, which has the same kind of momentum conservation as the BBM,
and which corresponds isomorphically to a simple CA rule.

\section{A Soft Sphere Model}

Suppose we set things up exactly as we did for the BBM, with balls on
a grid, moving so that they stay on the grid, but we change the
collision, making the balls very compressible.  In
Figure~\ref{fig.ebm}a, we illustrate the elastic collision of two
balls in the resulting Soft Sphere Model (SSM).  If the springiness of
the balls is just right (i.e., we choose an appropriate interaction
potential), then the balls find themselves back on the grid after the
collision.  If only one or the other ball comes in, they go straight
through.  Notice that the output paths are labeled exactly as in the
BBM model, except that the \AB{} paths are deflected inwards rather
than outwards (cf. Appendix to \cite{bbmca}).  If we add BBM-style
hard-collisions with mirrors,\footnote{All of the $90^\circ$ turns
that we use in our SSM circuits can also be achieved by soft mirrors
placed at slightly different locations.} then this model can compute
in the same manner as the BBM, with the same kind of momentum
conservation aiding synchronization.

In Figure~\ref{fig.ebm}b, we have drawn an arrow in each grid cell
corresponding to the velocity of the center of a ball at an integer
time.  The pair of colliding balls is taken to be a single particle,
and we also draw an arrow at its center.  We've colored the arrows
alternately gray and black, corresponding to successive positions of
an incoming pair of logic values.  We can now interpret the arrows as
describing the dynamics of a simple lattice gas, with the sites of the
lattice taken to be the corners of the cells of the grid.

In a lattice gas, we alternately move particles and let them interact.
In this example, at each lattice site we have room for up to eight
particles (1's): we can have one particle moving up-right, one
down-right, one up-left, one down-left, one right, one left, one up
and one down.  In the movement step, all up-right particles are
simultaneously moved one site up and one site to the right, while all
down-right particles are moved down and to the right, etc.  After all
particles have been moved, we let the particles that have landed at
each lattice site interact---the interaction at each lattice site is
independent of all other lattice sites.

In the lattice gas pictured in Figure~\ref{fig.ebm}b, we see on the
left particles coming in on paths {\A} and {\B} that are entering two
lattice sites (black arrows) and the resulting data that leaves those
sites (gray arrows).  Our inferred rule is that single diagonal
particles that enter a lattice site come out in the same direction
they came in.  At the next step, these gray arrows represent two
particles entering a single lattice site.  Our inferred rule is that
when two diagonal particles collide at right angles, they turn into a
single particle moving in the direction of the net momentum.  Now a
horizontal black particle enters the next lattice site, and our rule
is that it turns back into two diagonal particles.  If only one
particle had come in, along either {\A} or {\B}, it would have
followed our ``single diagonal particles go straight'' rule, and so
single particles would follow the dotted path in the figure.  Thus
our lattice gas exactly duplicates the behavior of the SSM at integer
times.

\fig{eba-rule}{%
\begin{array}{c@{\hspace{.7in}}c@{\hspace{.7in}}c}
\includegraphics[height=1.2in]{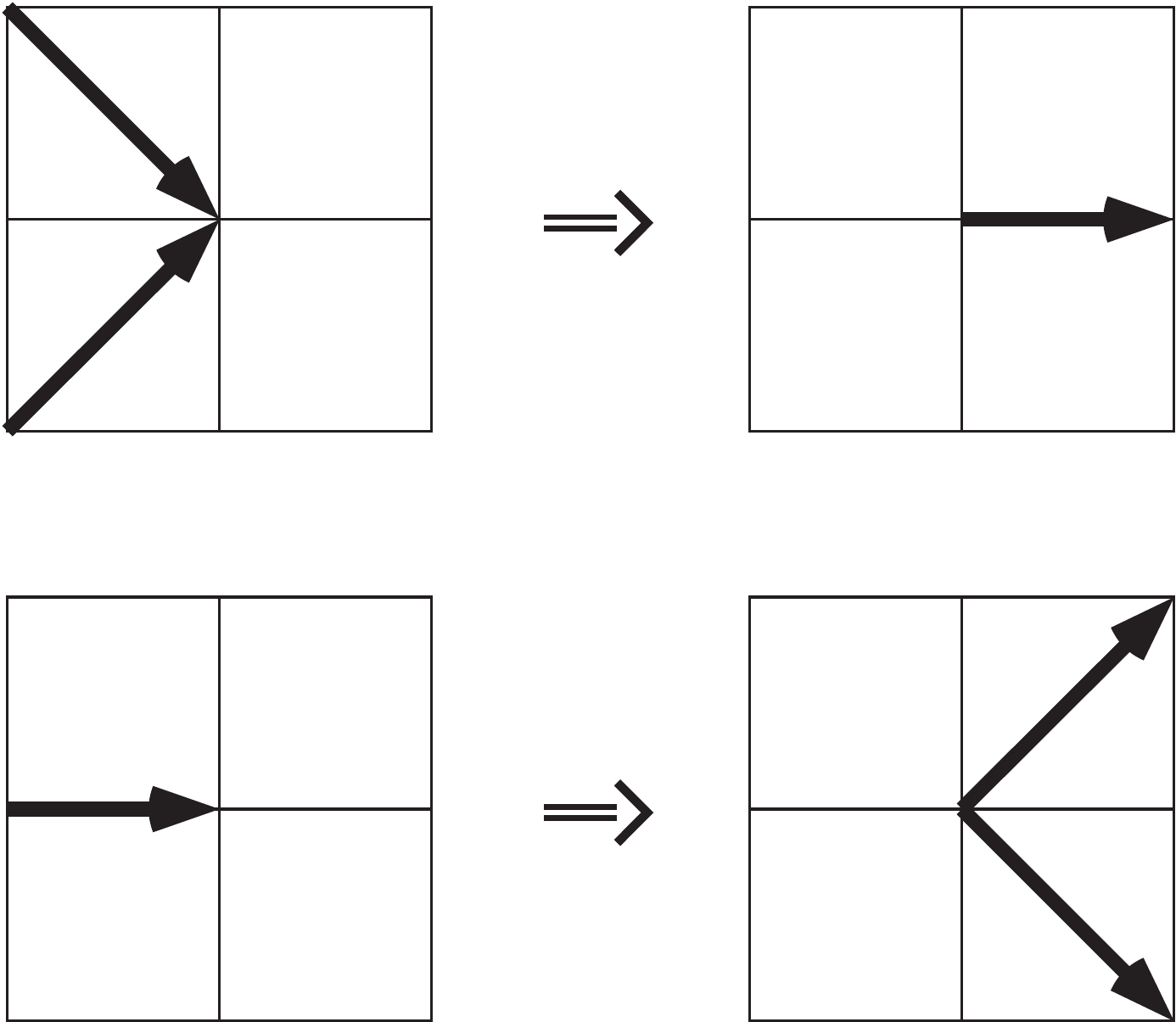} &
\includegraphics[height=1.2in]{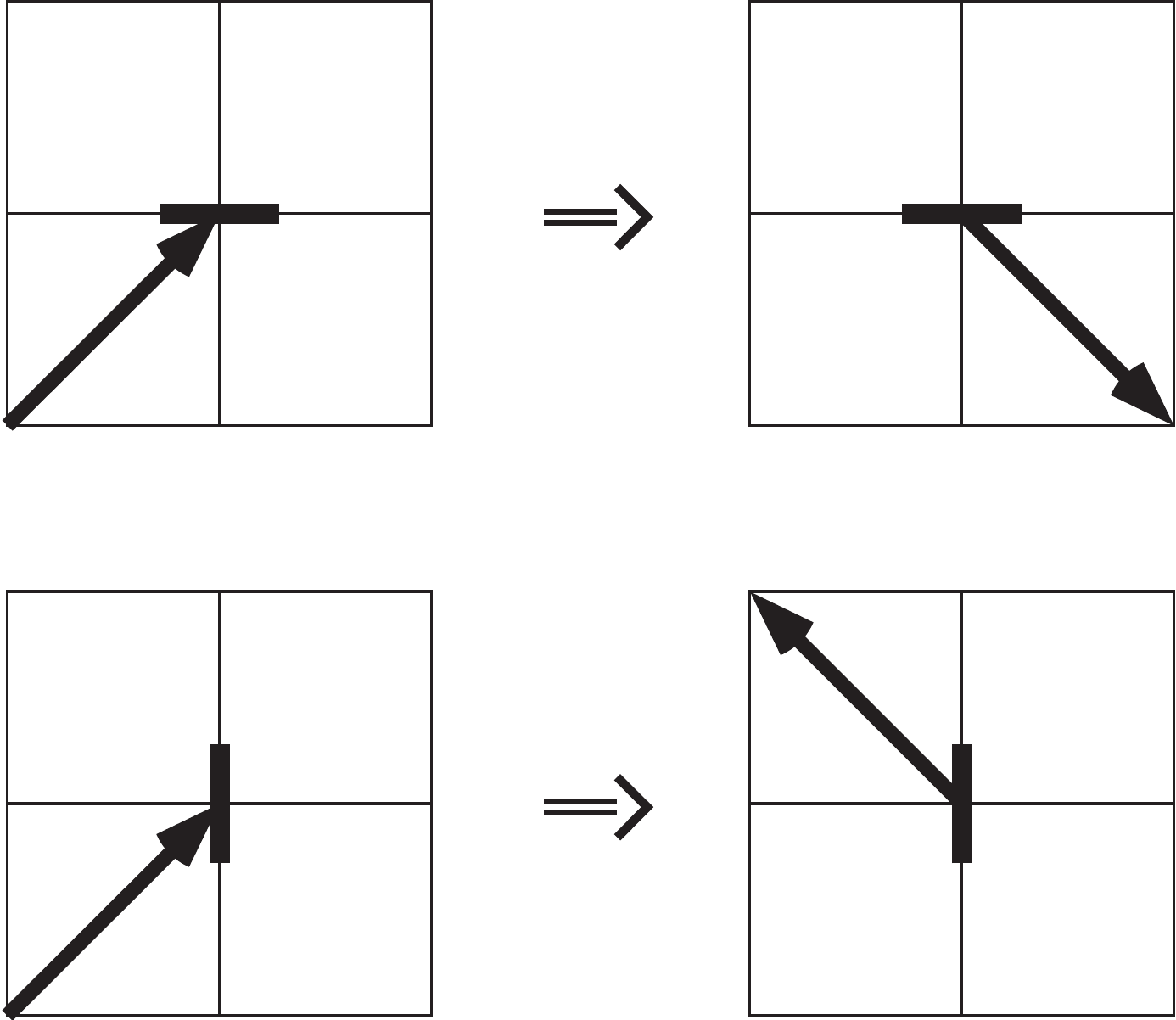} &
\includegraphics[height=1.2in]{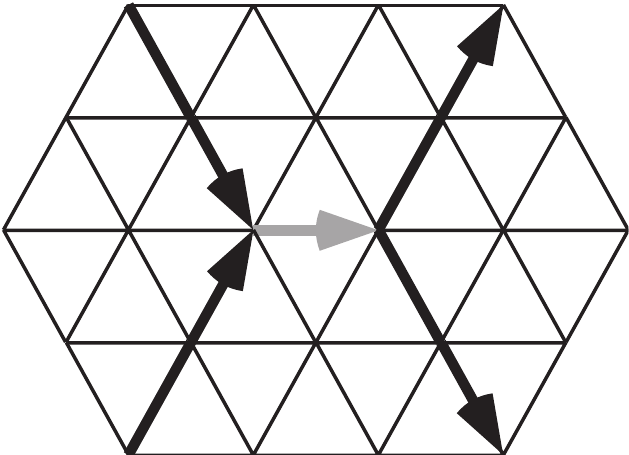} \\
\mbox{\bf (a)} & \mbox{\bf (b)} & \mbox{\bf (c)} \\
\end{array}}
{(a)~A simple lattice gas rule captures the dynamics of the soft
sphere collision.  Two particles colliding at right angles turn into a
single new particle of twice the mass for one step, which then turns
back into two particles.  A mirror deflects a particle through
$90^\circ$.  In all other cases, particles go straight. (b)~A soft
sphere collision on a triangular lattice.}

From Figure~\ref{fig.ebm}c we can infer the rule with the addition of
mirrors.  Along with particles at each lattice site, we allow the
possibility of one of two kinds of mirrors---horizontal mirrors and
vertical mirrors.  If a single particle enters a lattice site occupied
only by a mirror, then it is deflected as shown in the diagram.
Signal crossover takes more mirrors than in the BBM
(Figure~\ref{fig.ebm}d).  Our lattice gas rule is summarized in
Figure~\ref{fig.eba-rule}a.  For each case shown, $90^\circ$
rotations of the state shown on the left turn into the same rotation
of the state shown on the right.  In all other cases, particles go
straight.  This is a simple reversible rule, and (except in the
presence of mirrors) it exactly conserves momentum.  We will discuss a
version of this model later without mirrors, in which momentum is
always conserved.

The relationship between the SSM of Figure~\ref{fig.ebm}a and a
lattice gas can also be obtained by simply shrinking the size of the
SSM balls without changing the grid spacing.  With the right
time-constant for the two-ball impact process, tiny particles would
follow the paths indicated in Figure~\ref{fig.ebm}b, interacting at
grid-corner lattice sites at integer times.  The BBM cannot be turned
into a lattice gas in this manner, because the BBM depends upon the
finite extent of the interaction in space, rather than in
time.

Notice that in establishing an isomorphism between the integer-time
dynamics of this SSM and a simple lattice gas, we have added the
constraint to the SSM that we cannot place mirrors at half-integer
coordinates, as we did in order to route signals around in the BBM
model in Figure~\ref{fig.bbm}.  This means, in particular, that we
can't delay a signal by one time unit---as the arrangement of mirrors
in Figure~\ref{fig.ebm}c would if the spacing between all mirrors were
halved.  This doesn't impair the universality of the model, however,
since we can easily guarantee that all signal paths have an even
length.  To do this, we simply design our SSM circuits with mirrors at
half-integer positions and then rescale the circuits by an even factor
(four is convenient).  Then all mirrors land at integer coordinates.
The separation of outputs in the collision of Figure~\ref{fig.ebm}b
can be rescaled by a factor of four by adding two mirrors to cause the
two {\AB} outputs to immediately collide a second time (as in the
bottom image of Figure~\ref{fig.ebm}d).  We will revisit this issue
when we discuss mirror-less models in Section~\ref{sec.mom}.

\section{Other Soft Sphere Models}

\fig{3d}{%
\begin{array}{c@{\hspace{.5in}}c@{\hspace{.5in}}c}
\includegraphics[height=1.7in]{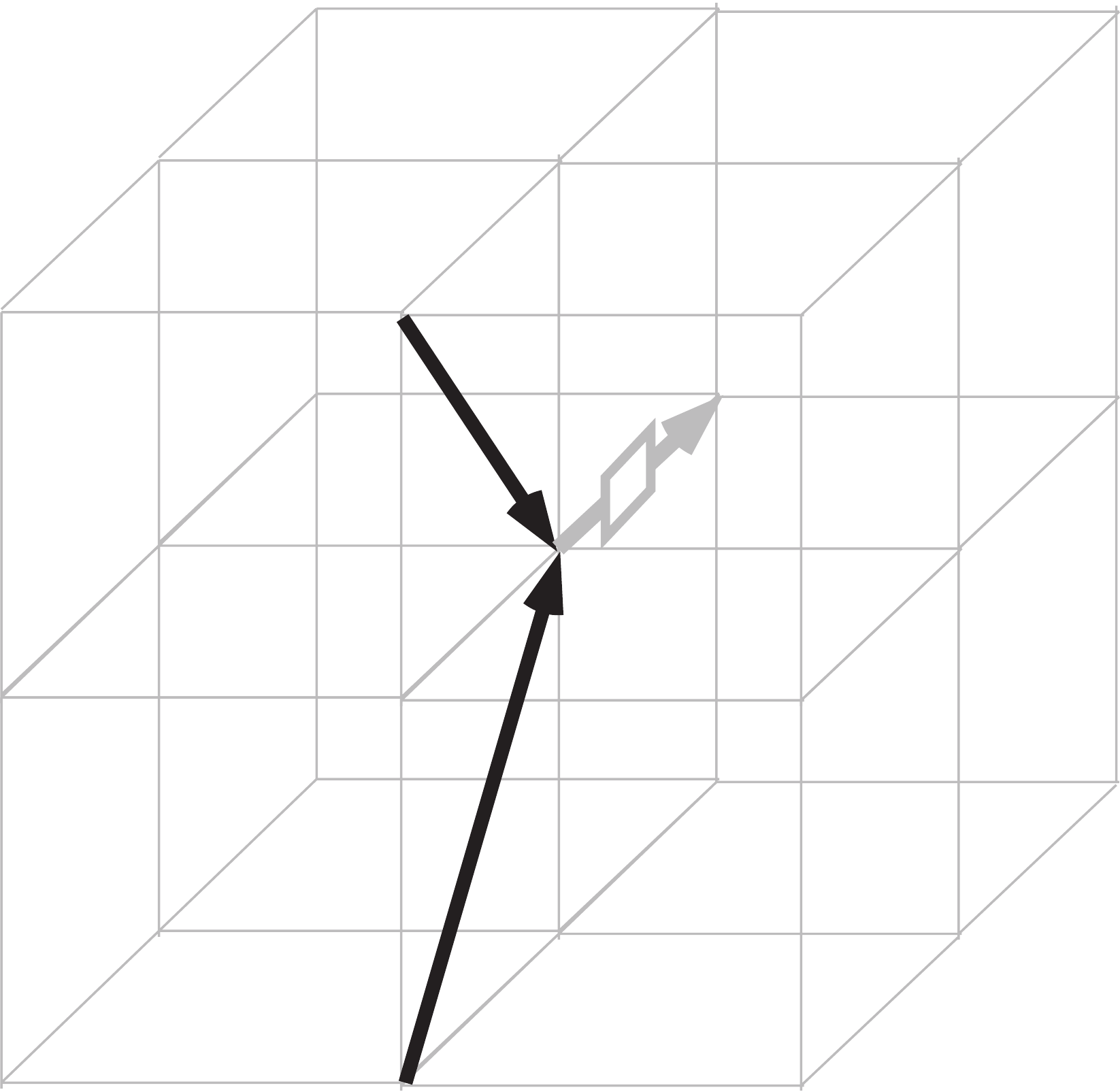} &
\includegraphics[height=1.7in]{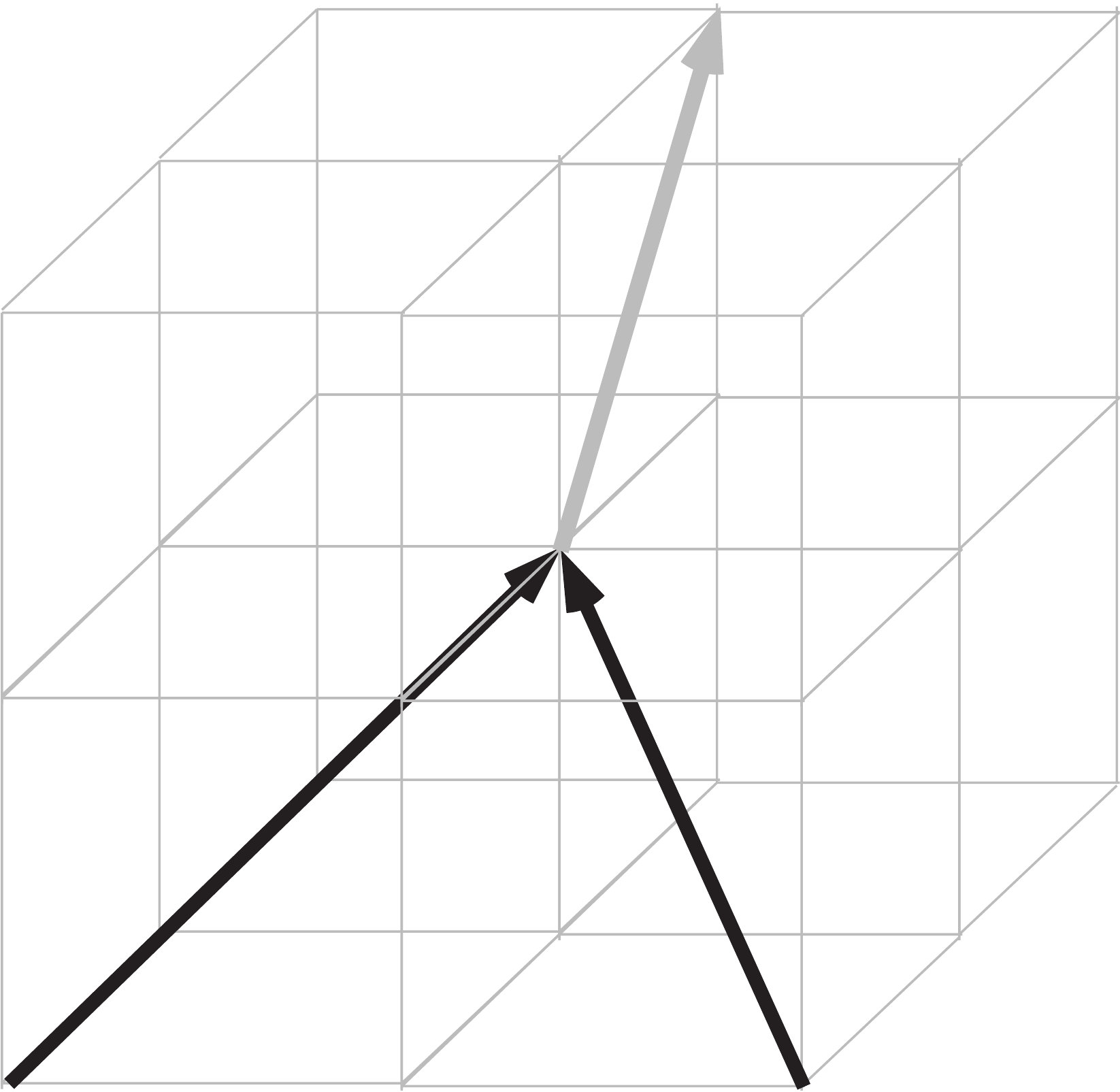} &
\includegraphics[height=1.7in]{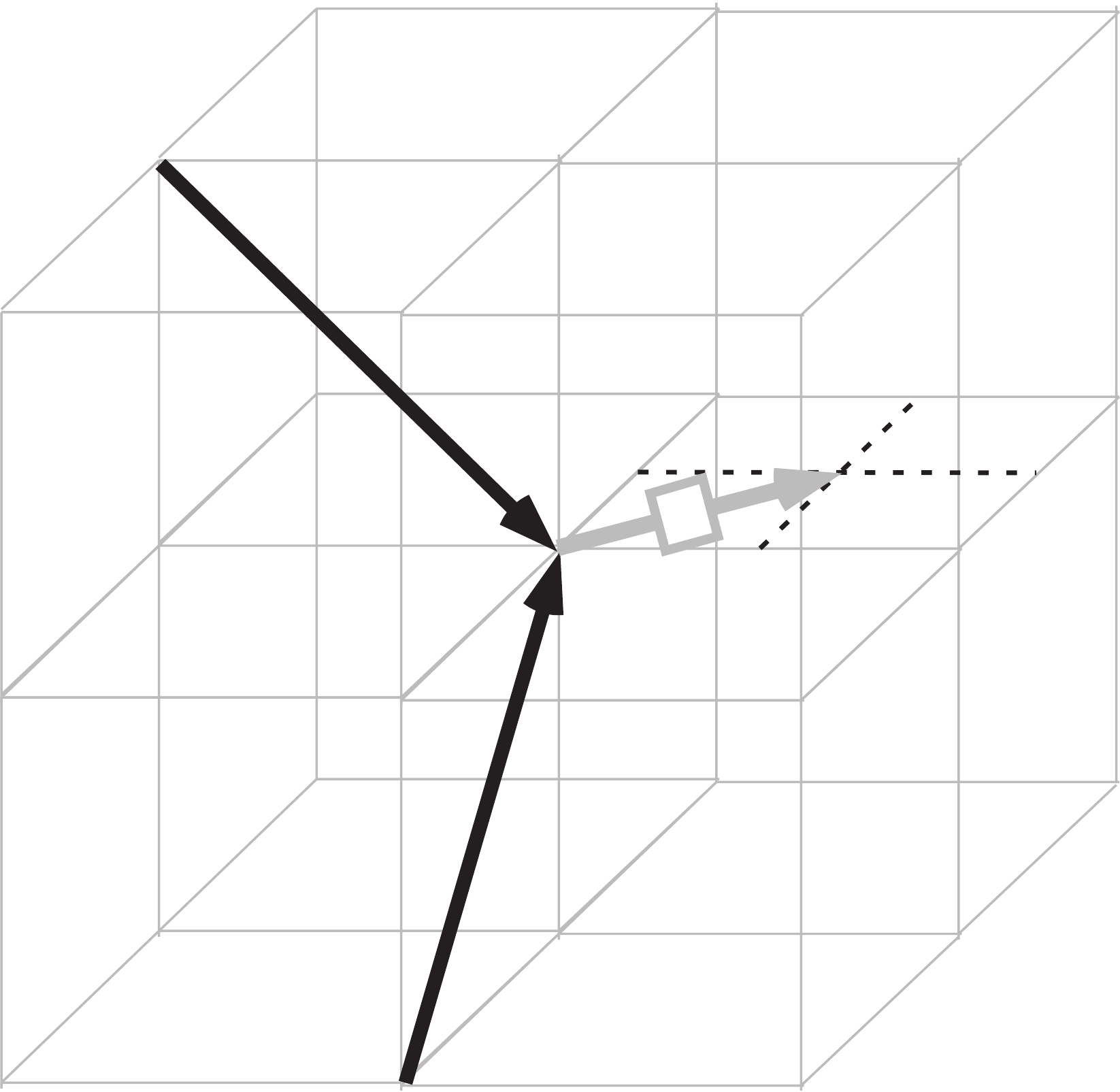} \\
\mbox{\bf (a)} & \mbox{\bf (b)} & \mbox{\bf (c)} \\
\end{array}}
{3D Soft Sphere Models.  (a)~Collisions using cube edges and cube-face
diagonals.  Each edge particle carries one bit of information about
which of two planes the diagonal particles that created it were in.
(b)~Collisions using face and body diagonals.  Two body-diagonal
particles collide only if they are both coplanar with a face-diagonal.
The resulting face-diagonal particle doesn't carry any extra planar
information, since there is a unique pair of body-diagonal particles
that could have produced it.  (c)~Collisions using only face
diagonals, with two speeds.  If particles are confined to a single
plane, this is equivalent to the triangular lattice model of
Figure~\ref{fig.eba-rule}b.  Again the slower particle must carry an
extra bit of collision-plane information.}

In Figure~\ref{fig.eba-rule}b, we show a mass- and momentum-conserving
SSM collision on a triangular lattice, which corresponds to a
reversible lattice gas model of computation in exactly the same manner
as discussed above.  Similarly, we can construct SSM's in 3D.  In
Figure~\ref{fig.3d}a, we see a mass and momentum conserving SSM
collision using the face-diagonals of the cubes that make up our 3D
grid.  The resulting particle (gray) carries one bit of information
about which of two possible planes the face-diagonals that created it
resided in.  In a corresponding diagram showing colliding spheres (a
3D version of Figure~\ref{fig.ebm}a), we would see that this
information is carried by the plane along which the spheres are
compressed.  This model is universal within a single plane of the 3D
space, since it is just the 2D square-lattice SSM discussed above.  To
allow signals to get out of a single plane, mirrors can be applied to
diagonal particles to deflect them onto cube-face diagonals outside of
their original plane.

A slightly simpler 3D scheme is shown in Figure~\ref{fig.3d}b.  Here
we only use body and face diagonals, and body diagonals only collide
when they are coplanar with a face diagonal.  Since each face diagonal
can only come from one pair of body diagonals, no collision-plane
information is carried by face-diagonal particles.  For mirrors, we
can restrict ourselves to reflecting each body diagonal into one of
the three directions that it could have been deflected into by a
collision with another body diagonal.  This is an interesting
restriction, because it means that we can potentially make a
momentum-conserving version of this model without mirrors, using only
signals to deflect signals.

Finally, the scheme shown in Figure~\ref{fig.3d}c uses only face
diagonals, with the heavier particle traveling half as fast as the
particles that collide to produce it.  As in Figure~\ref{fig.3d}a, the
slower particle carries a bit of collision-plane information.  To
accommodate the slower particles, the lattice needs to be twice as
fine as in Figures \ref{fig.3d}a and \ref{fig.3d}b, but we've only
shown one intermediate lattice site for clarity.  Noting that three
coplanar face-diagonals of a cube form an equilateral triangle, we see
that this model, for particles restricted to a single plane, is
exactly equivalent to the triangular-lattice model pictured in
Figure~\ref{fig.eba-rule}b.  As in the model pictured in
Figure~\ref{fig.3d}b, the deflection directions that can be obtained
from particle-particle collisions are sufficient for 3D routing, and
so this model is also a candidate for mirrorless momentum-conserving
computation in three dimensions.

\section{Momentum conserving \\ models}\label{sec.mom}

A rather unphysical property of the BBM, as well as of the related
soft sphere models we have constructed, is the use of immovable
mirrors.  If the mirrors moved even a little bit, they would spoil the
digital nature of these models.  To be perfectly immovable, as we
demand, these mirrors must be infinitely massive, which is not very
realistic.  In this section, we will discuss SSM gases which compute
without using mirrors, and hence are perfectly momentum conserving.


The issue of computation universality in momentum-conserving lattice
gases was discussed in
\cite{moore}, where it was shown that some 2D LGA's of physical
interest can compute any logical function.  This paper did not,
however, address the issue of whether such LGA's can be spatially
efficient models of computation, reusing spatial resources as ordinary
computers do.  There is also a new question about the generation of
entropy (undesired information) which arises in the context of
reversible momentum conserving computation models, and which we will
address.  With mirrors, any reversible function can be computed in the
SSM (or BBM) without leaving any intermediate results in the
computer's memory\cite{conservative-logic}.  Is this still true
without mirrors, where even the routing of signals requires an
interaction with other signals?  We will demonstrate mirrorless
momentum-conserving SSM's that are just as efficient spatially as an
SSM {\em with} mirrors, and that don't need to generate any more
entropy than an SSM with mirrors.  In the process we will illustrate
some of the general physical issues involved in efficiently routing
signals without mirrors.

\subsection{Reflections without mirrors}

\fig{1mirrors}{%
\begin{array}{c@{\hspace{1in}}c}
\includegraphics[height=1.5in]{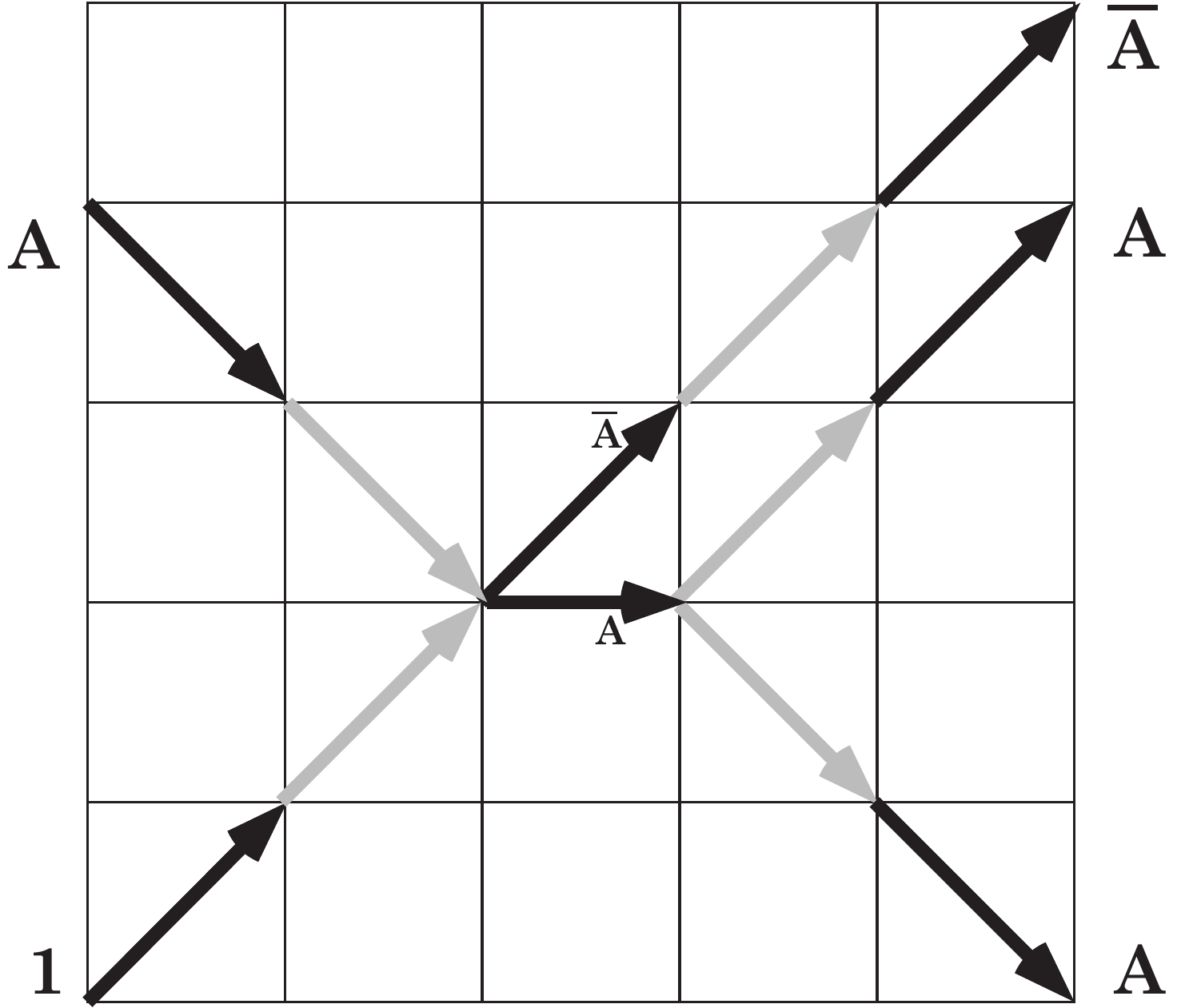} &
\includegraphics[height=1.5in]{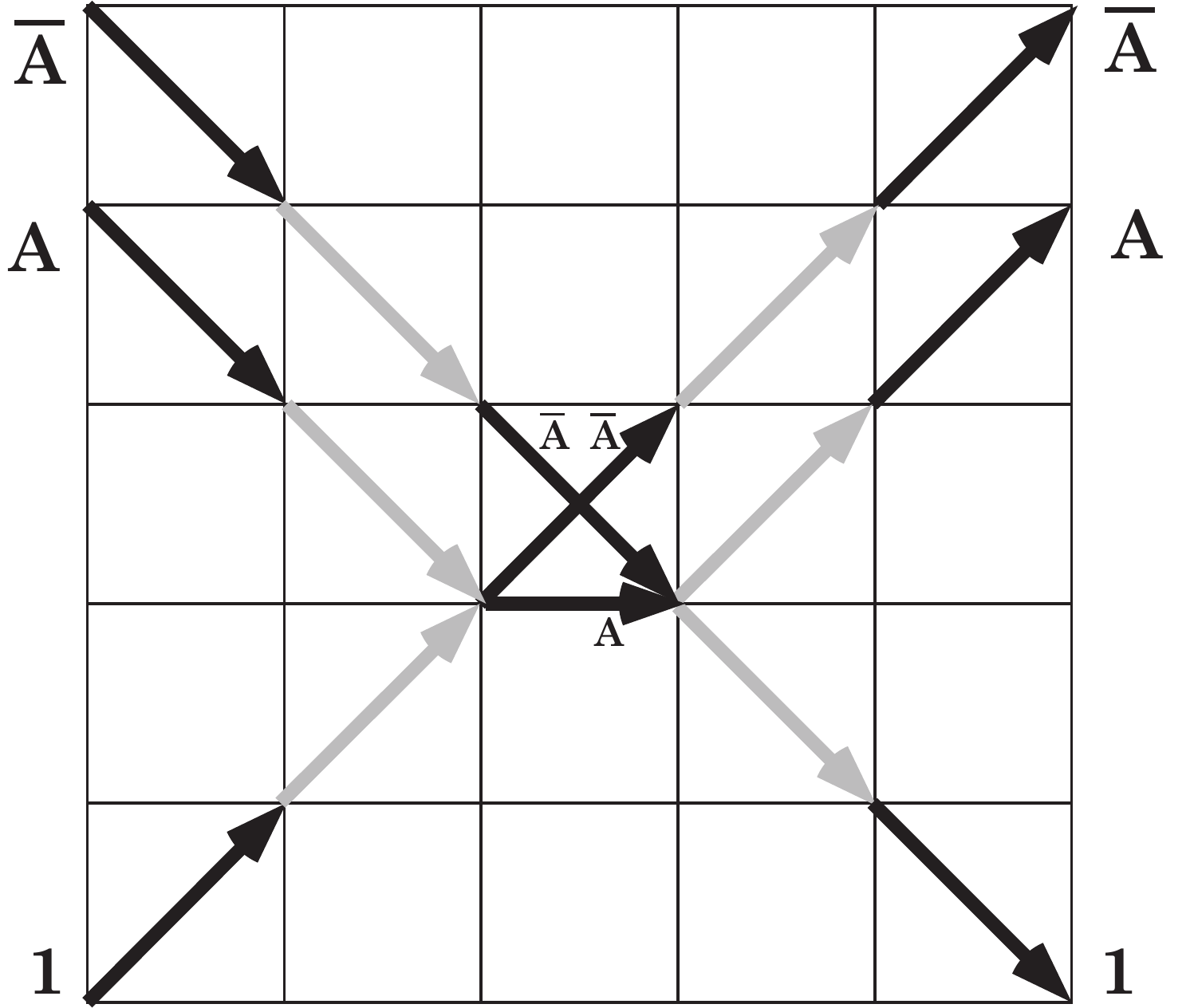} \\
\mbox{\bf (a)} & \mbox{\bf (b)} \\
\end{array}}
{Using streams of balls as mirrors. (a)~A stream of 1's (balls)
diverts a signal {\bf A}, but also makes two copies of the signal.
(b)~If dual-rail (complementary) signalling is used, signals can be
cleanly reflected.}

We begin our discussion by replacing a fixed mirror with a constant
stream of particles (ones), aimed at the position where we want a
signal reflected.  This is illustrated in Figure~\ref{fig.1mirrors}a.
Here we show the 2D square-lattice SSM of Figure~\ref{fig.ebm}a, with
a signal {\A} being deflected by the constant stream.  Along with the
desired reflection of {\A}, we also produce two undesired copies of
{\A} (one of them complemented).  This suggests that perhaps every
bend in every signal path will continuously generate undesired
information that will have to be removed from the computer.

Figure~\ref{fig.1mirrors}b shows a more promising deflection.  The
only thing that has changed is that we have brought in {\Abar} along
with {\A}, and so we now get a 1 coming out the bottom regardless of
what the value of {\A} was.  Thus signals that are represented in
complementary form (so-called ``dual-rail'' signals) can be deflected
cleanly.  This makes sense, since each signal now carries one unit of
momentum regardless of its value, and so the change of momentum in the
deflecting mirror stream can now also be independent of the signal
value.

\subsection{Signal crossover}

\fig{cross}{%
\begin{array}{c@{\hspace{1in}}c}
\includegraphics[height=1.5in]{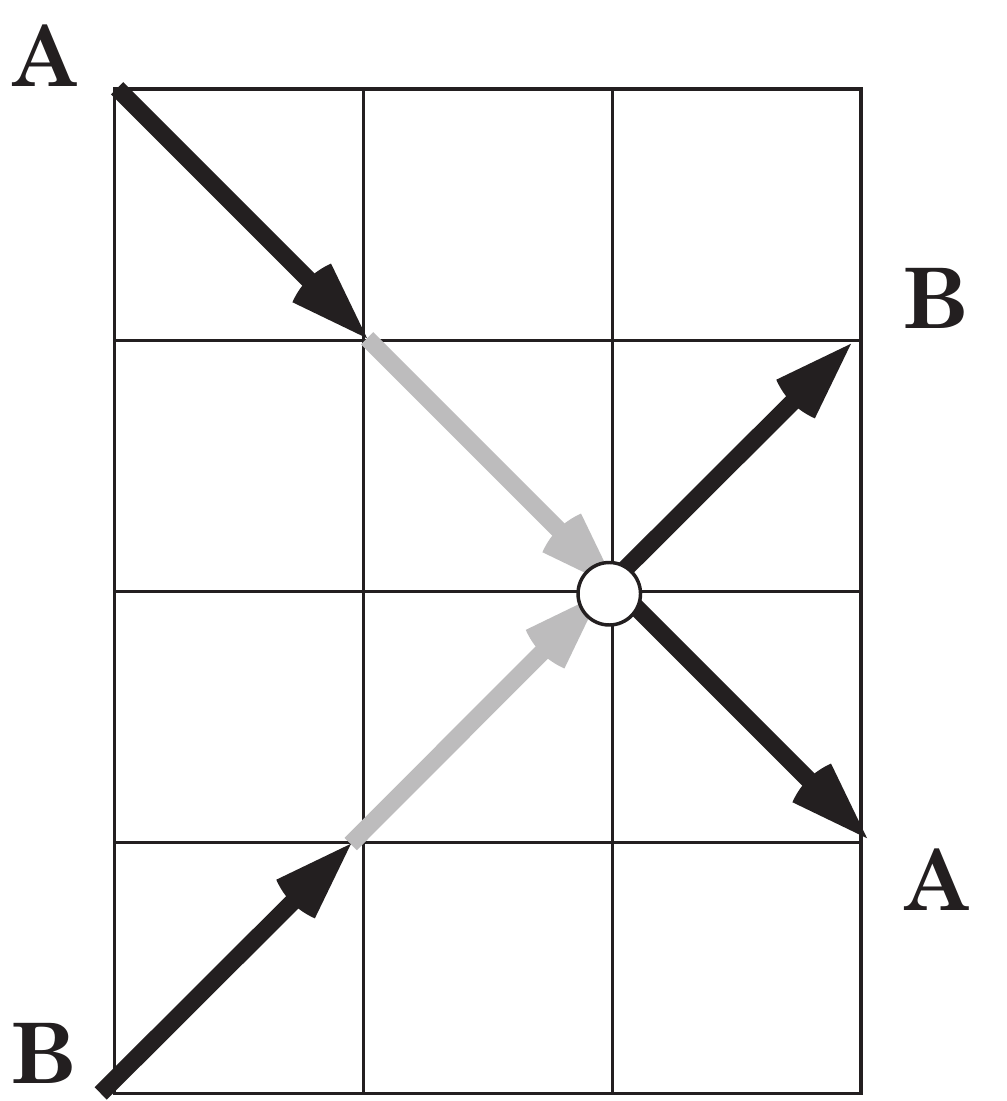} &
\includegraphics[height=1.5in]{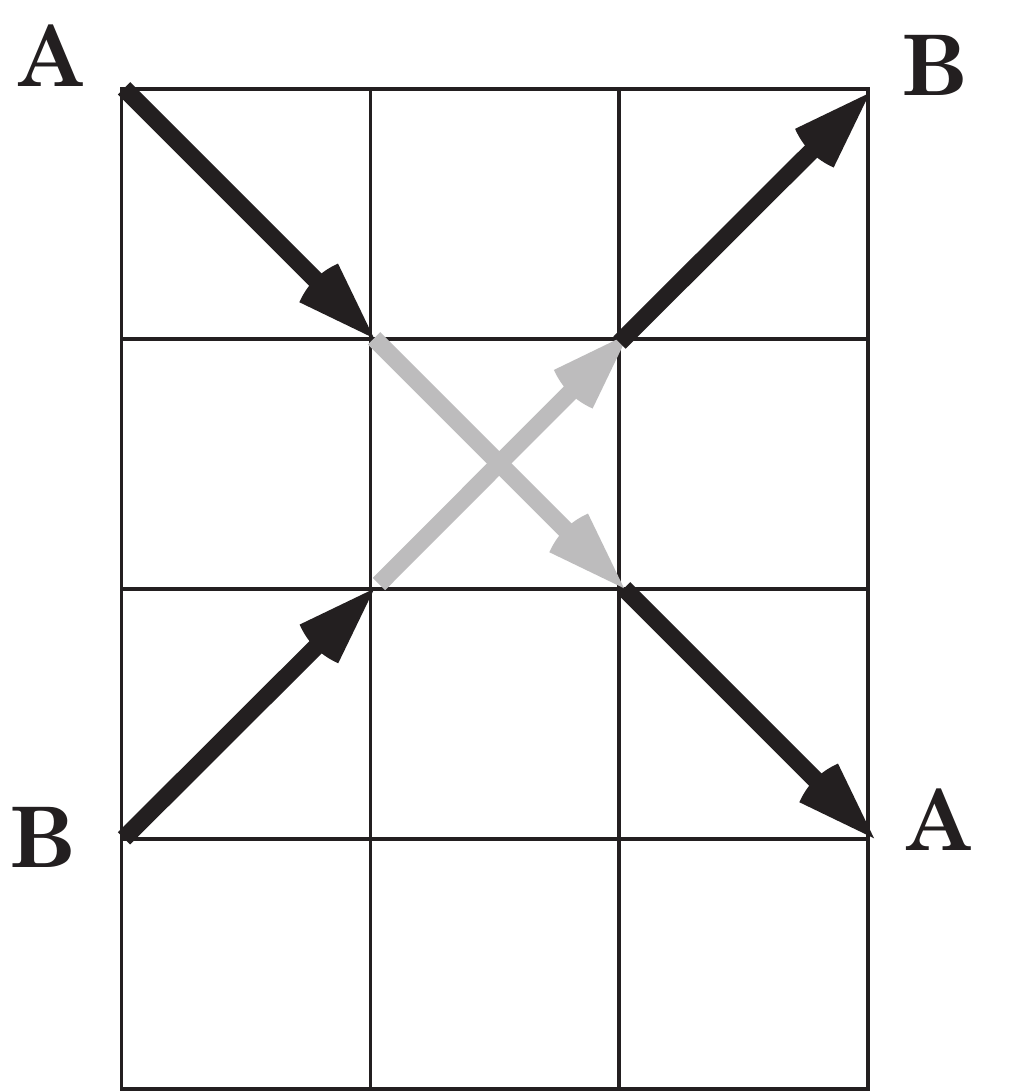} \\
\mbox{\bf (a)} & \mbox{\bf (b)} \\
\end{array}}
{Signals that cross.  (a)~The circle indicates a rest particle.  Two
signals cross at a rest particle without interacting.  (b)~Signals can
also cross between lattice sites, where no interaction is possible.}

An important use of mirrors in the BBM and in SSM's is to allow
signals to cross each other without interacting.  While signals can
also be made to cross by leaving regular gaps in signal streams and
delaying one signal stream relative to the other, this technique
requires the use of mirrors to insert compensating delays that
resynchronize streams.  If we're using streams of balls to act as
mirrors, we have a problem when these mirror streams have to cross
signals, or even each other.

We can deal with this problem by extending the non-interacting portion
of our dynamics.  In order to make our SSM's unconditionally digital,
we already require that balls pass through each other when too many
try to pile up in one place.  Thus it seems natural to also use the
presence of extra balls to force signals to cross.  The simplest way
to do this is to add a rest particle to the model---a particle that
doesn't move.  At a site ``marked'' by a rest particle, signals will
simply pass through each other.  This is mass and momentum conserving,
and is perfectly compatible with continuous classical mechanics.
Notice that we don't actually have to change our SSM collision rule to
include this extra non-interacting case, since we gave the rule in the
form, ``these cases interact, and in all other cases particles go
straight.''  Figure~\ref{fig.cross}a shows an example of two signal
paths crossing over a rest particle (indicated by a circle).

Figure~\ref{fig.cross}b shows an example of a signal crossover that
doesn't require a rest particle in the lattice gas version of the SSM.
Since LGA particles only interact at lattice sites, which are the
corners of the grid, two signals that cross as in this Figure cannot
interact.  Such a crossover occurs in Figure~\ref{fig.1mirrors}b, for
example.  Without the LGA lattice to indicate that no interaction can
take place at this site, this crossover would also require a rest
particle.  To keep the LGA and the continuous versions of the model
equivalent, we will consider a rest particle to be present implicitly
wherever signals cross between lattice sites.

\subsection{Spatially-efficient computation}

With the addition of rest particles to indicate signal crossover, we
can use the messy deflection of Figure~\ref{fig.1mirrors}a to build
reusable circuitry and so perform spatially-efficient computation.
The paths of the incoming ``mirror streams'' can cross whatever
signals are in their way to get to the point where they are needed,
and then the extra undesired ``garbage'' output streams can be led
away by allowing them to cross any signals that are in their way.
Since every mirror stream (which brings in energy but no information)
and every garbage stream (which carries away both energy and entropy)
crosses a surface that encloses the circuit, the number of such
streams that we can have is limited by the area of the enclosing
surface.  Meanwhile, the number of circuit elements (and hence also
the demand for mirror and garbage streams) grows as the volume of the
circuit\cite{conservative-logic,bennett-thermo,mpf}.  This is the
familiar surface to volume ratio problem that limits heat removal in
ordinary heat-generating physical systems: the rate of heat generation
is proportional to the volume, but the rate of heat removal is only
proportional to the surface area.  We have the same kind of problem if
we try to bring free energy (i.e., energy without information) into a
volume.

Using dual-rail signalling, we've seen that we have neat collisions
available that don't corrupt the deflecting mirror streams.  We do
not, however, avoid the surface to volume problem unless these clean
mirror-streams can be reused: otherwise each reflection involves
bringing in a mirror stream all the way from outside of the circuit,
using it once, and then sending the reflected mirror stream all the
way out of the circuit.  Thus if we can't reuse mirror streams, the
maximum number of circuit elements we can put into a volume of space
grows like the surface area rather than like the volume!  We will show
that (at least in 2D) mirror streams can be reused, and consequently
momentum conservation doesn't impair the spatial efficiency of
computations.

\subsection{Signal Routing}

\fig{mirror-shift}{%
\begin{array}{c@{\hspace{.7in}}c@{\hspace{.7in}}c}
\includegraphics[height=1.4in]{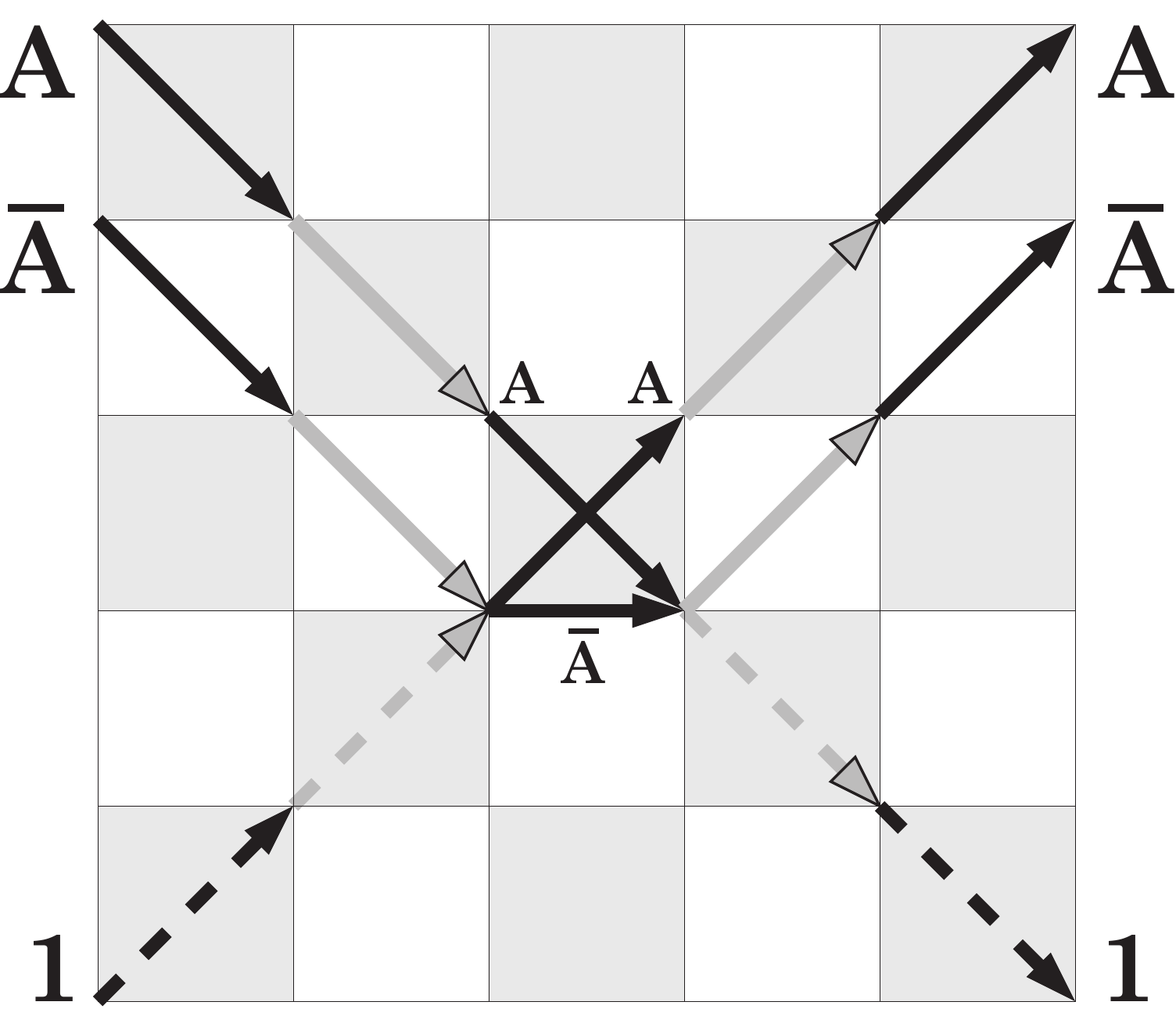} &
\includegraphics[height=1.4in]{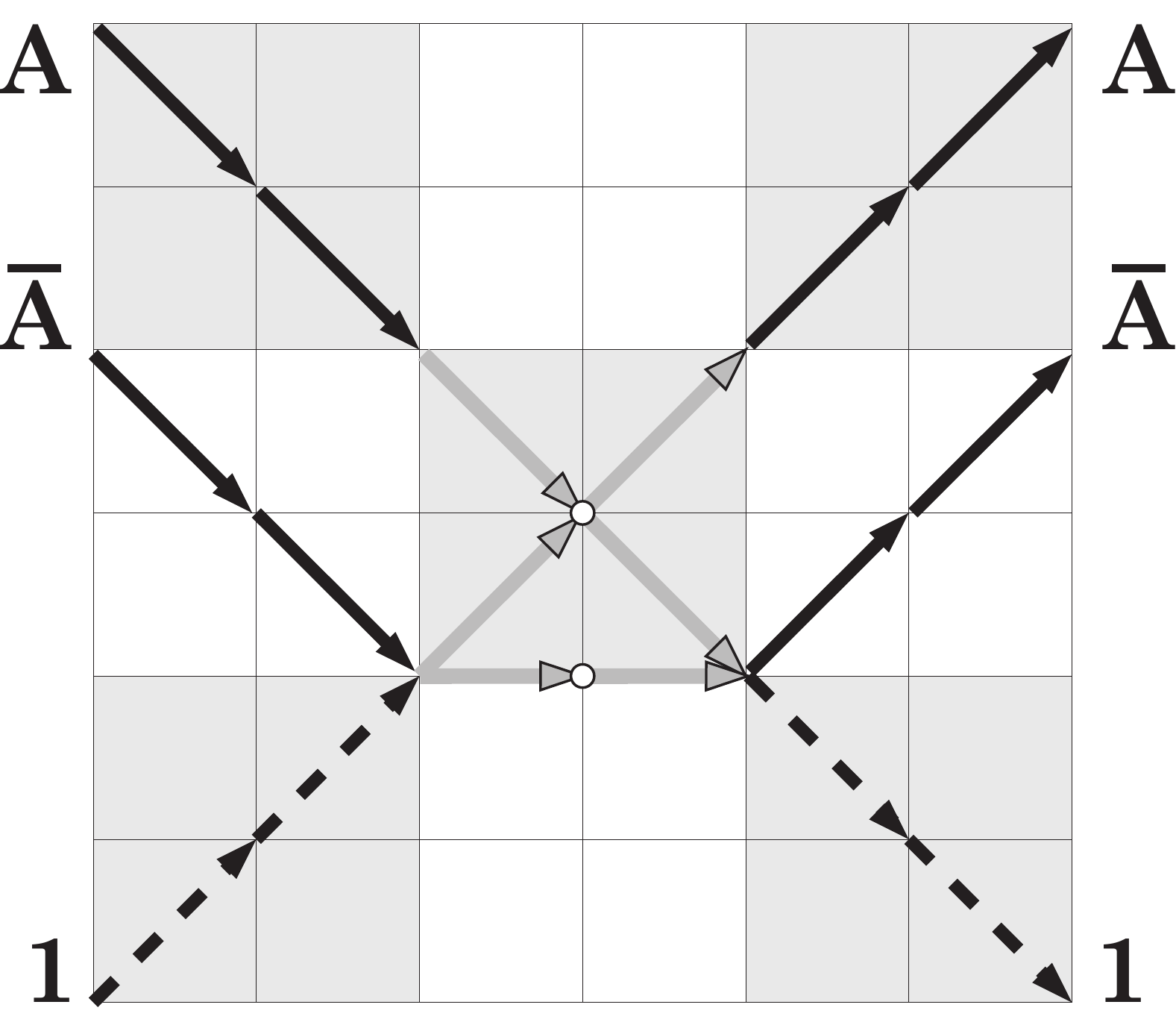} &
\includegraphics[height=1.4in]{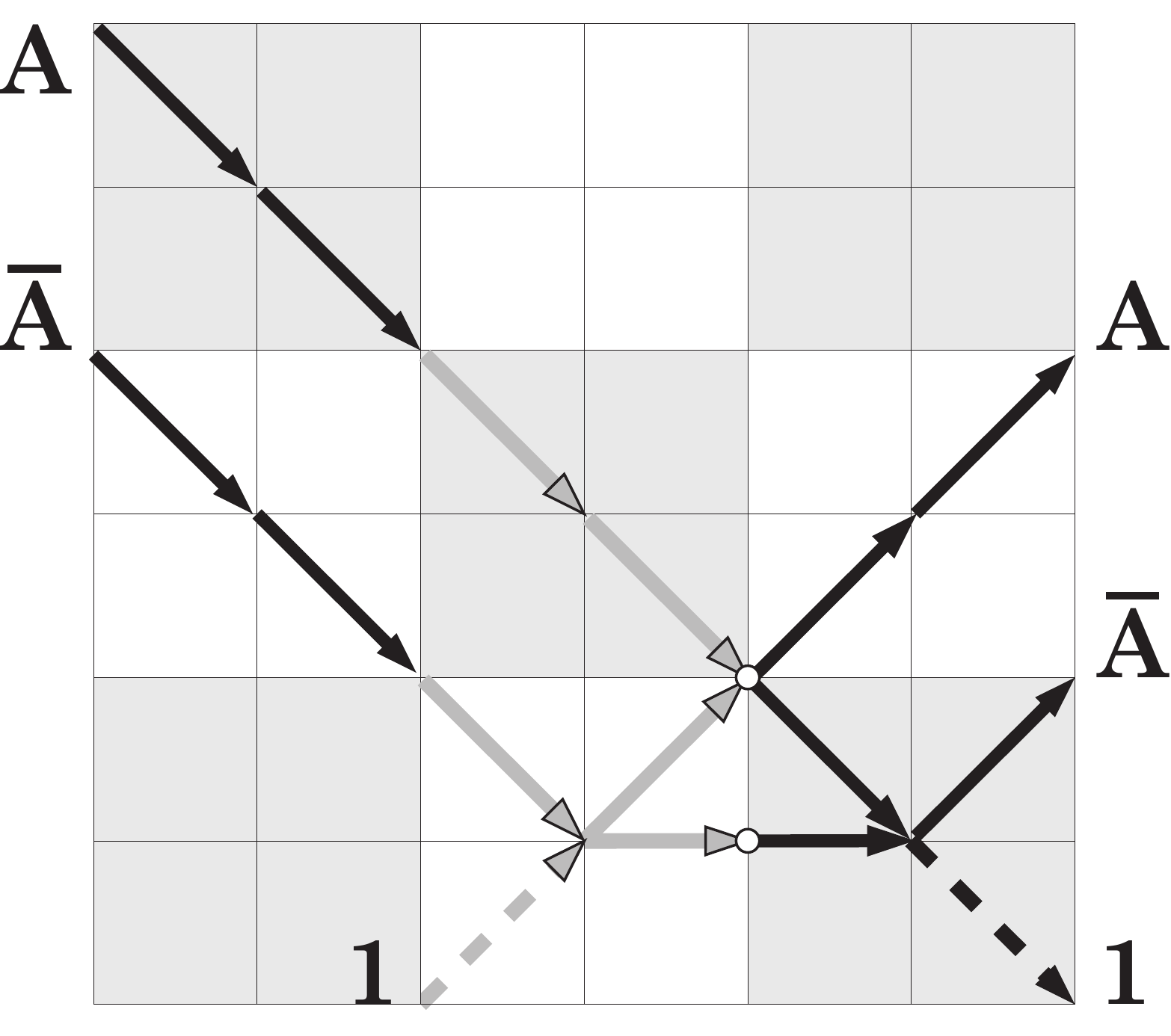} \\
\mbox{\bf (a)} & \mbox{\bf (b)} & \mbox{\bf (c)} \\
\end{array}}
{A signal routing constraint. (a)~When signal pairs are deflected by
a stream of 1's, each component of the pair remains on the same
checkerboard region of the space.  (b)~If we spread the signals so
that pairs are twice as far apart, we can also rescale the mirror
collision.  (c)~After rescaling, we can move the ``mirror'' to what
would have originally been a half-integer position, and so avoid this
constraint.}

Even though we can reflect dual-rail signals and make them cross, we
still have a problem with routing signals (actually two problems, but
we'll discuss the second problem when we confront it).
Figure~\ref{fig.mirror-shift}a illustrates a problem that stems from
not being able to reflect signals at half-integer locations.  Every
reflection leaves the top {\A} signal on the dark checkerboard we've
drawn---it can't connect to an input on the light checkerboard.  We
can fix this by rescaling the circuit, spreading all signals twice as
far apart (Figure~\ref{fig.mirror-shift}b).  Now the implicit
crossover in the middle of Figure~\ref{fig.mirror-shift}a must be made
explicit.  Notice also that the horizontal particle must be
stretched---it too goes straight in the presence of a rest particle.
Now we can move the reflection to a position that was formerly a
half-integer location (Figure~\ref{fig.mirror-shift}c), and the {\A}
signal is deflected onto the white checkerboard.

\subsection{Dual-rail logic}

\fig{switch}{%
\begin{array}{c@{\hspace{.7in}}c@{\hspace{.7in}}c}
\includegraphics[height=1.5in]{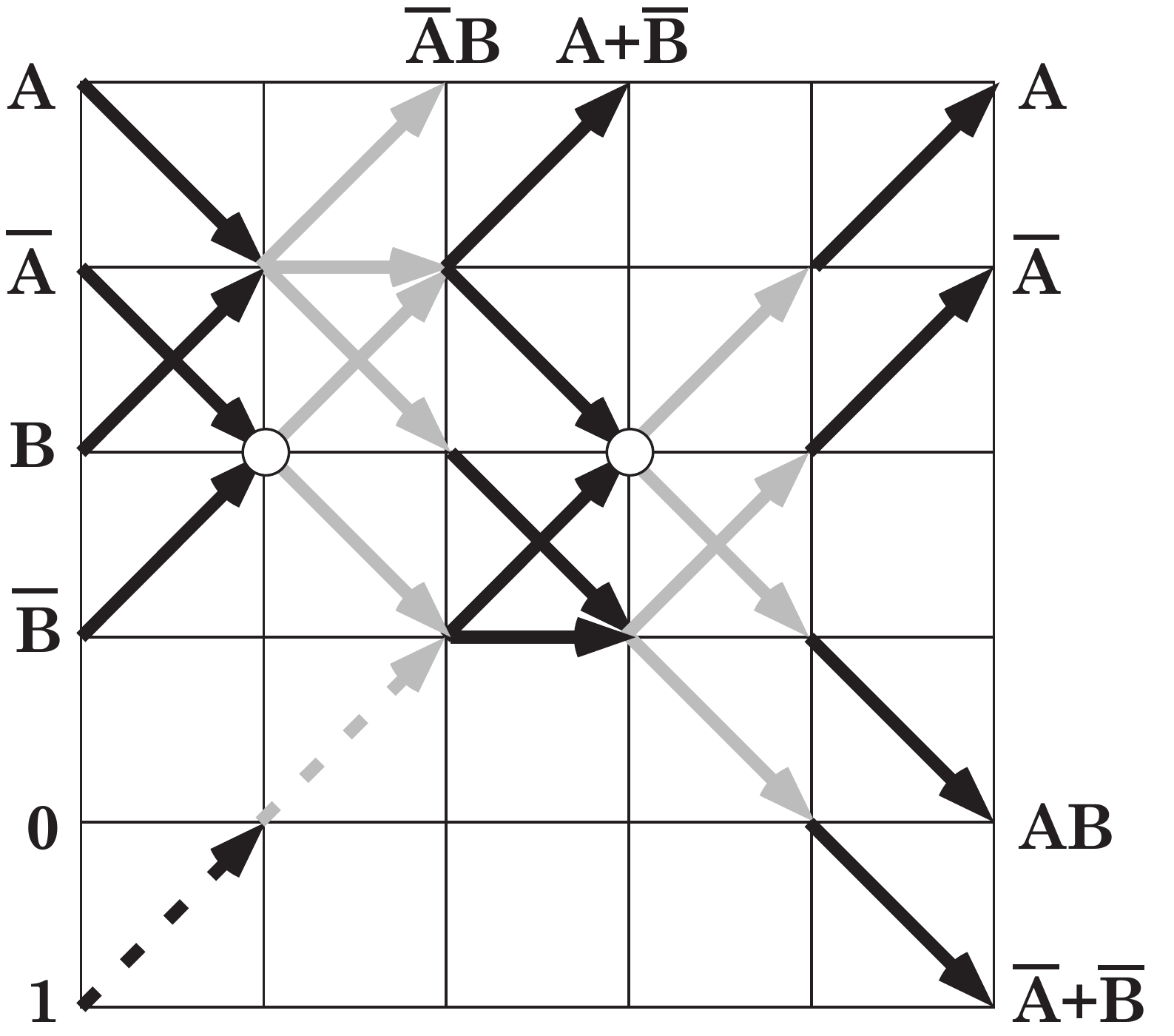} &
\includegraphics[height=1.5in]{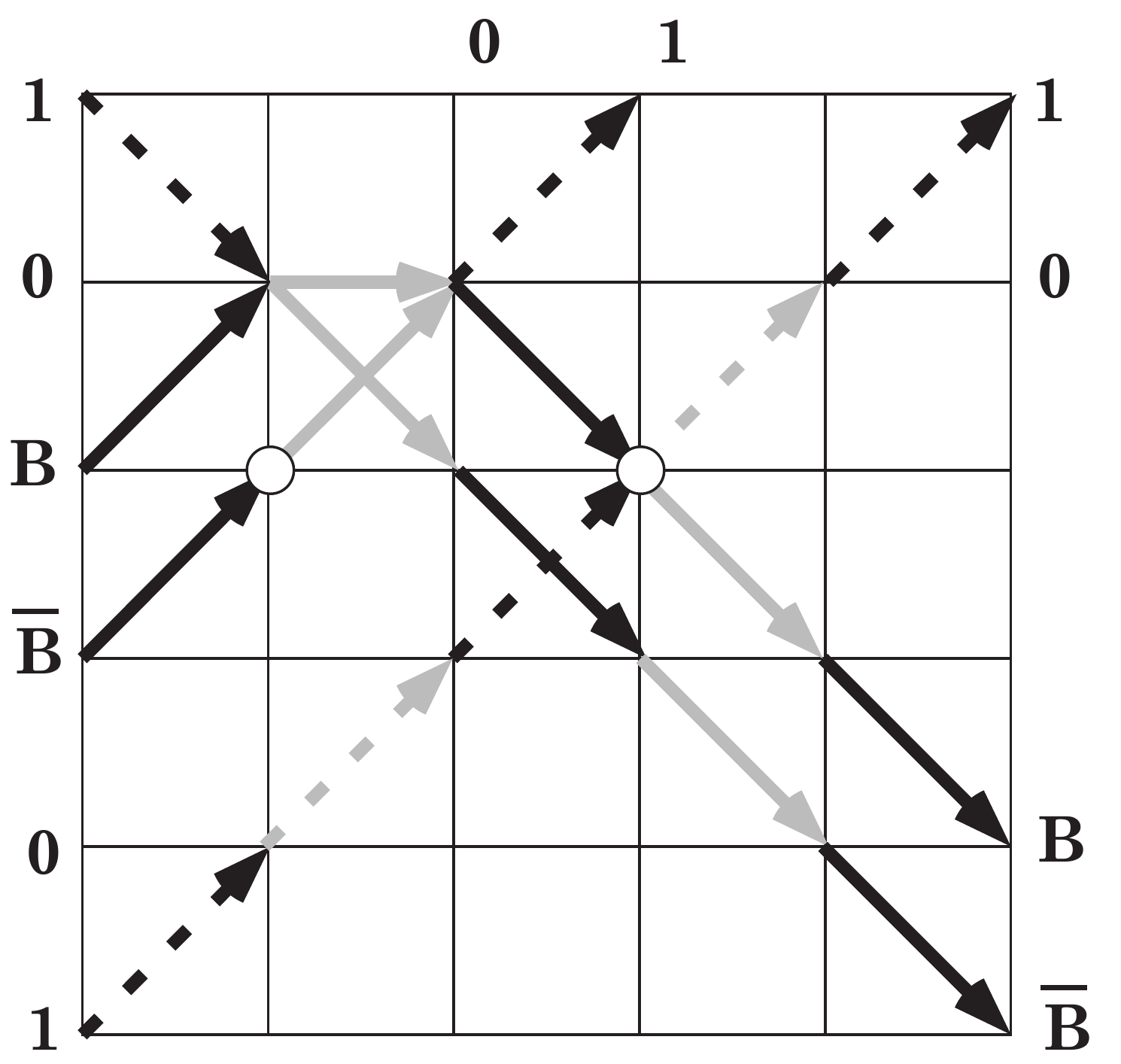} &
\includegraphics[height=1.5in]{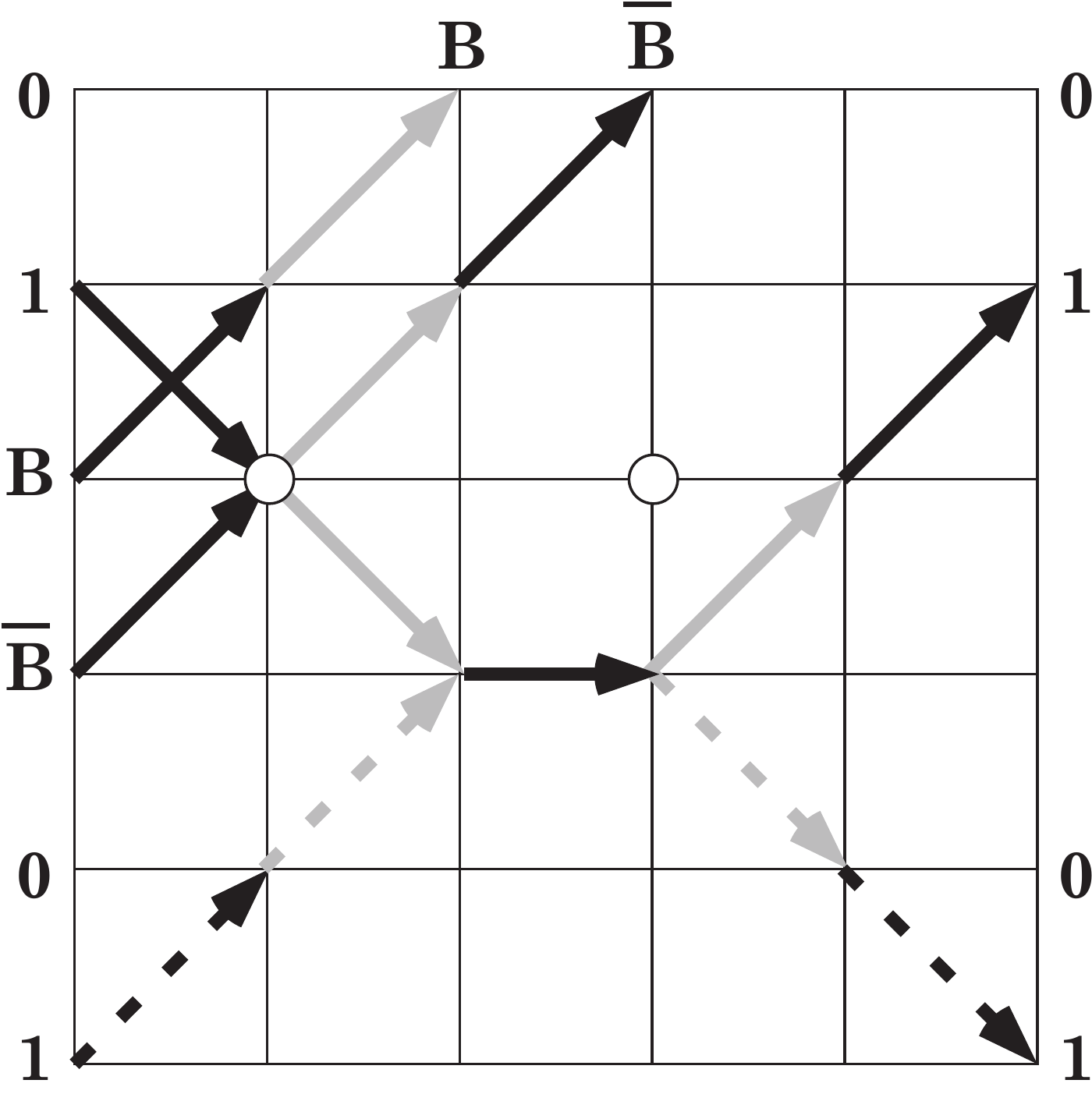} \\
\mbox{\bf (a)} & \mbox{\bf (b)} & \mbox{\bf (c)} \\
\end{array}}
{A switch gate using dual-rail signalling.  (a)~The general case.  The
{\A} signal either deflects {\B} and {\Bbar} or not, doing most of the
work.  We've highlighted the constant stream of ones by using dotted
lines.  (b)~The case {\A=1}. {\B} and {\Bbar} are reflected down, and
the one is reflected the opposite way.  (c)~The case {\A=0}.  There is
no interaction with {\B} or {\Bbar}, and they go straight.}

We've seen that dual-rail signals can be cleanly routed.  In order to
use such signals for computation, we need to be able to build logic
with dual-rail inputs and outputs.  We will now see that if we let two
dual-rail signals collide, we can form a
switch-gate\cite{conservative-logic}, as shown in
Figure~\ref{fig.switch}a.  The switch gate is a universal logic
element that leaves the control input {\A} unchanged, and routes the
controlled input {\B} to one of two places, depending on the value of
{\A}.  Since each dual rail signal contains a 1, and since all
collisions conserve the number of 1's, all dual-rail logic gates need
an equal number of inputs and outputs.  Thus our three output
switch-gate needs an extra input which is a dual-rail constant of 0.

The switch gate (Figure~\ref{fig.switch}a) is based on a reflection of
the type shown in Figure~\ref{fig.1mirrors}b.  If \A=1
(Figure~\ref{fig.switch}b), the {\B} and {\Bbar} pair are reflected
downward; if \A=0 there is no reflection and they go straight.  The
{\Abar} signal reflects off the constant-one input as in
Figure~\ref{fig.1mirrors}a, to regenerate the {\A} and {\Abar}
outputs.  Notice that if a rest particle were added in
Figure~\ref{fig.switch}a at the intersection of the {\A} and {\B}
signals, the switch would be stuck in the {\em off} position: {\B} and
{\Bbar} would always go straight through, and {\A} and {\Abar} would
get reflected by the constant-one, and come out in their normal
position.

\subsection{A Fredkin Gate}

In order to see that momentum conservation doesn't impair the spatial
efficiency of SSM computation, we first illustrate the issues involved
by showing how mirror streams can be reused in an array of Fredkin
gates\cite{conservative-logic}.

\fig{fgate}{%
\begin{array}{c@{\hspace{.6in}}c}
\includegraphics[height=2.5in]{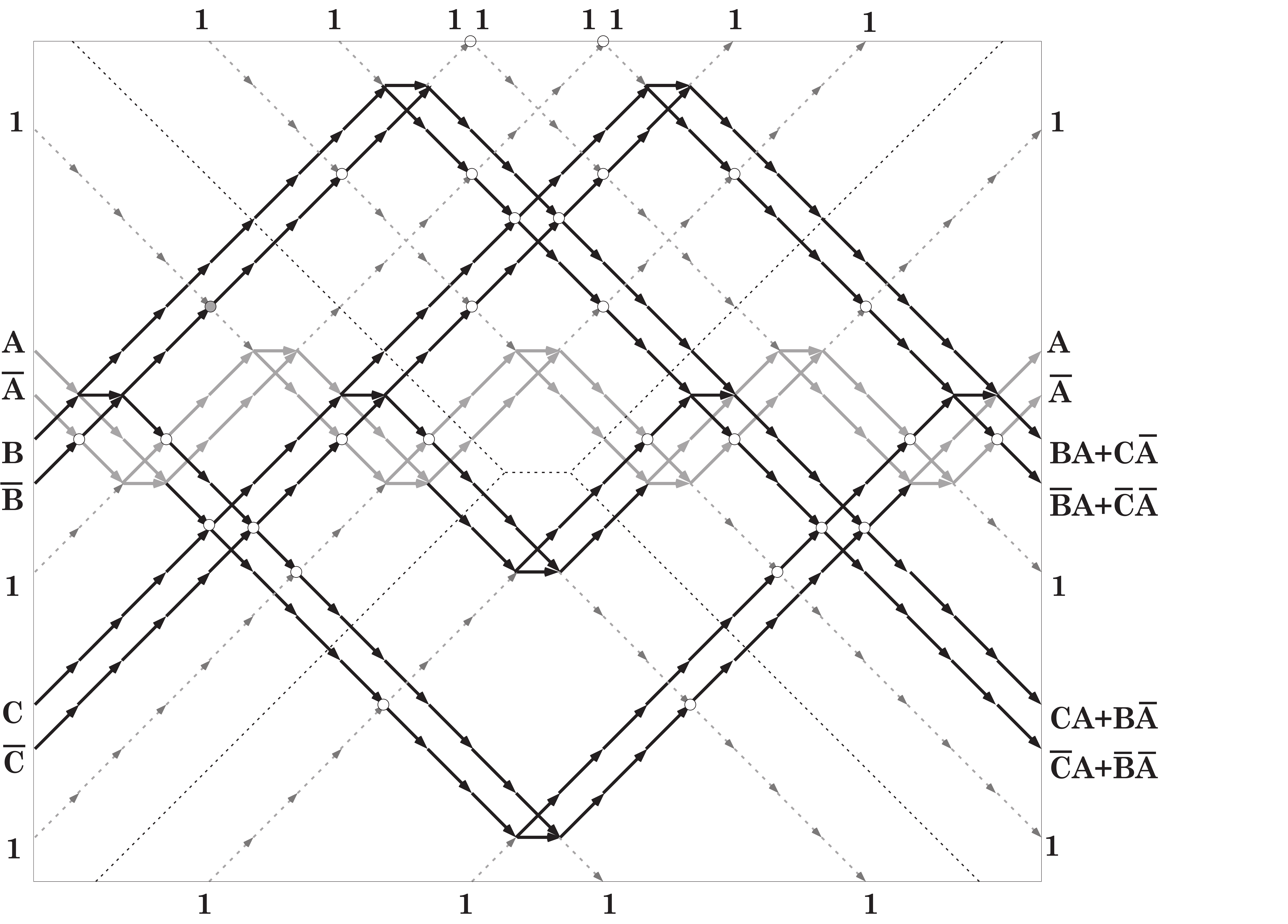} &
\includegraphics[height=2.5in]{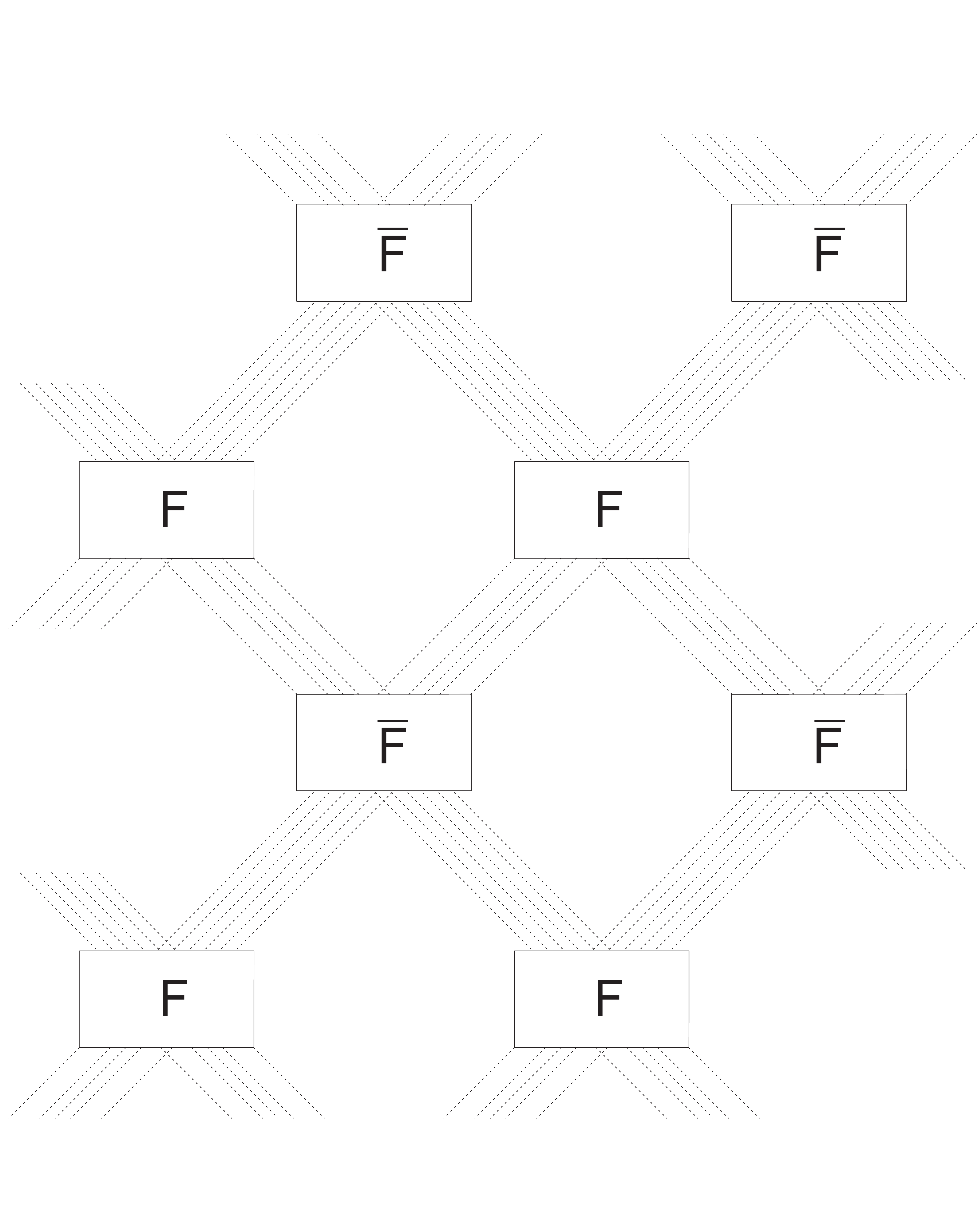} \\
\mbox{\bf (a)} & \mbox{\bf (b)} \\
\end{array}}
{(a)~A Fredkin gate.  We construct a Fredkin gate out of four switch
gates, two used forward, and two backward.  Constant 1's are drawn in
lightly using dotted arrows.  The path of the control signal {\bf A}
is shown in solid gray.  If we added constant streams of 1's along the
four paths drawn as dotted lines without arrows, then the constant
streams would be symmetrical about diagonal axes.  (b)~Because of the
diagonal symmetry of this Fredkin gate construction, we can make an
array of them, as indicated here, and reuse the constant streams of
1's that act as signal mirrors.  The upside down Fredkin gates are
also Fredkin gates, but with the sense of the control inverted.}

A Fredkin gate has three inputs and three outputs.  The {\A} input,
called the control, appears unchanged as the {\A} output.  The other
two inputs either appear unchanged at corresponding outputs (if \A=1),
or appear interchanged at the corresponding outputs (if \A=0).  We
construct a Fredkin gate out of four switch gates, as shown in
Figure~\ref{fig.fgate}a.  The first two switch gates are used
forward, the last two switch gates are used backward (i.e., flipped
about a vertical axis).  The control input {\A} is colored in solid
gray, and we see it wend its way through the four switch gates.
Constant 1's are shown using dotted gray arrows.  In the case \A=0,
all four switch gates pass their controlled signals straight through,
and so {\B} and {\C} interchange positions in the output.  In the case
\A=1, all four switch gates deflect their controlled signals, and so
{\B} and {\C} come out in the same positions they went in.

Now notice the bilateral symmetry of the Fredkin gate implementation.
We can make use of this symmetry in constructing an array of Fredkin
gates that reuse the constant 1 signals.  If we add an extra stream of
constant 1's along the four paths drawn as arrowless dotted lines
(making these lie on the lattice involves rescaling the circuit), then
the set of constant streams coming in or leaving along each of the
four diagonal directions is symmetric about some axis.  This means
that we can make a regular array of Fredkin gates and upside-down
Fredkin gates, as is indicated in Figure~\ref{fig.fgate}b, with the
constants all lining up.  These constants are passed back and forth
between adjacent Fredkin gates, and so don't have to be supplied from
outside of the array.  Since an upside-down Fredkin gate is still a
Fredkin gate, but with the sense of the control inverted, we have
shown that constant streams of ones can be reused in a regular array
of logic.

We still have not routed the inputs and outputs to the Fredkin gates,
and so we have another set of associated mirror-streams that need to
be reused.  The obvious approach is to create a regular pattern of
interconnection, thus allowing us to again solve the problem globally
by solving it locally.  But a regular pattern of interconnected logic
elements that can implement universal computation is just a universal
CA: we should simply implement a universal CA that doesn't have
momentum conservation!

\subsection{Implementing the BBMCA}

\fig{bbmca-array}{%
\begin{array}{c@{\hspace{1in}}c}
\includegraphics[height=1.5in]{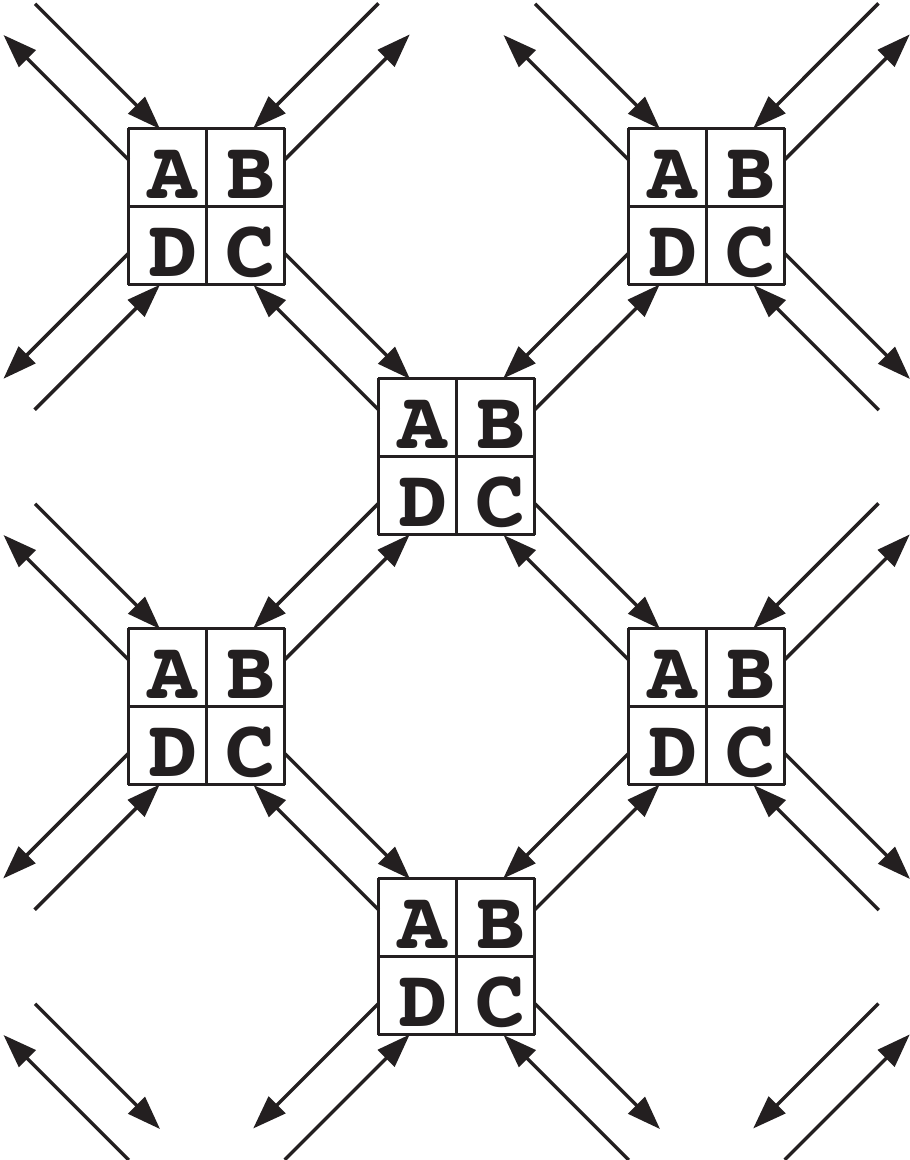} &
\includegraphics[height=1.5in]{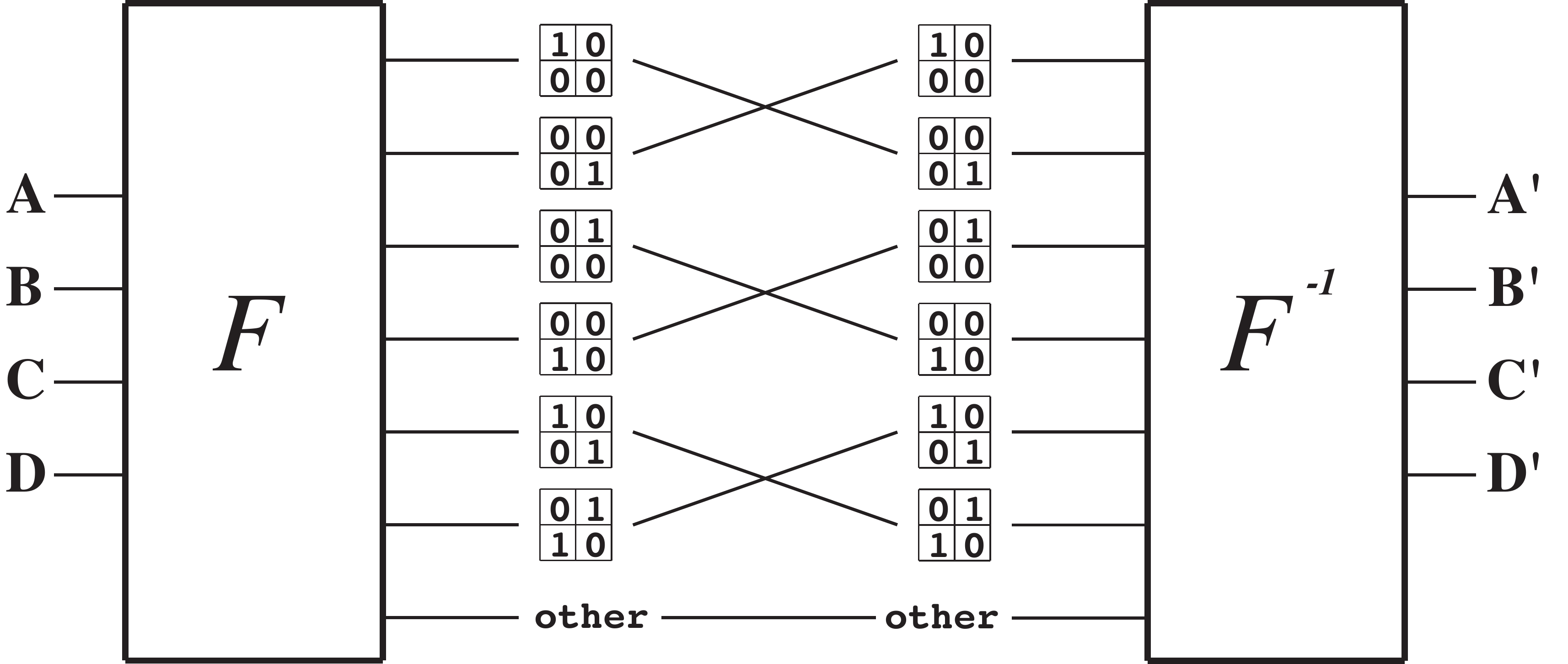} \\
\mbox{\bf (a)} & \mbox{\bf (b)} \\
\end{array}}
{Emulating the BBMCA using an SSM.  (a)~The BBMCA can be implemented
as a 2D array of identical blocks of logic, each of which processes
four bits at a time.  The bits that are grouped together in one step
go to four different diagonally adjacent blocks in the next step.
(b)~We construct a circuit out of switch gates to implement the BBMCA
logic block.  The first half of the circuit ($F$) produces a set of
outputs that are each one only if the four BBMCA bits have some
particular pattern.  The second half ($F^{-1}$) is the mirror image of
the first.  In between, the cases that interchange are wired to each
other.}

The BBMCA is a simple reversible CA based on the BBM, with fixed
mir\-rors\cite{bbmca,marg-thesis,cc}.  It can be implemented as a
regular array of identical logic blocks, each of which takes four bits
of input, and produces four bits of output
(Figure~\ref{fig.bbmca-array}a).  Each logic block exchanges one bit
of data with each of the four blocks that are diagonally adjacent.
The four bits of input can be thought of as a pattern of data in a
2$\times$2 region of the lattice, and the four outputs are the next
state for this region.  According to the BBMCA rule, certain patterns
are turned into each other, while other patterns are left unchanged.
This rule can be implemented by a small number of switch gates, as is
indicated schematically in Figure~\ref{fig.bbmca-array}b.  We first
implement a demultiplexer $F$, which produces a value of 1 at a given
output if and only if a corresponding 2$\times$2 pattern appears in
the inputs.  Patterns that don't change under the BBMCA dynamics only
produce 1's in the outputs labeled ``other.''  The demultiplexer is a
combinational circuit (i.e., one without feedback).  The inverse
circuit $F^{-1}$ is simply the mirror image of $F$, obtained by
reflecting $F$ about a vertical axis.  In between $F$ and $F^{-1}$ we
wire together the cases that need to interchange.  This gives us a
bilaterally symmetric circuit which implements the BBMCA logic block
in the same manner that our circuit of Figure~\ref{fig.fgate}a
implemented the Fredkin gate.  Note that the overall circuit is its
own inverse, as any bilaterally symmetric combinational SSM circuit
must be.

\fig{bbmca-symm}{%
\begin{array}{c@{\hspace{.5in}}c@{\hspace{.5in}}c}
\includegraphics[height=1.4in]{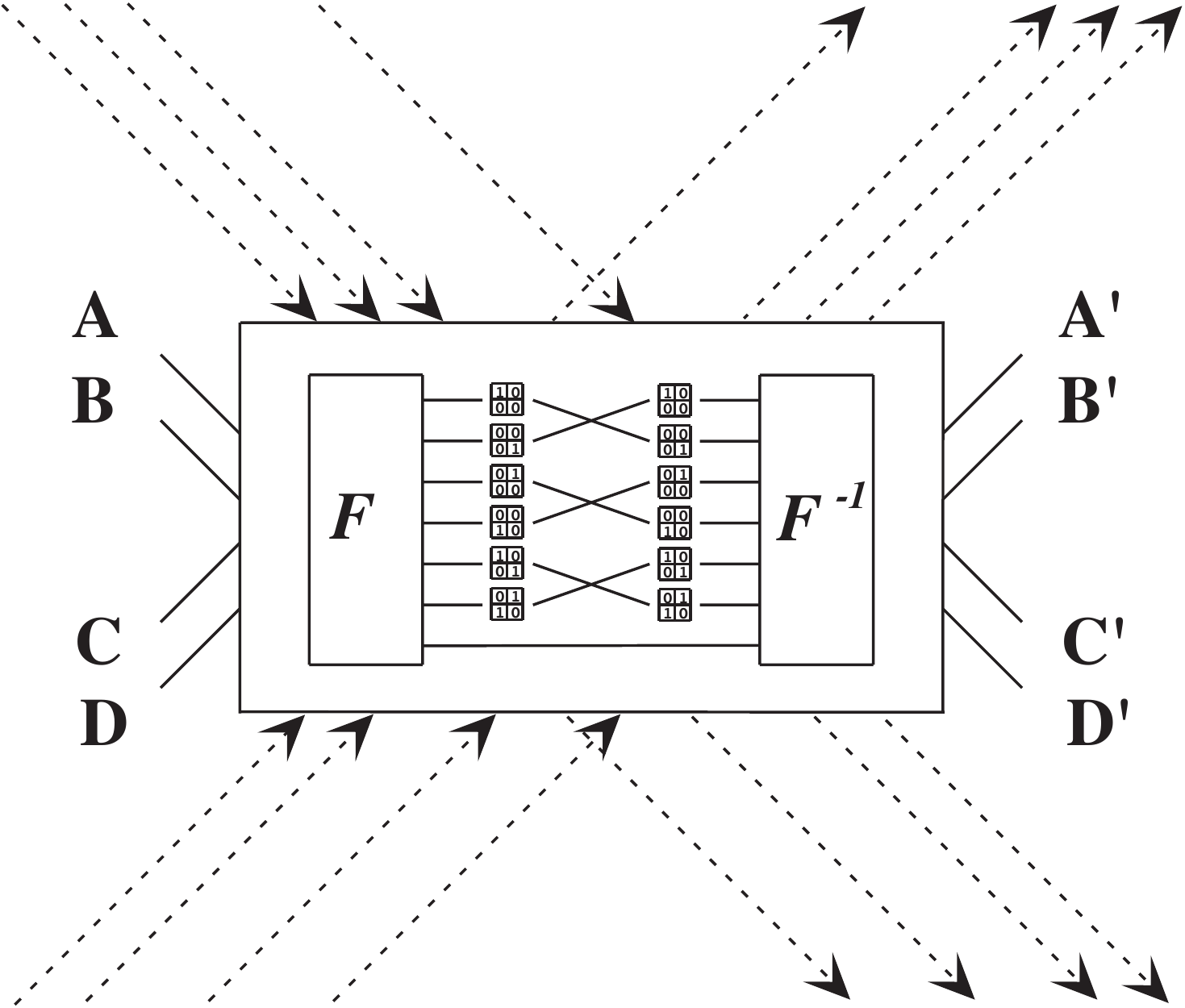} &
\includegraphics[height=1.1in]{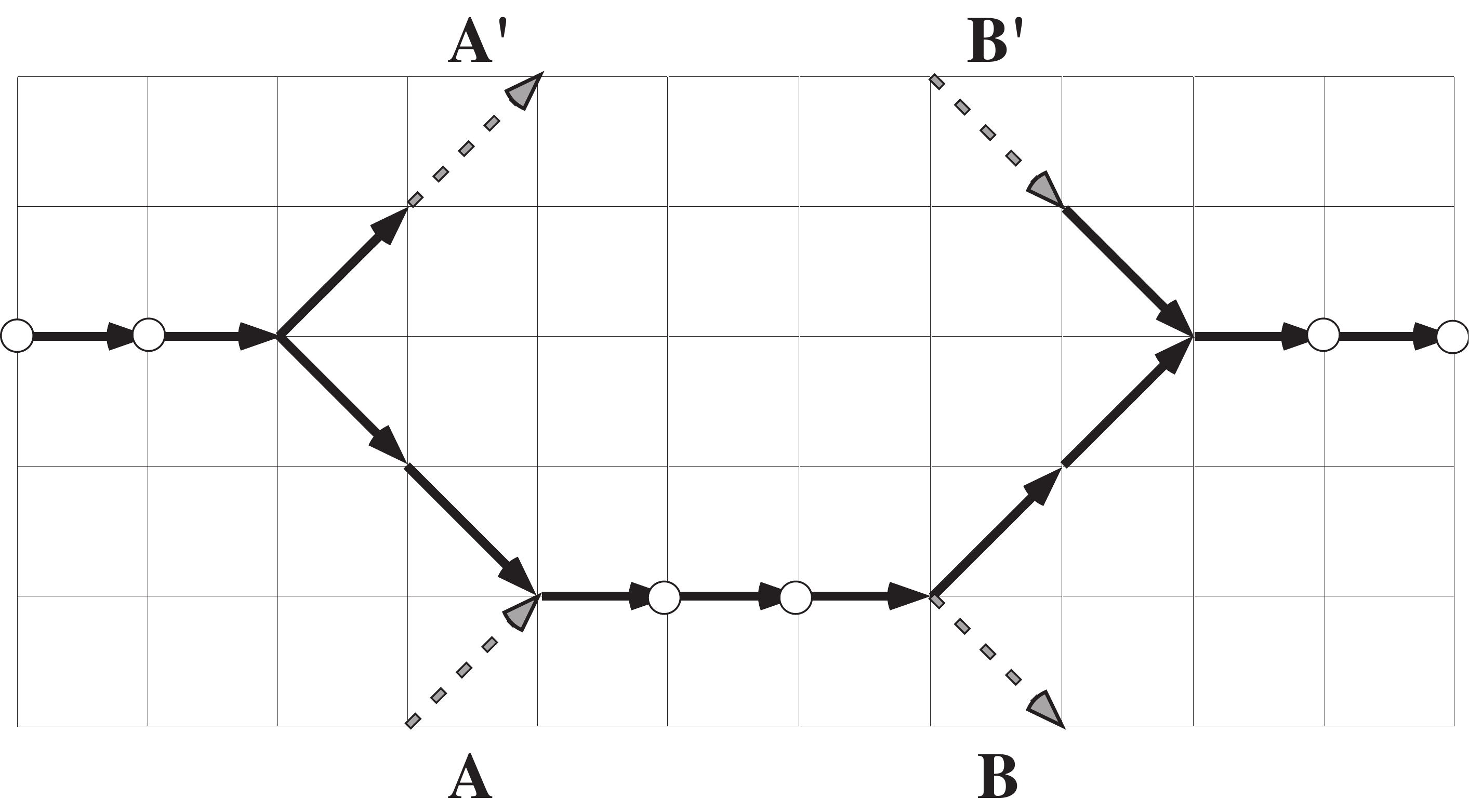} &
\includegraphics[height=1.4in]{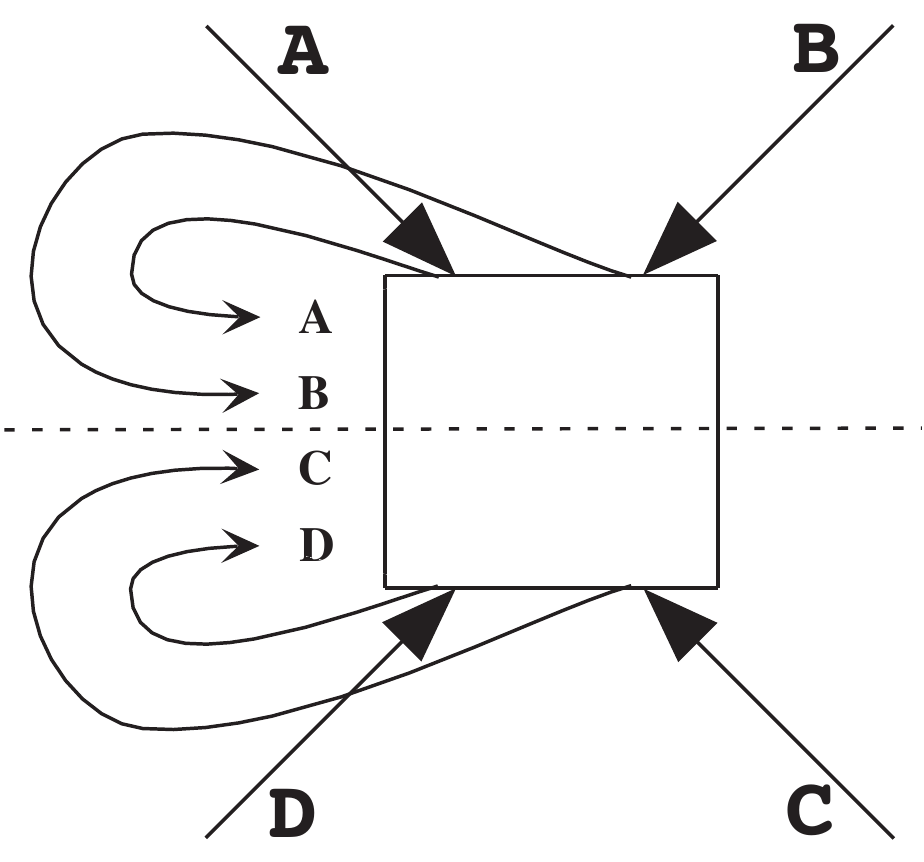} \\
\mbox{\bf (a)} & \mbox{\bf (b)} & \mbox{\bf (c)} \\
\end{array}}
{Symmetrizing signal paths so that adjacent BBMCA logic blocks can
share their mirror constants.  (a)~The BBMCA block circuit is
bilaterally symmetric, with an equal number of constants flowing in or
out along each of the four diagonal directions.  (b)~Symmetric pairs
of constant 1's can be shifted vertically in order to align the
``mirror streams'' so that the blocks can be arrayed.  (c)~The wiring
of the four BBMCA signal inputs (and outputs) to each block is also
bilaterally symmetric, so the same alignment techniques should apply.}

Now we would like to connect these logic blocks in a uniform array.
We will first consider the issue of sharing the mirror streams
associated with the individual logic blocks, and then the issue of
sharing the mirror streams associated with interconnecting the four
inputs and outputs.  In Figure~\ref{fig.bbmca-symm}a we see a
schematic representation of our BBMCA block.  It is a combinational
circuit, with signals flowing from left to right.  The number of
signal streams flowing in along one diagonal direction is equal to the
number flowing out along the same direction---this is true overall
because it's true of every collision!  In particular, since the four
inputs and outputs are already matched in the diagram, the mirror
streams must also be matched---there are an equal number of streams of
constant 1's coming in and out along each direction.  The input
streams will not, however, in general be aligned with the output
streams.  If we can align these, then we can make a regular array of
these blocks, with mirror-stream outputs of one connected to the
mirror-stream inputs of the next.

In Figure~\ref{fig.bbmca-symm}b we show how to align streams of ones.
Due to the bilateral symmetry of the BBMCA circuit, every incoming
stream that we would like to shift up or down on one side is matched
by an outgoing stream that needs to be shifted identically on the
other side.  Thus we will shift streams in pairs.  To understand the
diagram, suppose that {\A} and {\B} are constant streams of ones, with
{\B} going into a circuit below the diagram, and {\A} coming out of
it.  Now suppose that we would like to raise {\A} and {\B} to the
positions labeled {\Ap} and {\Bp}.  If a constant stream of horizontal
particles is provided midway in between the two vertical positions,
then we can accomplish this as shown.  The constant horizontal stream
splits at the first position without a rest particle.  It provides the
shifted {\Ap} signal, and a matching stream of ones collides with the
original {\A} signal.  The resulting horizontal stream is routed
straight across until it reaches {\B}, where an incoming stream of
ones is needed.  Here it splits, with the extra stream of ones
colliding with the incoming {\Bp} signal to restore the original
horizontal stream of ones, which can be reused in the next block of
the array of circuit blocks to perform the same function.  The net
effect is that the mirror streams {\A} and {\B} coming out of and into
a circuit have been replaced with new streams that are shifted
vertically.  By reserving some fraction of the horizontal channels for
horizontal constants that stream across the whole array, and reserving
some channels for horizontal constants that connect pairs of streams
being raised, we can adjust the positions of the mirror streams as
needed.  Note that a mirror pair can be raised by several smaller
shifts rather than just one large shift, in case there are conflicts
in the use of horizontal constants.  Exactly the same arrangement can
be used to lower {\Ap} and {\Bp} going into and out of a circuit above
the diagram.  If we flip the diagram over, we see how to handle pairs
of streams going in the opposite directions.

Now we note that the wiring of the four signal inputs and outputs in
our BBMCA array also has bilateral symmetry, about a horizontal axis
(Figure~\ref{fig.bbmca-symm}c).  Thus it seems that we should be able
to apply the same technique to align the mirror streams associated
with this routing, in order to complete our construction.  But there
is a problem.

\subsection{Signal routing revisited}

\fig{problem-delay}{%
\begin{array}{c@{\hspace{.8in}}c@{\hspace{.8in}}c}
\includegraphics[height=1.6in]{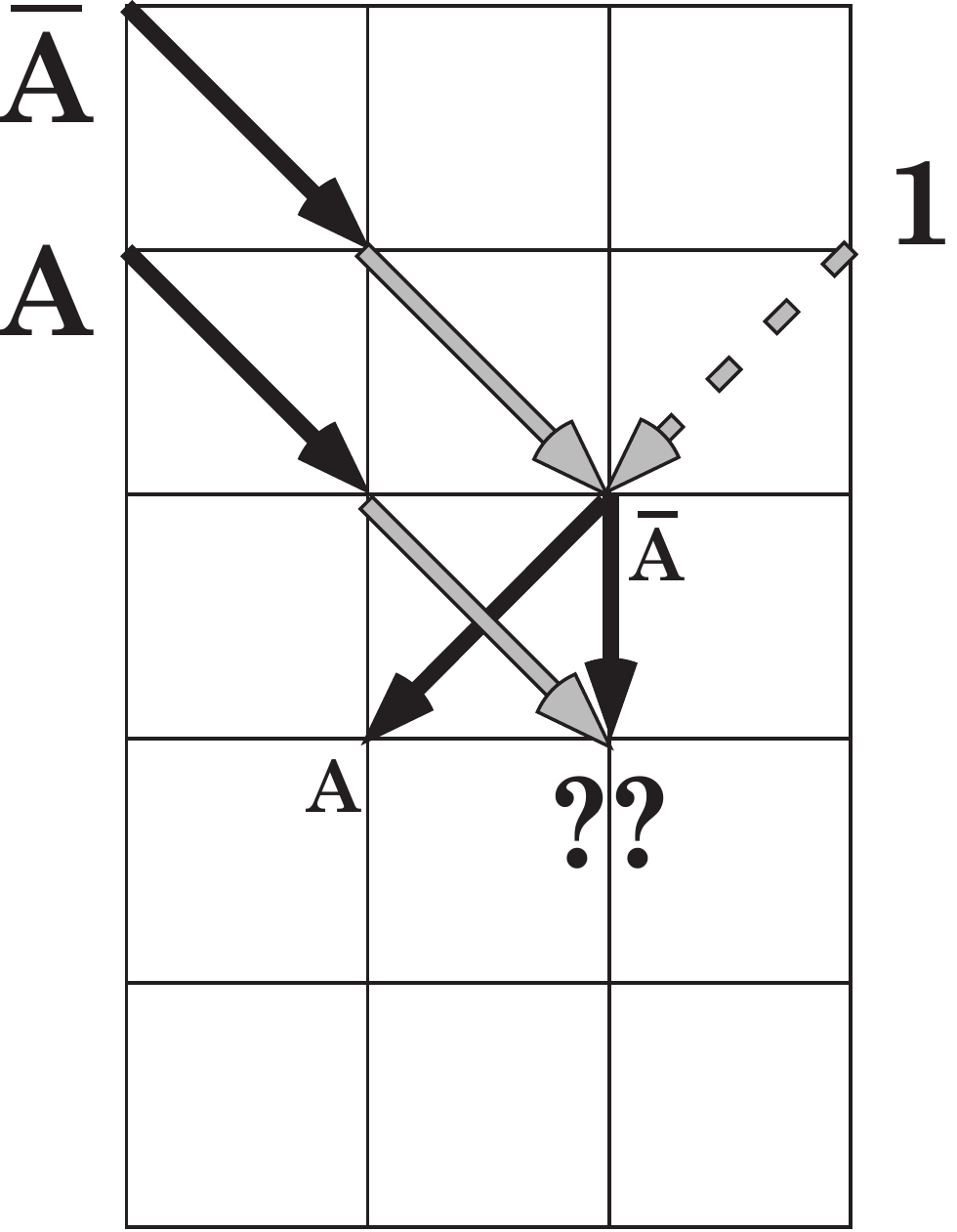} &
\includegraphics[height=1.6in]{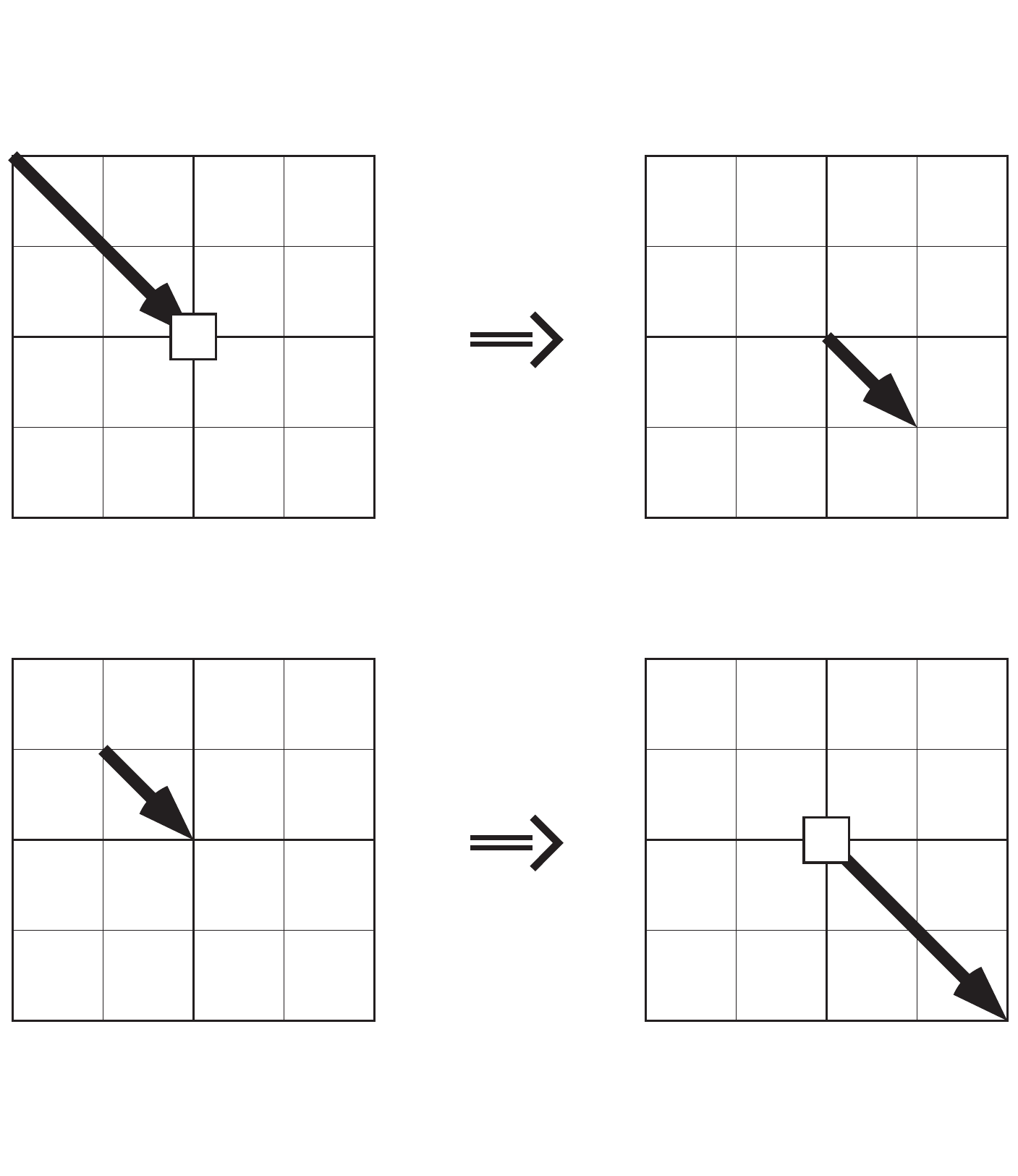} &
\includegraphics[height=1.6in]{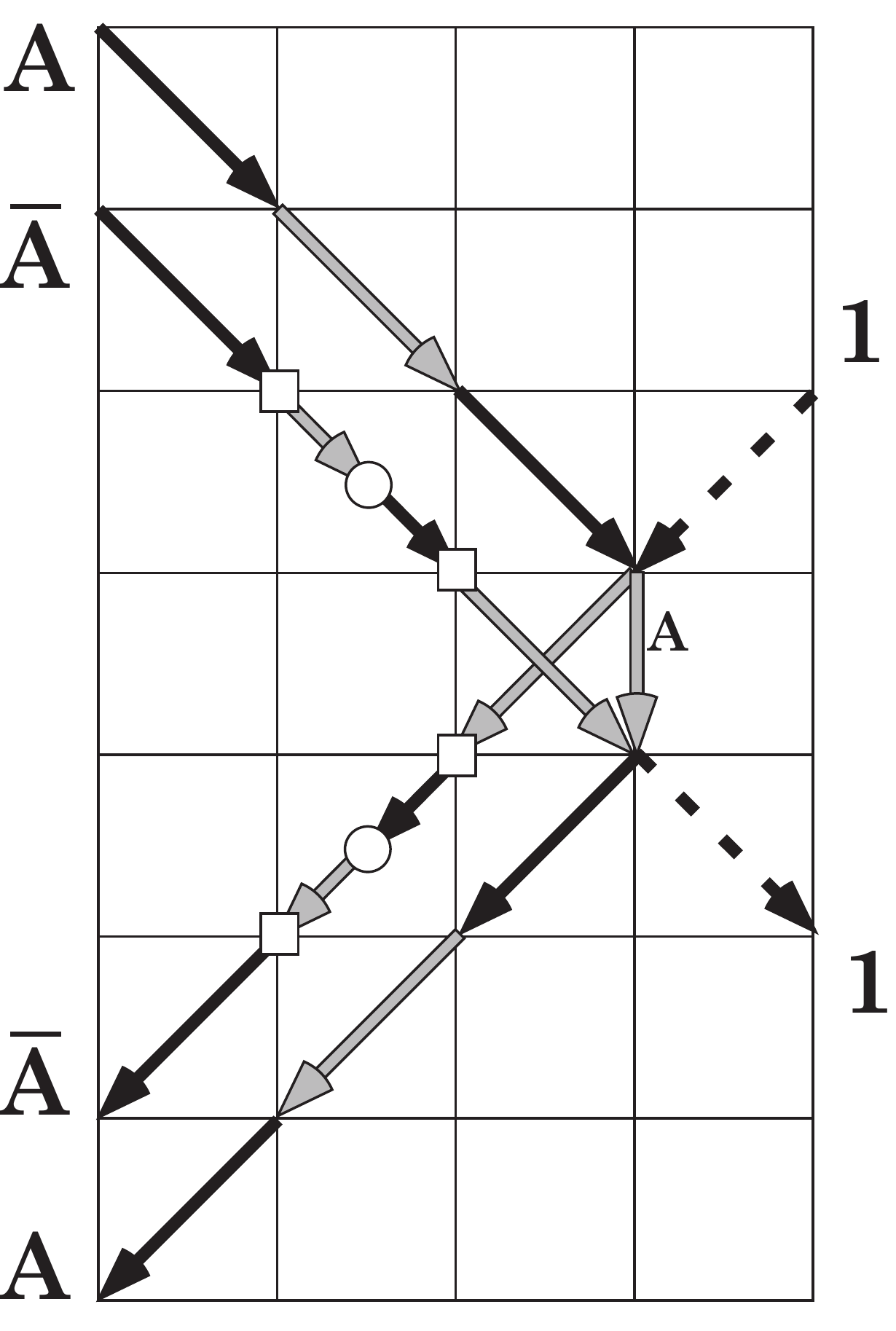} \\
\mbox{\bf (a)} & \mbox{\bf (b)} & \mbox{\bf (c)} \\
\end{array}}
{Reflecting signals back.  (a)~Dual rail pairs that are synchronized
by column only reflect correctly off ``horizontal mirrors.''  If we
try to bounce them off a ``vertical mirror,'' signals from different
times interact.  (b)~This can be fixed by providing a way to slow down
a signal.  The rule for adding slower particles involves refining the
lattice to admit a half-speed double-mass diagonal particle, and
adding a single-mass rest particle (square block in the diagram).
When a soft sphere collides with an equally massive sphere at rest,
the first slows down as the second speeds up (giving a net speed of
1/2), and then the sphere that was at rest proceeds.  (c)~Using our
``no-interaction'' rest particle to extend the lifetime of the
half-speed diagonal particle, we can change a column-synchronized pair
into a row-synchronized pair, and then back.}

So far, we have only constructed circuits without feedback---all
signal flow has been left to right.  Because of the $90^\circ$
rotational symmetry of the SSM, we might expect that feedback isn't a
problem.  When we decided to use dual-rail signalling, however, we
broke this symmetry.  The timing of the dual rail signal pairs is
aligned vertically and not horizontally.  In
Figure~\ref{fig.problem-delay}a, we see the problem that we encounter
when we try to reflect a right-moving signal back to the left.  A
signal that passed the input position labeled {\A} at an even
time-step collides with an unrelated signal that passed input {\Abar}
at an odd time-step.  These two signals need to be complements of each
other in order to reconstitute the reflecting mirror stream.  Thus we
only know how to reflect signals vertically, not horizontally!

We will discuss two ways of fixing this problem.  Both involve using
additional collisions in the SSM.  The first method we describe is
more complicated, since it adds additional particles and velocities to
the model, but is more obvious.  The idea is that we can resynchronize
dual-rail pairs by delaying one signal.  We do this by introducing an
interacting rest particle (distinct from our previously introduced
non-interacting rest particle) with the same mass as our
diagonally-moving particles.  The picture we have in mind is that if
there is an interacting rest particle in the path of a
diagonally-moving particle, then we can have a collision in which the
moving particle ends up stationary, and the stationary particle ends
up moving.  During the finite interval while the particles are
colliding, the mass is doubled and so (from momentum conservation) the
velocity is halved.  By picking the head-on impact interval
appropriately, the new stationary particle can be deposited on a
lattice site, so that the model remains digital.  This is illustrated
in Figure~\ref{fig.problem-delay}b.  Here the square block indicates
the interacting rest particle.  This is picked up in a collision with
a diagonal-moving particle to produce a half-speed double-mass
particle indicated by the short arrow.  Note that adding this delay
collision requires us to make our lattice twice as fine to accommodate
the slower diagonal particles.  It adds five new particle-states to
our model (four directions for the slow particle, and an interacting
rest particle).  The model remains, however, physically ``realistic''
and momentum conserving.

Figure~\ref{fig.problem-delay}c illustrates the use of this delay to
reflect a rightgoing signal leftwards.  We insert a delay in the
{\Abar} path both before the mirror-stream collision and afterward, in
order to turn the plane of synchronization $180^\circ$, turning it
$90^\circ$ at a time.  Notice that we use a non-interacting rest
particle (round) to extend the lifetime of the half-speed diagonal
particle.

In addition to complicating the model, this delay technique adds an
extra complication to showing that momentum conservation doesn't
impair spatial efficiency.  Signals are delayed by picking up and
later depositing a rest particle.  In order to reuse circuitry, we
must include a mechanism for putting the rest particle back where it
started before the next signal comes through.  Since we can pick this
particle up from any direction, this should be possible by using a
succession of constant streams coming from various directions, but
these streams must also be reused.  We won't try to show that this can
be done here---we will pursue an easier course in the next Section.

It would be simpler if the moving particle was deposited at the same
position that the particle it hit came from, so that no cleanup was
needed.  Unfortunately, this necessarily results in no delay.  Since
the velocity of the center of mass of the two particles is constant,
if we end up with a stationary particle where we started, the other
particle must exactly take the place of the one that stopped.

\subsection{A simpler extension}

\fig{extra-collisions}{%
\begin{array}{c@{\hspace{.8in}}c@{\hspace{.8in}}c}
\includegraphics[height=.5in]{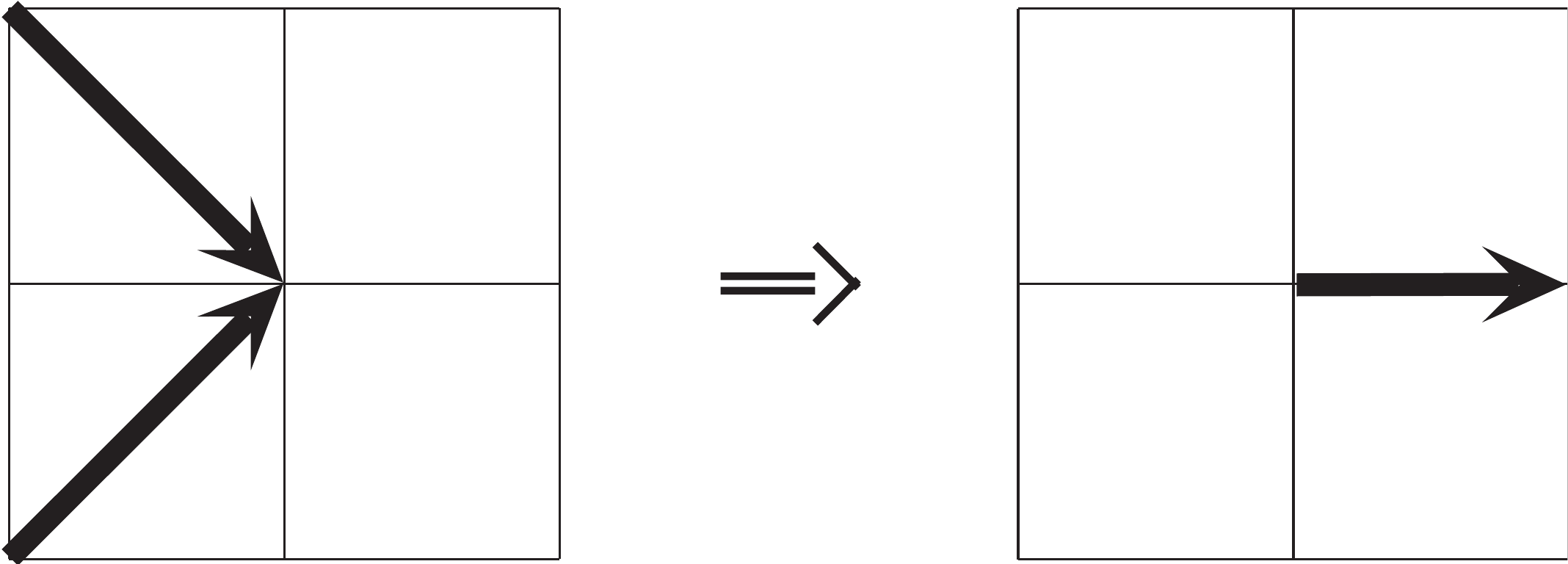} &
\includegraphics[height=.5in]{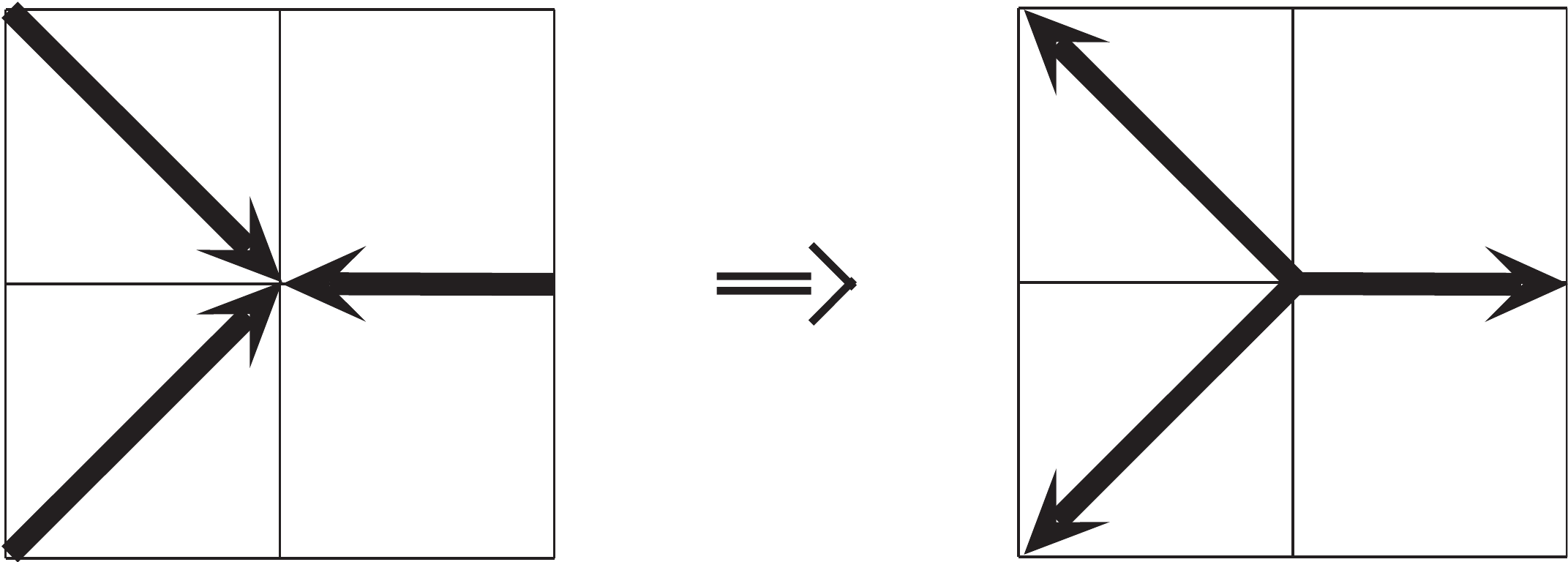} &
\includegraphics[height=.5in]{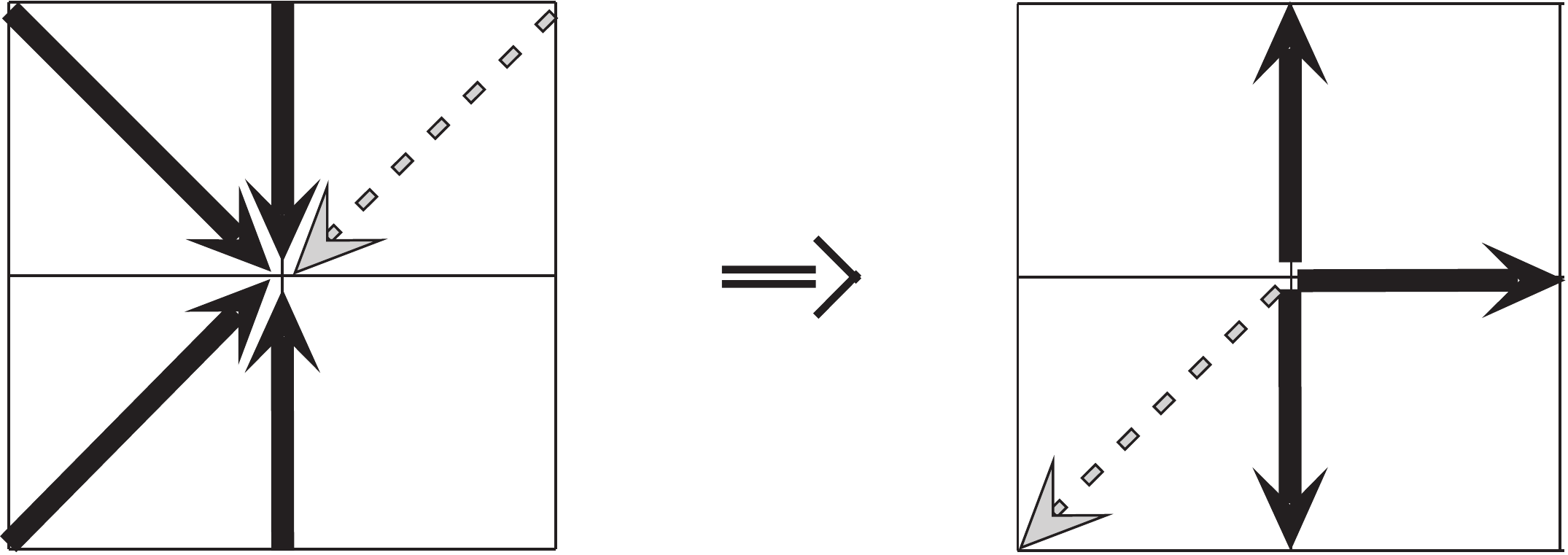} \\
\mbox{\bf (a)} & \mbox{\bf (b)} & \mbox{\bf (c)} \\
\end{array}}
{A 2D square-lattice SSM.  Particles go straight unless they interact.
One sample orientation is shown for each interacting case.  The
inverse cases also apply.  (a)~Basic collision.  (b)~Same collision
and its inverse operating in two opposite directions simultaneously.
(c)~Same collision as (a), but with ``spectator particles'' present.
The dotted-arrow particle may or may not be present, and may come from
below instead (i.e., flipped orientation).  Spectators go straight.}

We can complete our construction without adding any extra particles or
velocities to the model.  Instead, we simply add some cases to the SSM
in which our normal soft-sphere collisions happen even when there are
extra particles nearby.  The cases we will need are shown in
Figure~\ref{fig.extra-collisions}.  In this diagram, we show each
forward collision in one particular orientation---collisions in other
orientations and the corresponding inverse collisions also apply.  The
first case is the SSM collision with nothing else around.  The second
case is a forward and backward collision simultaneously---this will
let us bounce signals back the way they came.  The third case has at
least two spectators, and possibly a third (indicated by a dotted
arrow).  The collision proceeds normally, and all spectators pass
straight through.  This case will allow us to separate forward and
backward moving signals.  As usual, all other cases go straight.  In
particular, we will depend for the first time on head-on colliding
particles going straight.  We have not used any head-on collisions in
our circuits thus far, and so we are free to define their behavior
here.

\fig{problem-reverse}{%
\begin{array}{c@{\hspace{.6in}}c@{\hspace{.6in}}c}
\includegraphics[height=1.3in]{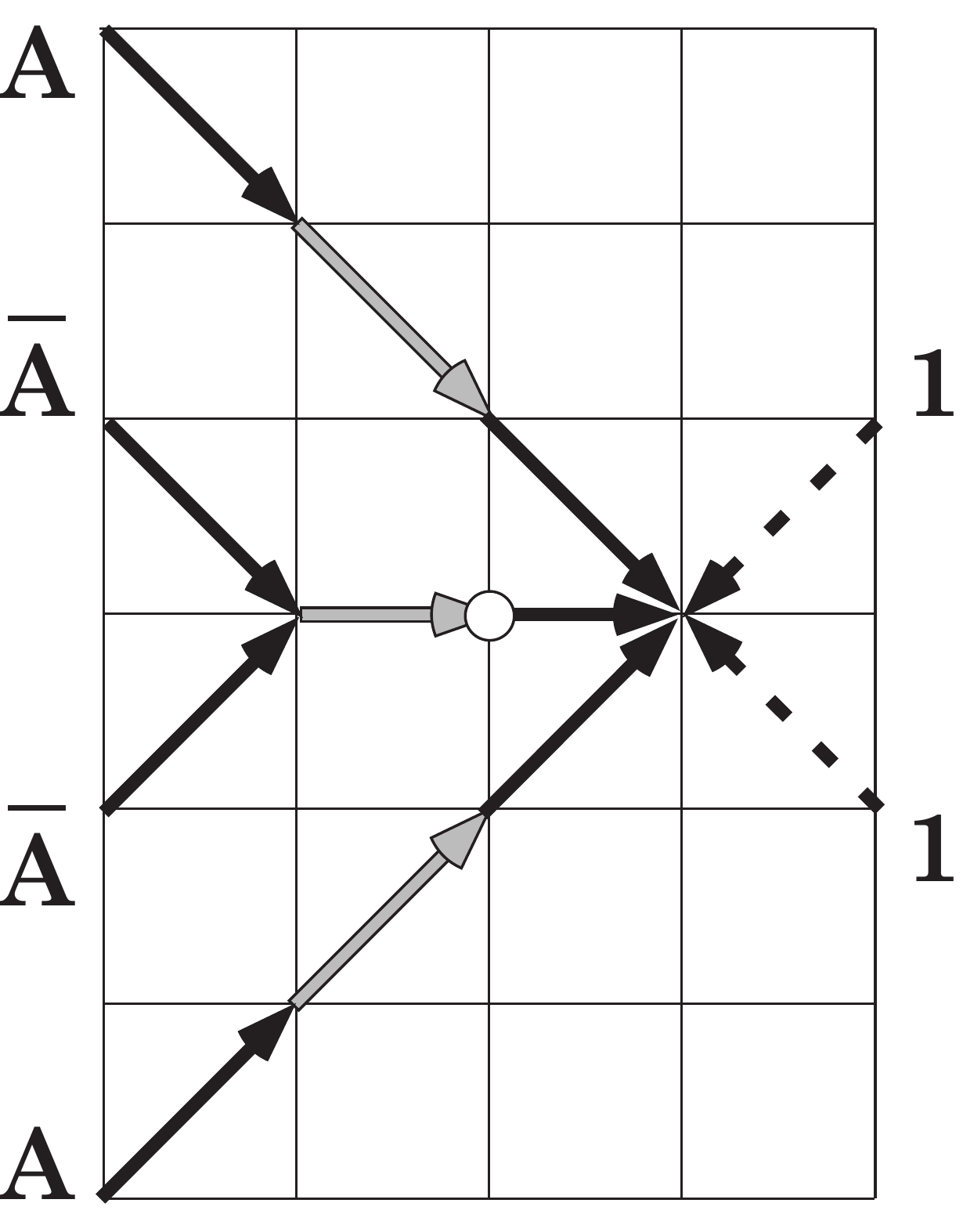} &
\includegraphics[height=1.3in]{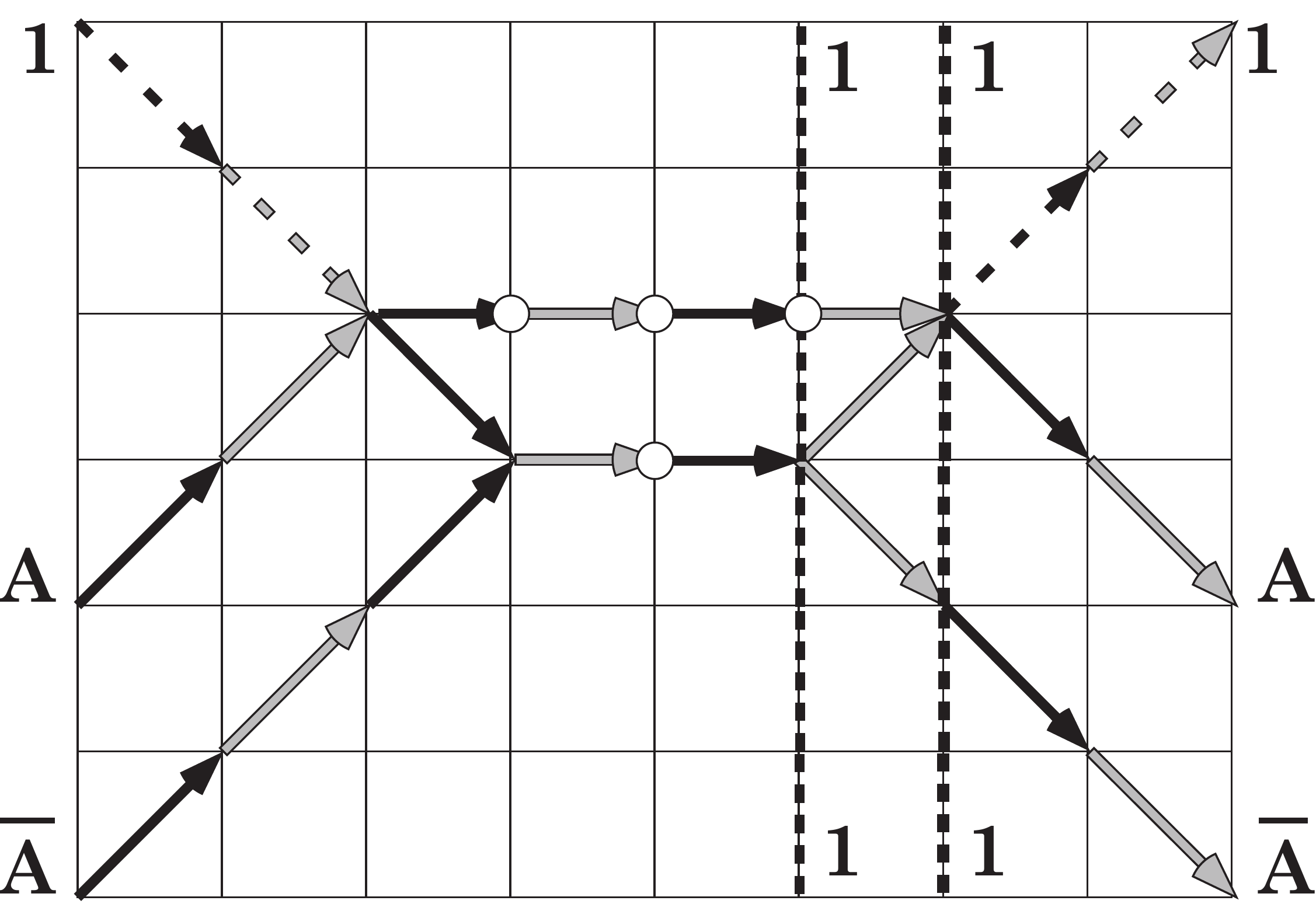} &
\includegraphics[height=1.3in]{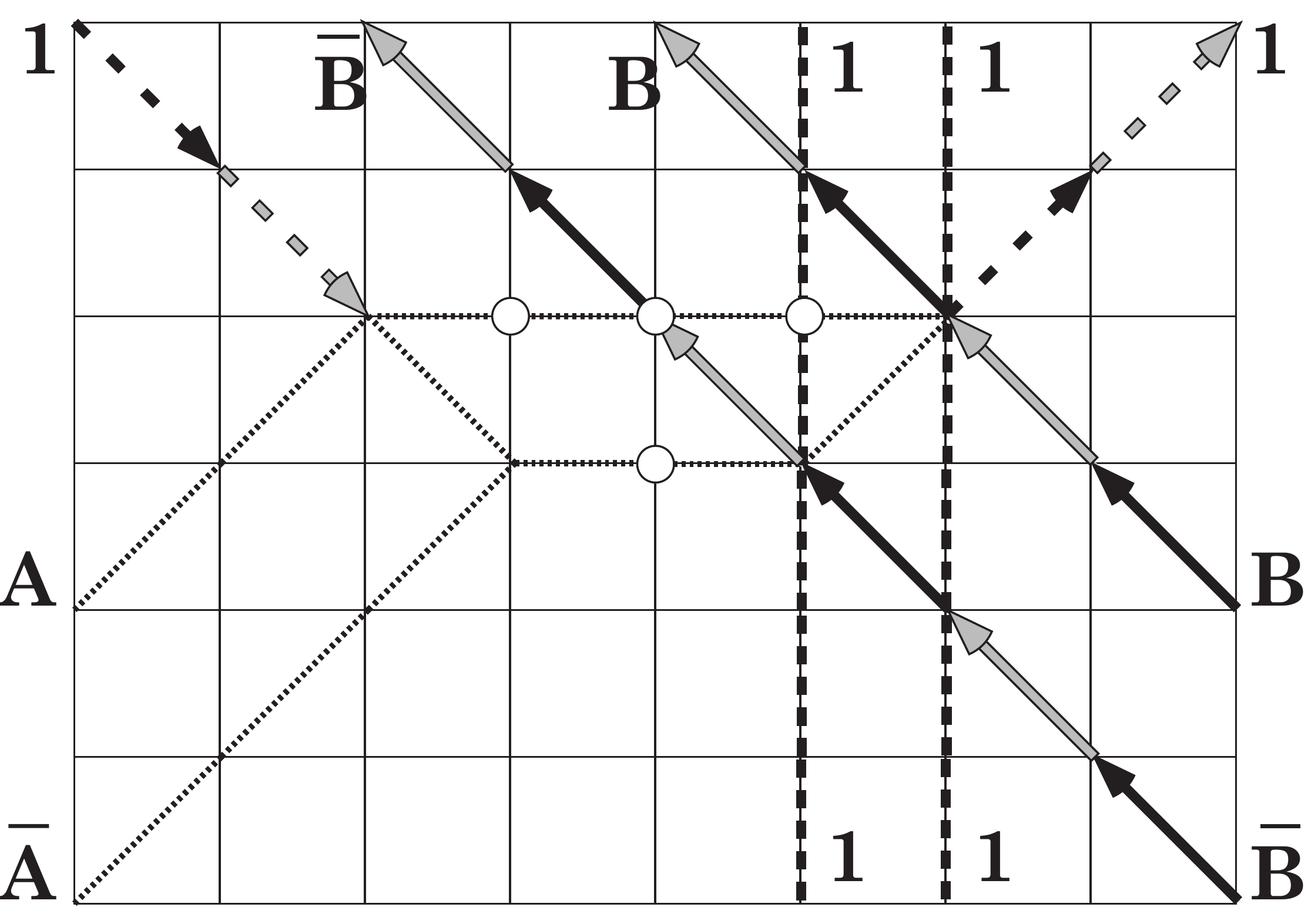} \\
\mbox{\bf (a)} & \mbox{\bf (b)} & \mbox{\bf (c)} \\
\end{array}}
{A way to reflect signals back, without refining the lattice or adding
extra particles.  (a)~If we have two dual rail signal-pairs (one the
complement of the other), then they can be bounced straight backward
along the same paths they came in on by bringing all the signals to
one point at which two mirror signals impinge.  In either case (\A=0
and \A=1), the constant 1's that reflect these signals are also
reversed along their paths.  (b)~The thick vertical dotted
line-segments indicate constants of one that are moving in both
directions at the indicated locations.  This is otherwise a normal
reflection of a signal---the extra vertical streams don't interfere
with the operation of the ``mirror.''  (c)~If a backward propagating
signal comes in from the right (\B{} and \Bbar{}), then it is not
reflected by this forward mirror---such a mirror separates the
backward moving stream from the forward stream.}

Figure~\ref{fig.problem-reverse}a shows how two complementary sets of
dual-rail signals can be reflected back the way they came.  We show
the signals up to the moment where they come to a point where they
collide with the two constant streams.  In the case where \A=1, we
have four diagonal signals colliding at a point, and so everything
goes straight through.  In particular, the constant streams have
particles going in both directions (passing through each other), and
the signal particles go back up the {\A} paths without interacting
with oncoming signals.  In the case where \A=0, we use our new
``both-directions'' collision, which again sends all particles back
the way they came.  Thus we have succeeded in reversing the direction
of a signal stream.

Figure~\ref{fig.problem-reverse}b shows a mirror with all signals
moving from left to right.  We've added in vertical constant-streams
in two places, which don't affect the operation of the ``mirror.''
These paths have a continual stream of particles in both the up and
down directions, and so these particles all go straight (head-on
collisions).  In Figure~\ref{fig.problem-reverse}c, we've just shown
signals coming into this mirror backward (with the forward paths
drawn in lightly).  This mirror doesn't reflect these backward-going
signals, and so they go straight through.  The vertical constants were
needed to break the symmetry, so that it's unambiguous which signals
should interact.  This separation uses the extra spectator-particle
cases added to our rule in Figure~\ref{fig.extra-collisions}c.  As we
will discuss, in a triangular-lattice SSM the separation at mirrors
doesn't require any vertical constants at the mirrors (see
Section~\ref{sec.hex}).

\fig{upside-down}{%
\begin{array}{c@{\hspace{.5in}}c@{\hspace{.5in}}c@{\hspace{.5in}}c}
\includegraphics[height=1.3in]{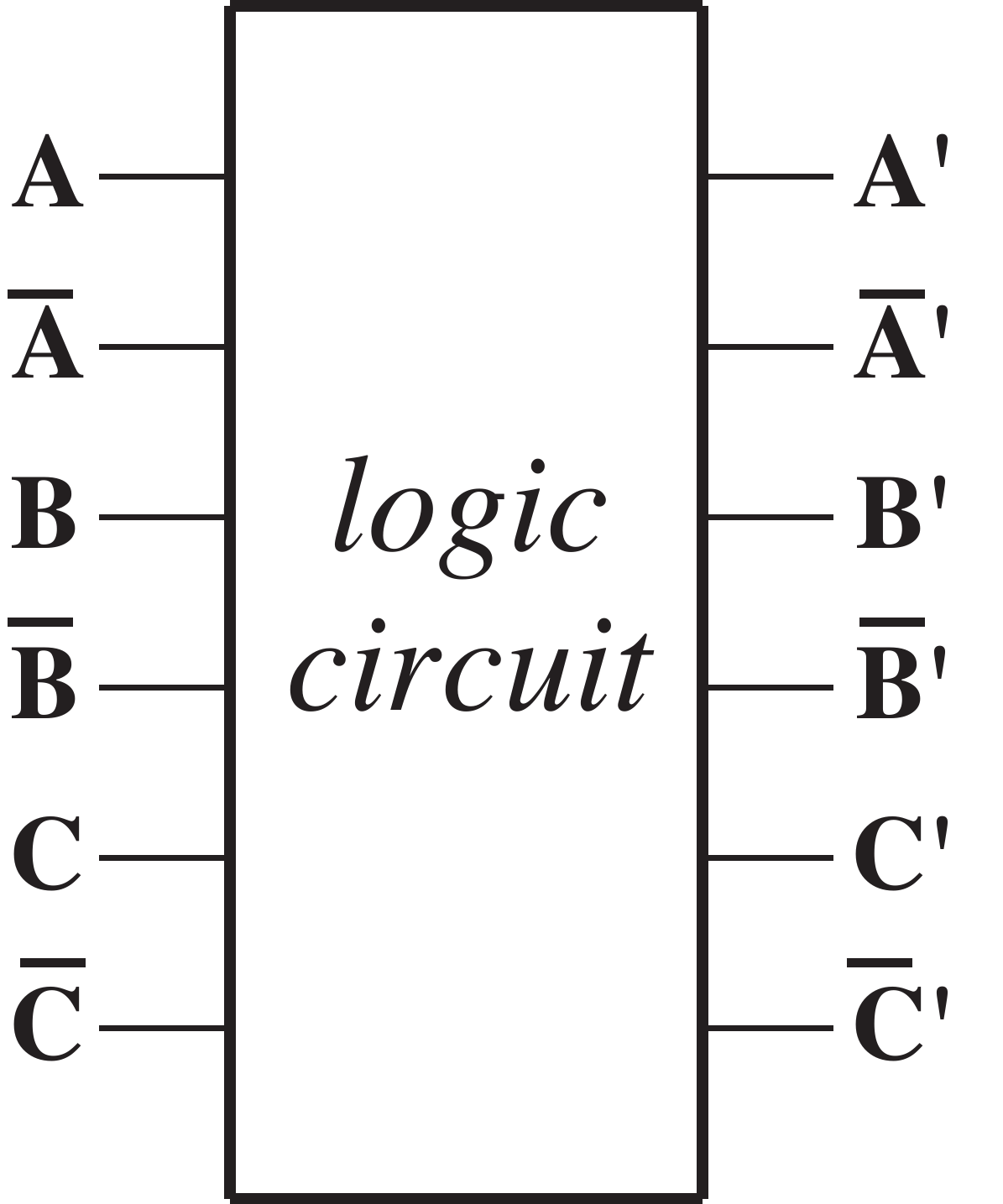} &
\includegraphics[height=1.3in]{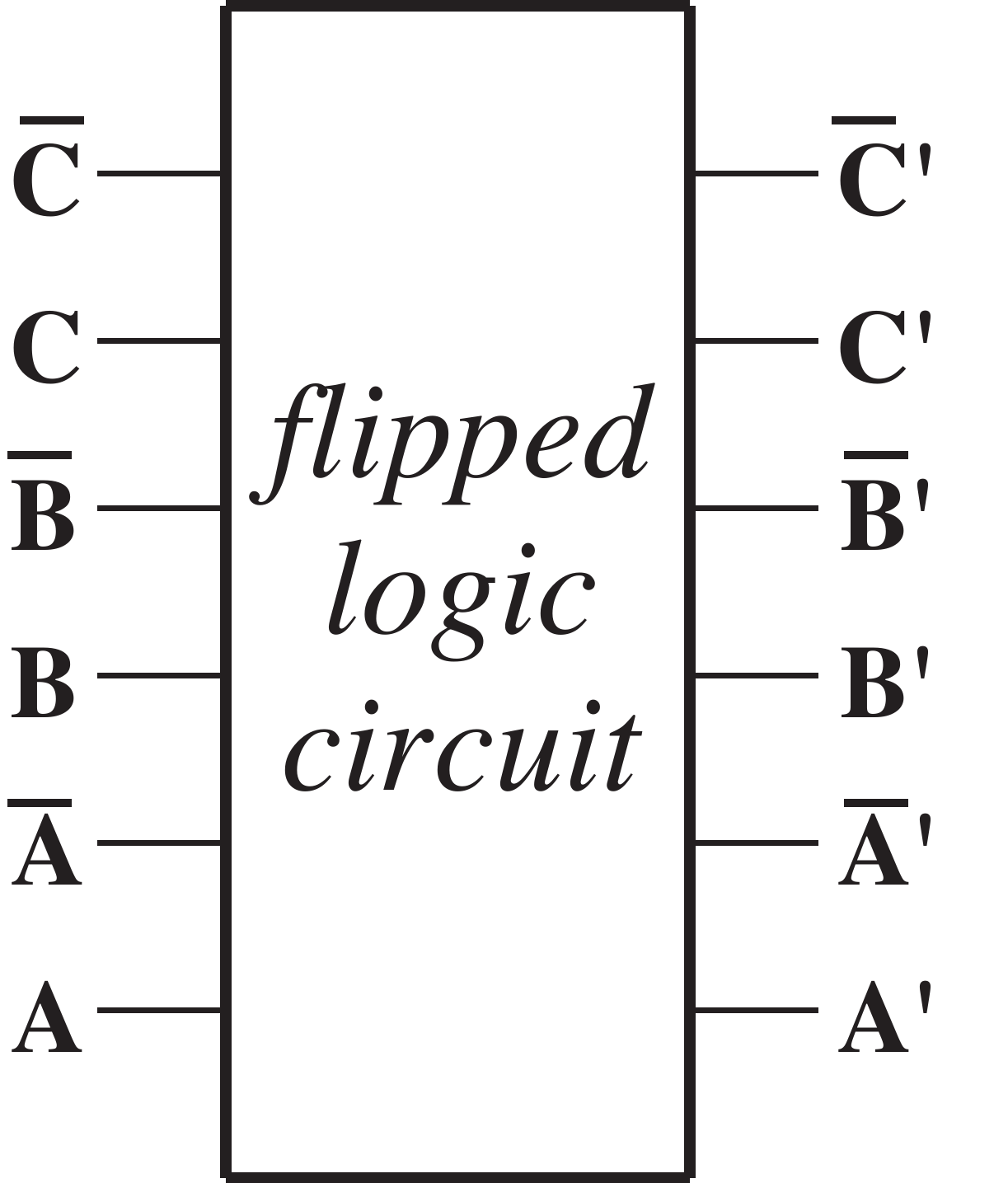} &
\includegraphics[height=1.3in]{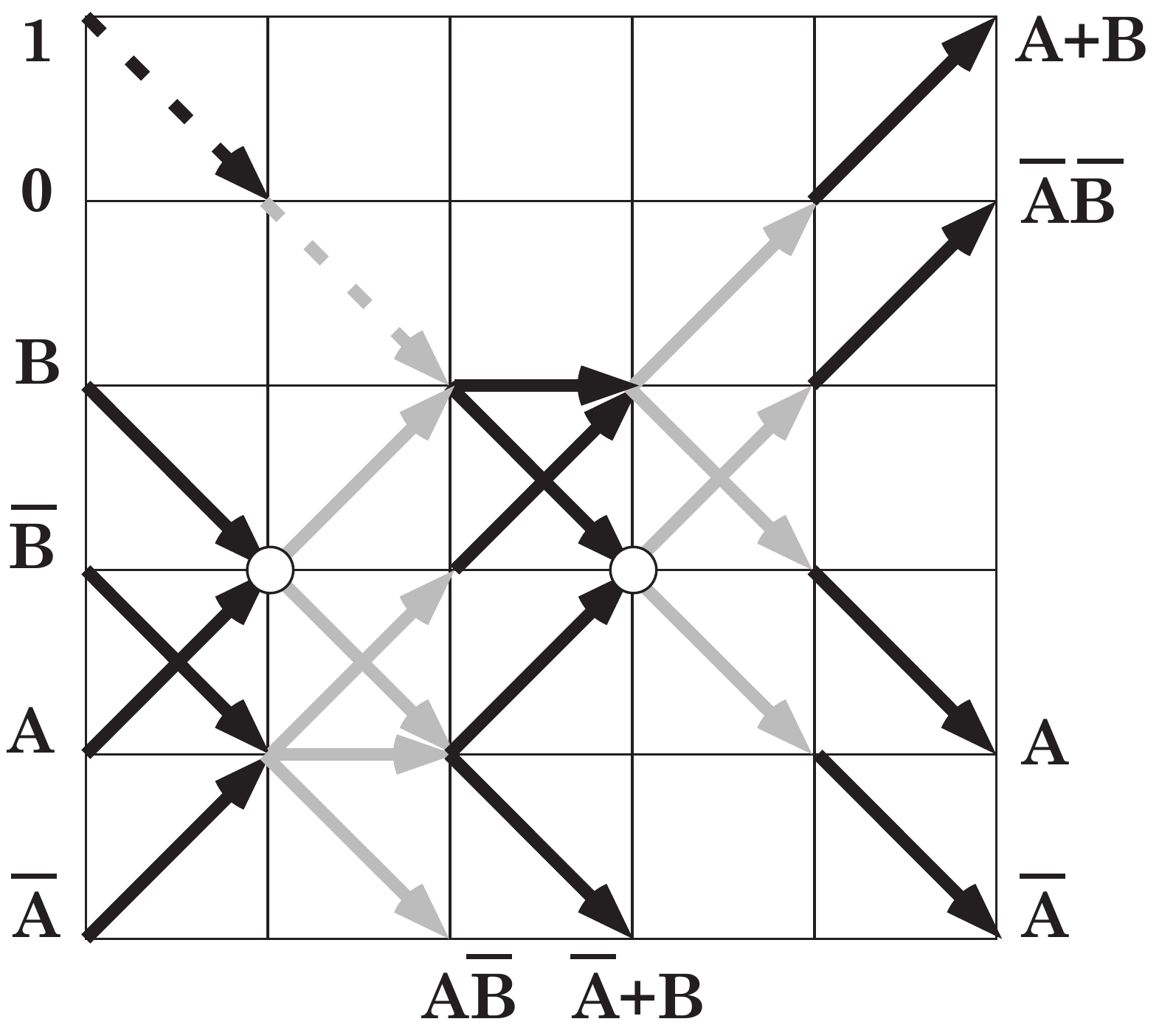} &
\includegraphics[height=1.3in]{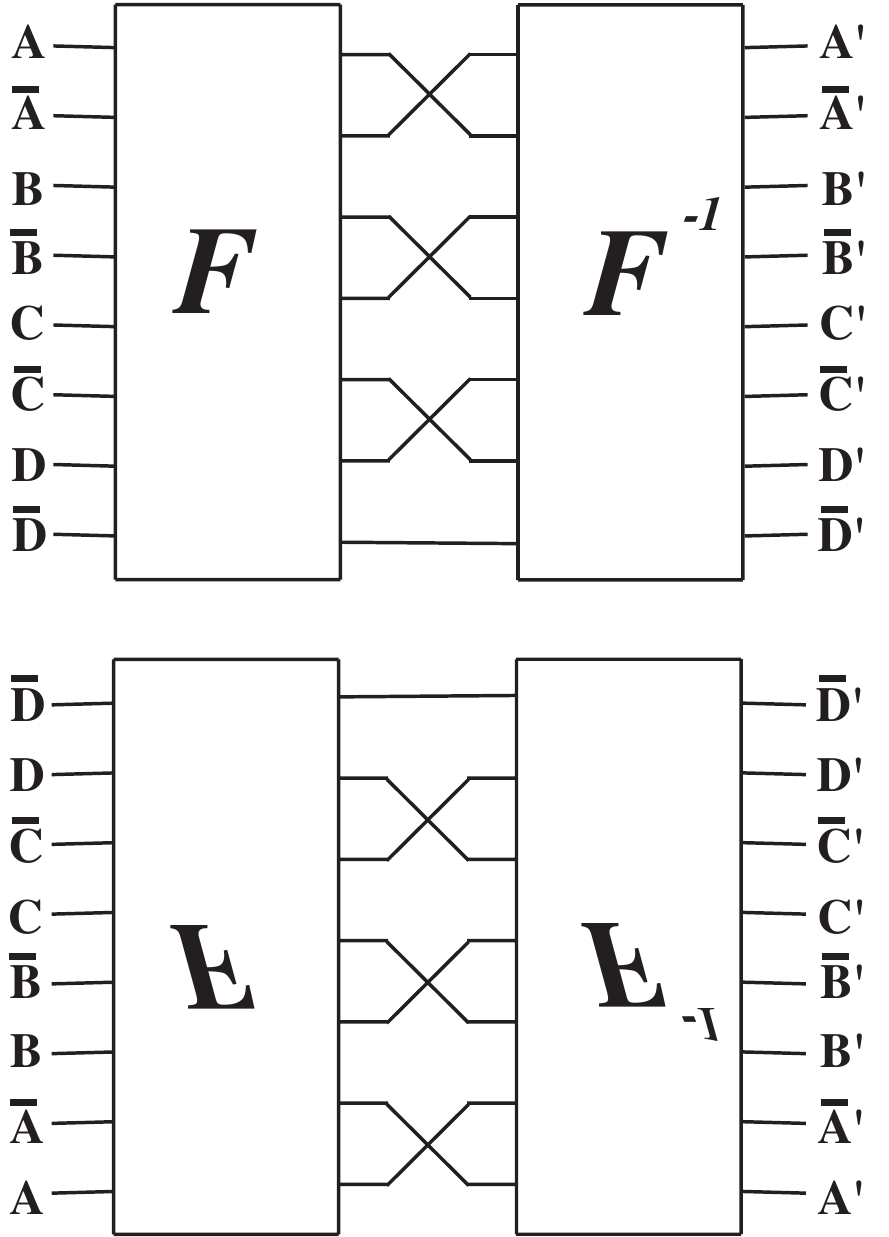} \\
\mbox{\bf (a)} & \mbox{\bf (b)} & \mbox{\bf (c)} & \mbox{\bf (d)} \\
\end{array}}
{DeMorgan inversion.  (a)~A logic circuit with dual-rail inputs.  In
each input pair, the complemented signal lies below the uncomplemented
one.  (b)~If this circuit is flipped vertically, the operation of the
circuit is unchanged, but it operates upon inputs that are
complemented (according to our conventions) and produces outputs that
are also complemented.  (c)~The switch gate of
Figure~\ref{fig.switch}a, flipped vertically.  Inputs and outputs have
been relabeled to call the top signal in each dual-rail pair
``uncomplemented.''  (d)~The BBMCA logic circuit of
Figure~\ref{fig.bbmca-array}b has been mirrored vertically to produce
a vertically symmetric circuit which has complementary pairs of
dual-rail pairs.}

\fig{hex-version}{%
\begin{array}{c@{\hspace{.6in}}c@{\hspace{.6in}}c@{\hspace{.6in}}c}
\includegraphics[height=1.6in]{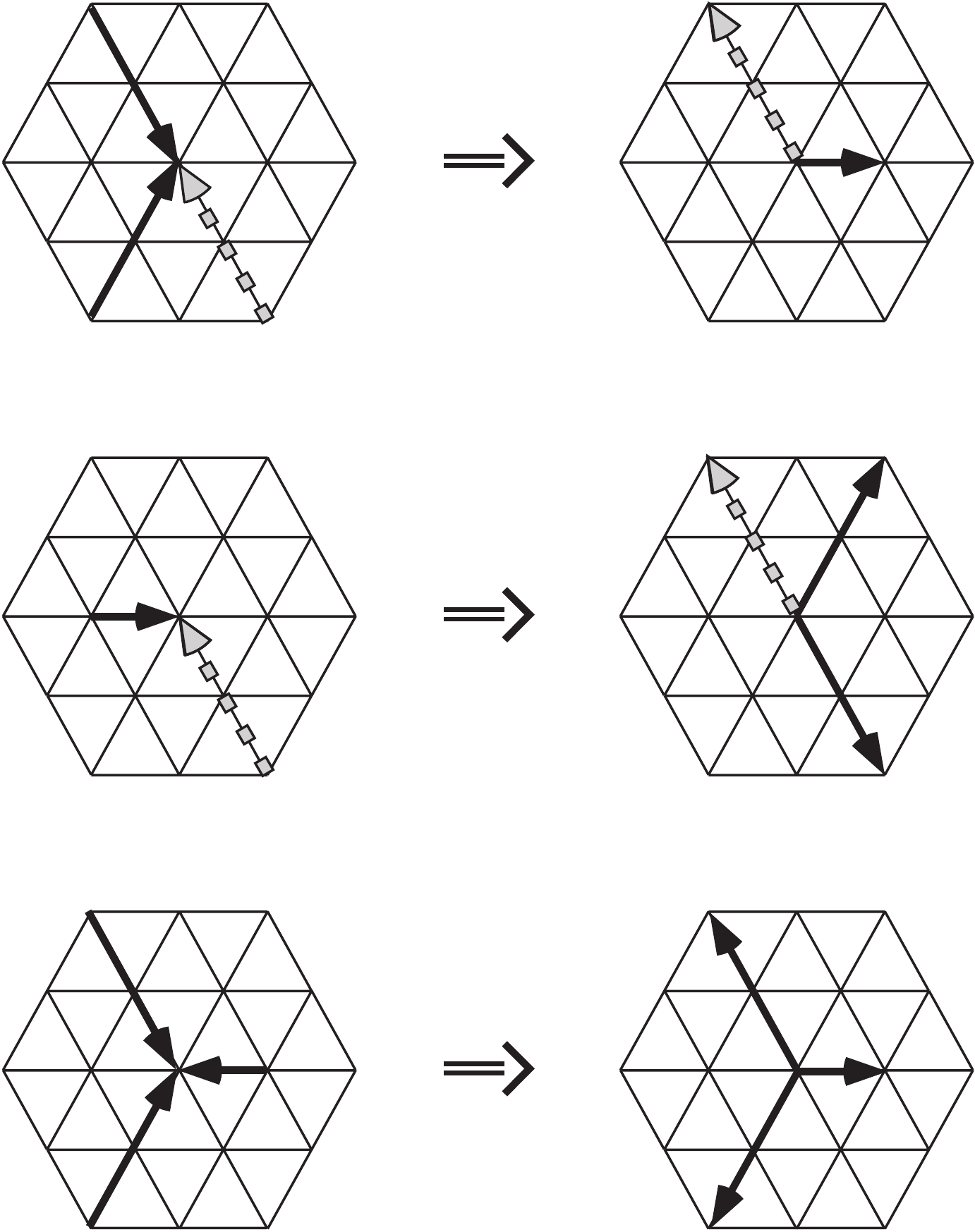} &
\includegraphics[height=1.6in]{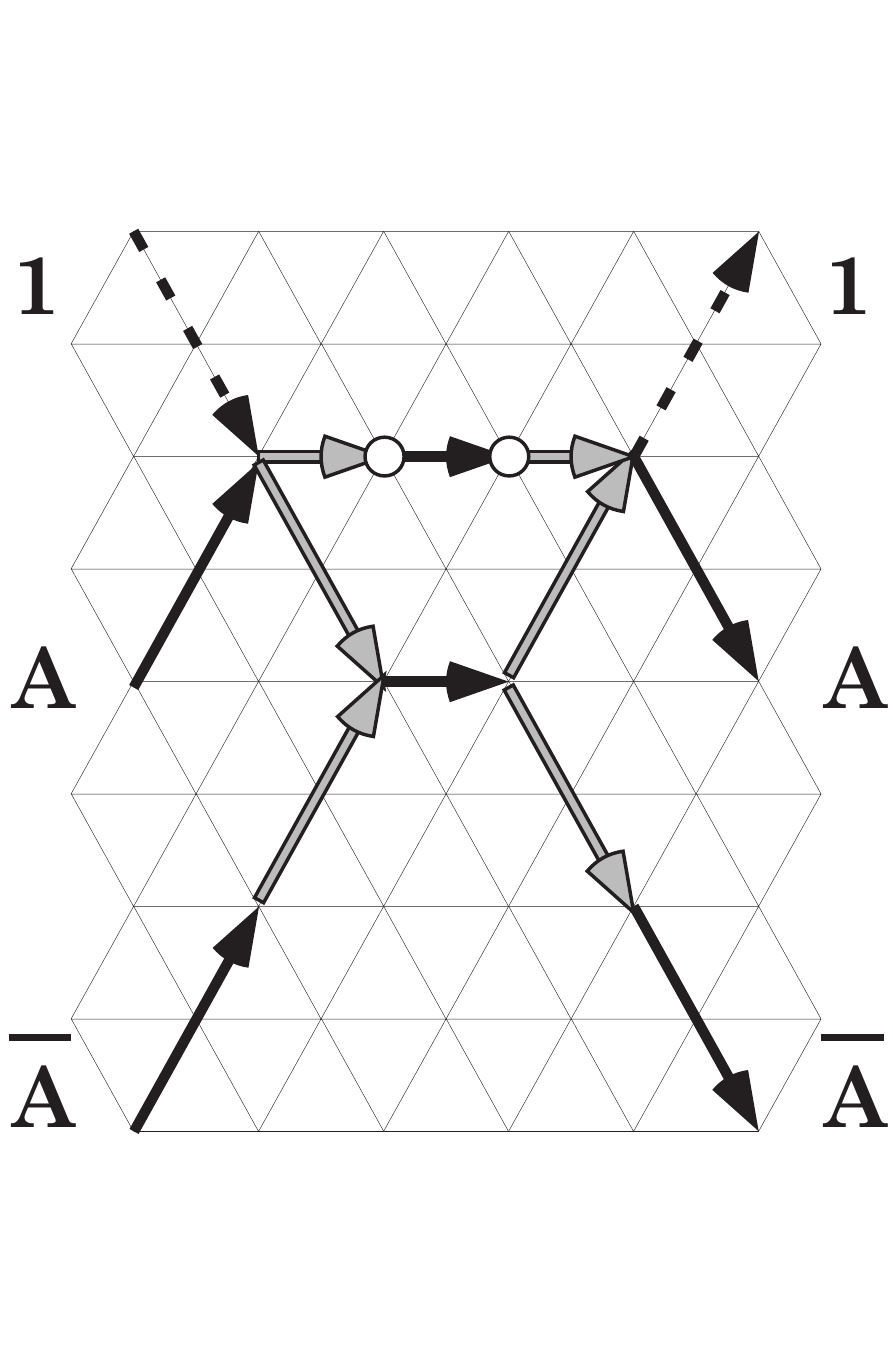} &
\includegraphics[height=1.6in]{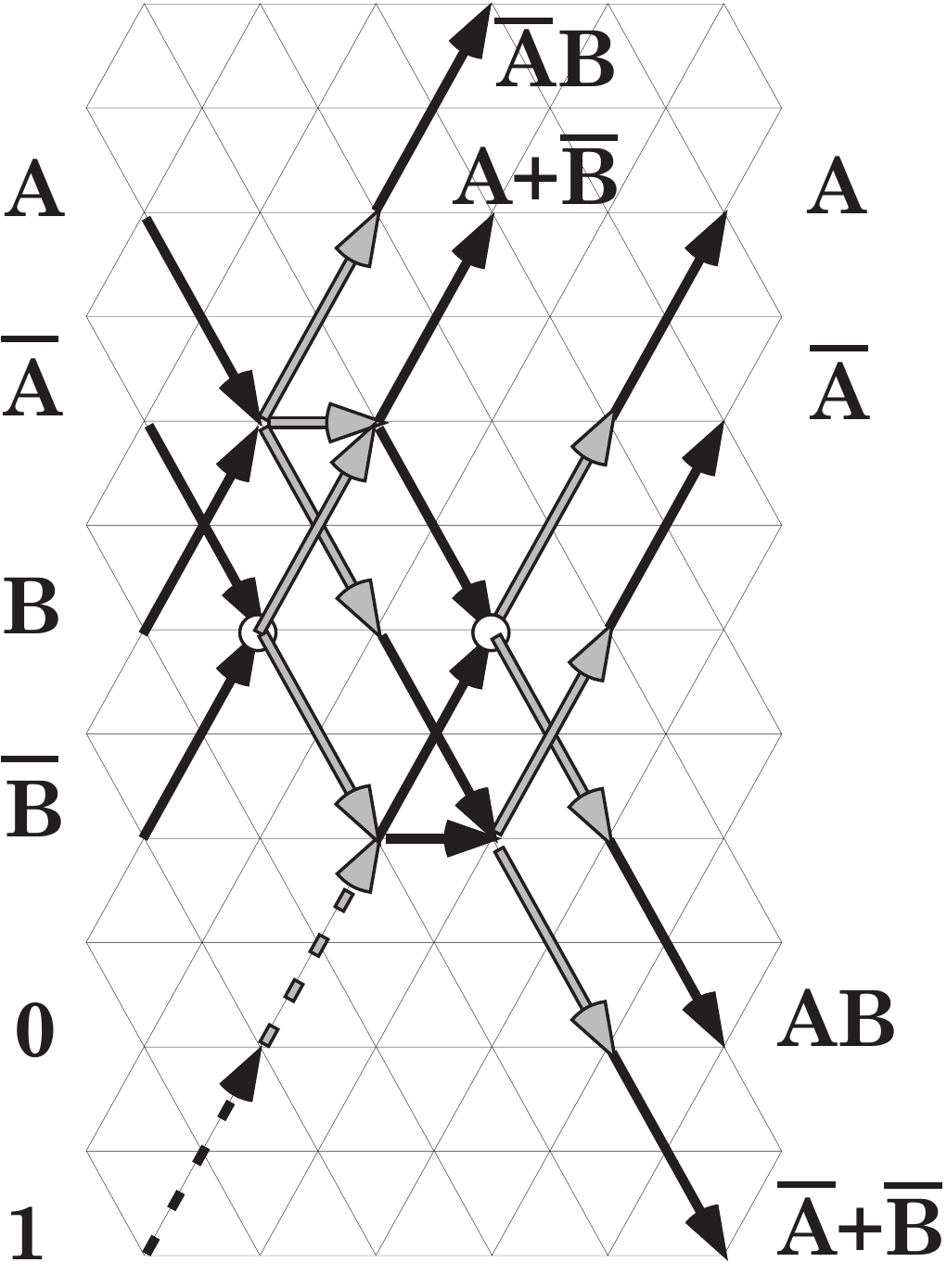} &
\includegraphics[height=1.6in]{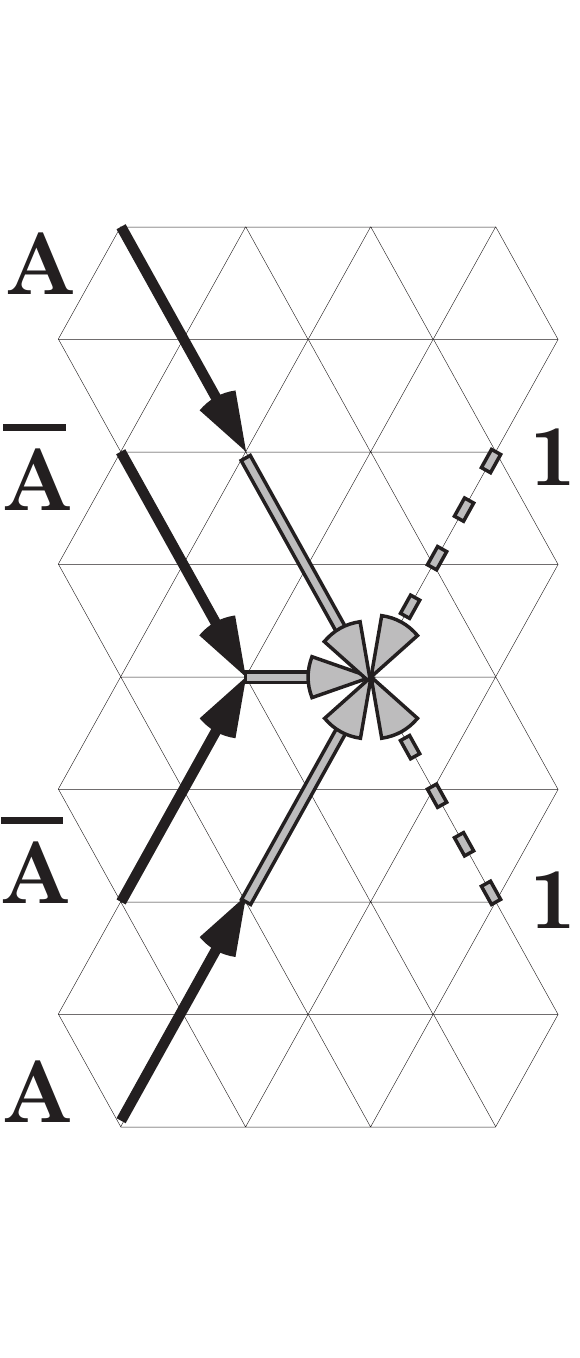} \\
\mbox{\bf (a)} & \mbox{\bf (b)} & \mbox{\bf (c)} & \mbox{\bf (d)} \\
\end{array}}
{An SSM gas on the triangular lattice which allows signal feedback.
(a)~Two speed-2 particles collide and turn into a speed-1 particle
with twice the mass.  This decays back into two speed-2 particles.  If
an extra ``spectator'' speed-2 particle comes in as shown with a
dotted-arrow (or the flip of these cases), it passes straight through.
Collisions can happen both forward and backward simultaneously.  In
all other cases, particles go straight.  (b)~Constants act as mirrors
for dual-rail signals.  (c)~This is a switch gate.  Other
combinational circuits from the square-lattice SSM can be similarly
stretched vertically to fit onto the triangular lattice.  (d)~The
third collision case in the rule makes signals bounce back the way
they came.  Backward-going signals will separate at a mirror such as
is shown in (b).}

Finally, Figure~\ref{fig.upside-down} shows how we can arrange to
always have two complementary dual-rail pairs collide whenever we need
to send a signal backward.  Figure~\ref{fig.upside-down}a shows an
SSM circuit with some number of dual-rail pairs.  In each pair, the
signals are synchronized vertically, with the uncomplemented signal on
top.  Figure~\ref{fig.upside-down}b shows the same gate flipped
vertically.  The collisions that implement the circuit work perfectly
well upside-down, but both the inputs and the outputs are complemented
by this inversion.  For example, in Figure~\ref{fig.upside-down}c, we
have turned a switch-gate upside down.  If we relabel inputs and
outputs in conventional order, then we see that this gate performs a
logical OR where the original gate performed an AND.  In
Figure~\ref{fig.upside-down}d, we take our BBMCA logic block of
Figure~\ref{fig.bbmca-array}b and add a vertically reflected copy.
This pair of circuits, taken together, has both vertical and
horizontal symmetry.  Given quad-rail inputs (dual rail inputs along
with their dual-rail complements), it produces corresponding quad-rail
outputs, which can be reflected backward using the collision of
Figure~\ref{fig.problem-reverse}a, and separated at mirrors, as shown
in Figure~\ref{fig.problem-reverse}c.  Now note that the
constant-lifting technique of Figure~\ref{fig.bbmca-symm}b works
equally well even if all of the constant streams have 1's flowing in
both directions simultaneously, by virtue of the bidirectional
collision case of Figure~\ref{fig.extra-collisions}b.  Thus we are
able to apply the constant-symmetrizing technique to mirror streams
that connect the four signals between our BBMCA logic blocks
(Figure~\ref{fig.bbmca-symm}c), and complete our construction.

\subsection{Other lattices}\label{sec.hex}

All of this works equally well for an SSM on the triangular lattice,
and is even slightly simpler, since we don't need to add extra
constant streams at mirrors where forward and backward moving signals
separate (as we did in Figure~\ref{fig.problem-reverse}c).  The
complete rule is given in Figure~\ref{fig.hex-version}a: the dotted
arrow indicates a position where an extra ``spectator'' particle may
or may not come in.  If present, it passes straight through and
doesn't interfere with the collision.  In Figures
\ref{fig.hex-version}b and \ref{fig.hex-version}c, we see how mirrors
and switch-gates (and similarly any other square-lattice SSM
combinational circuit) can simply be stretched vertically to fit onto
the triangular lattice.  A back-reflection, where signals are sent
back the way they came, is shown in Figure~\ref{fig.hex-version}d.

This of course also means that the corresponding 3D model
(Figure~\ref{fig.3d}c) can perform efficient momentum-conserving
computation, at least in a single plane.  If we have a dual-rail pair
in one plane of this lattice, and its dual-rail complement directly
below it in a parallel plane, this combination can be deflected
cleanly in either of two planes by a pair of constant mirror-streams.
Thus it seems plausible that this kind of discussion may be
generalized to three dimensions, but we won't pursue that here.

\section{Relativistic CA's}

We have presented examples of reversible lattice gases that support
universal computation and that can be interpreted as a discrete-time
sampling of the classical-mechanical dynamics of compressible balls.
We would like to present here an alternative interpretation of the
same models as a discrete-time sampling of relativistic classical
mechanics, in which kinetic energy is converted by collisions into
rest mass and then back into kinetic energy.

For a relativistic collision of some set of particles, both
relativistic energy and relativistic momentum are conserved, and so:
$$
\sum_i E_i = \sum_i E^\prime_i, \quad
\sum_i E_i \vec{v}_i = \sum_i E^\prime_i \vec{v}^\prime_i,
$$
where the unprimed quantities are the values for each particle before
the collision, and the primed quantities are after the collision.
These equations are true regardless of whether the various particles
involved in the collision are massive or massless.  Now we note that
for {\em any mass and momentum conserving lattice gas},
$$
\sum_i m_i = \sum_i m^\prime_i, \quad
\sum_i m_i \vec{v}_i = \sum_i m^\prime_i \vec{v}^\prime_i,
$$
and so we need only reinterpret what is normally called ``mass'' in
these models as relativistic energy in order to interpret the
collisions in such a lattice gas as being relativistic.  If all
collisions are relativistically conservative, then the overall
dynamics exactly conserves relativistic energy and momentum,
regardless of the frame of reference in which the system is analyzed.
Normal non-relativistic systems have separate conservations of mass
and non-relativistic energy---a property that the collisions in most
momentum-conserving lattice gases lack.  Thus we might argue that the
relativistic interpretation is more natural in general.

In the collision of Figure~\ref{fig.ebm}b, for example, we might call
the incoming pair of particles ``photons,'' each with unit energy and
unit speed.  The two photons collide, and the vertical components of
their momenta cancel, producing a slower moving ($v=1/\sqrt{2}$)
massive particle ($m=\sqrt{2}$) with energy 2 and with the same
horizontal component of momentum as the original pair.  After one
step, the massive particle decays back into two photons.  At each
step, relativistic energy and momentum are conserved.

As is discussed elsewhere\cite{bridge,cc}, macroscopic relativistic
invariance could be a key ingredient in constructing CA models with
more of the macroscopic richness that Nature has.  If a CA had
macroscopic relativistic invariance, then every complex macroscopic
structure (assuming there were any!)  could be set in motion, since
the macroscopic dynamical laws would be independent of the state of
motion.  Thus complex macroscopic structures could move around and
interact and recombine.

Any system with macroscopic relativistic symmetry is guaranteed to
also have the relativistic conservations of energy and momentum that
go along with it.  As Fredkin has pointed out, a natural approach to
achieving macroscopic symmetries in CA's is to start by putting the
associated microscopic conservations directly into the CA rule---we
certainly can't put the continuous symmetries there!  Momentum and
mass conserving LGA models effectively do this.

Of course, merely reinterpreting the microscopic dynamics of lattice
gases relativistically doesn't make their macroscopic dynamics any
richer.  One additional microscopic property that we can look for is
the ability to perform computation, using space as efficiently as is
possible: this enables a system to support the highest possible level
of complexity in a finite region.  Microscopically, SSM gases have
both a relativistic interpretation and spatial efficiency for
computation.  What we would really like is a dynamics in which both of
these properties persist at the macroscopic scale.

\fig{rel-example}{%
\begin{array}{c@{\hspace{.4in}}c@{\hspace{.4in}}c@{\hspace{.4in}}c}
\includegraphics[height=1.4in]{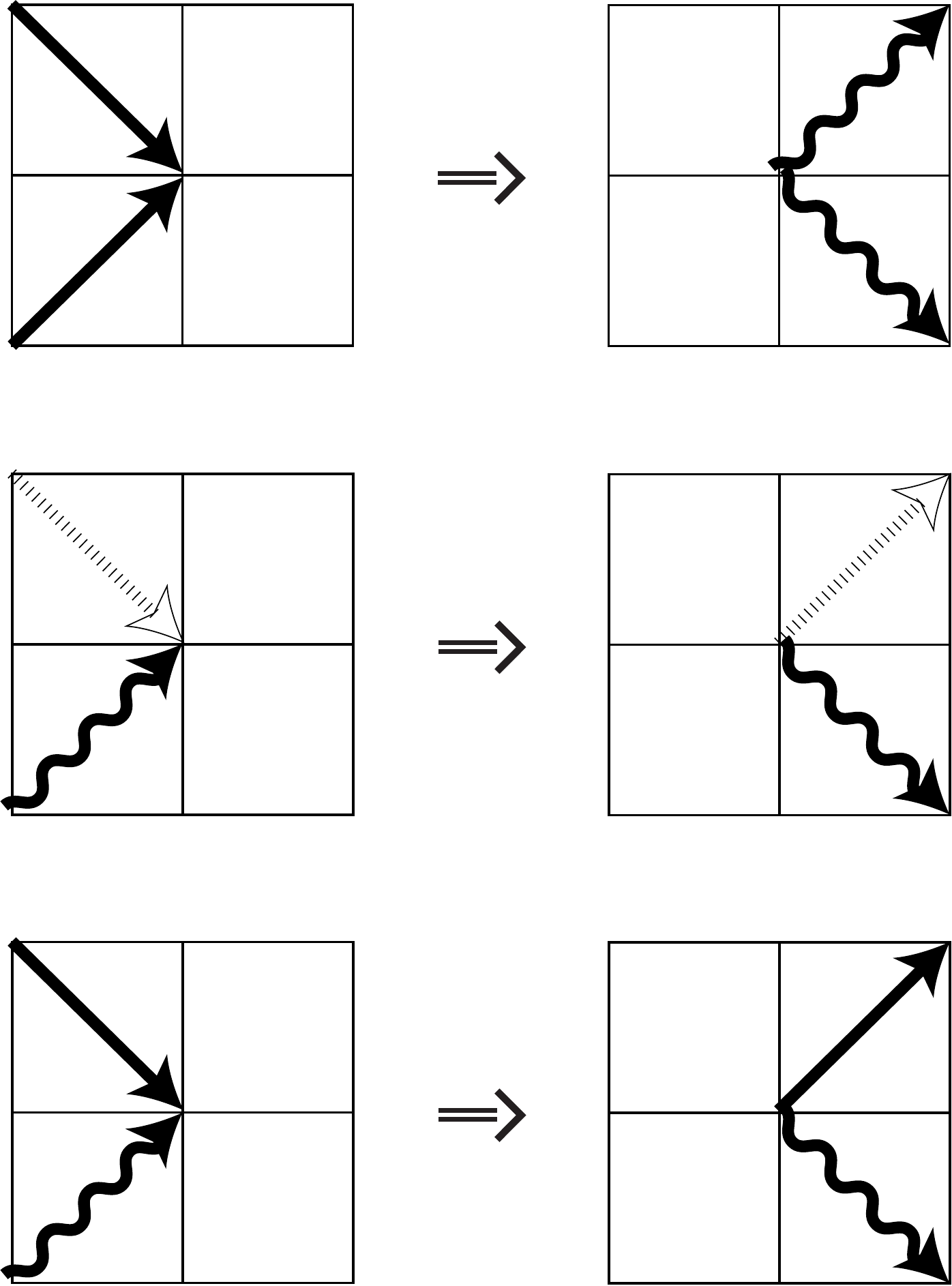} &
\includegraphics[height=1.4in]{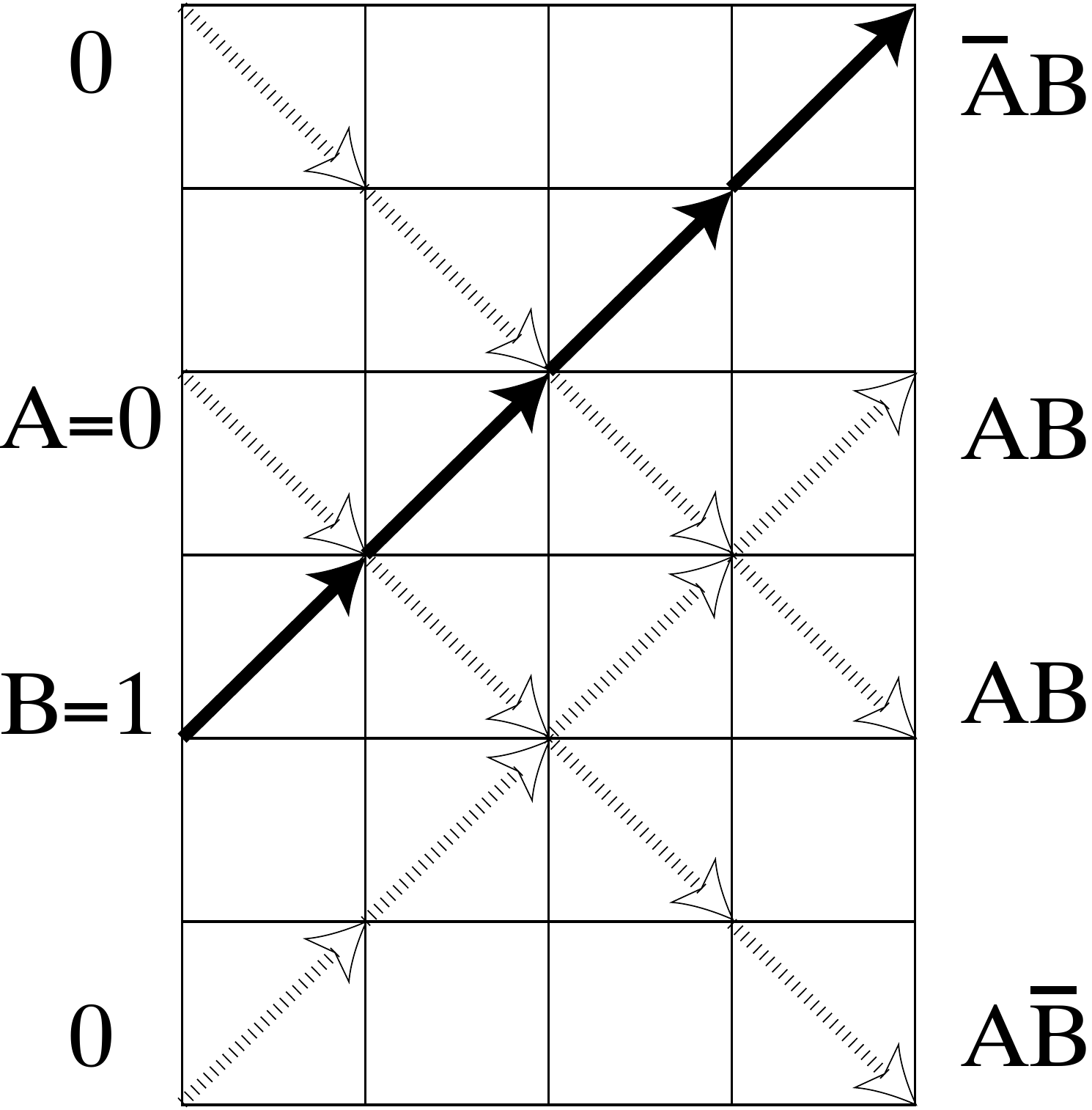} &
\includegraphics[height=1.4in]{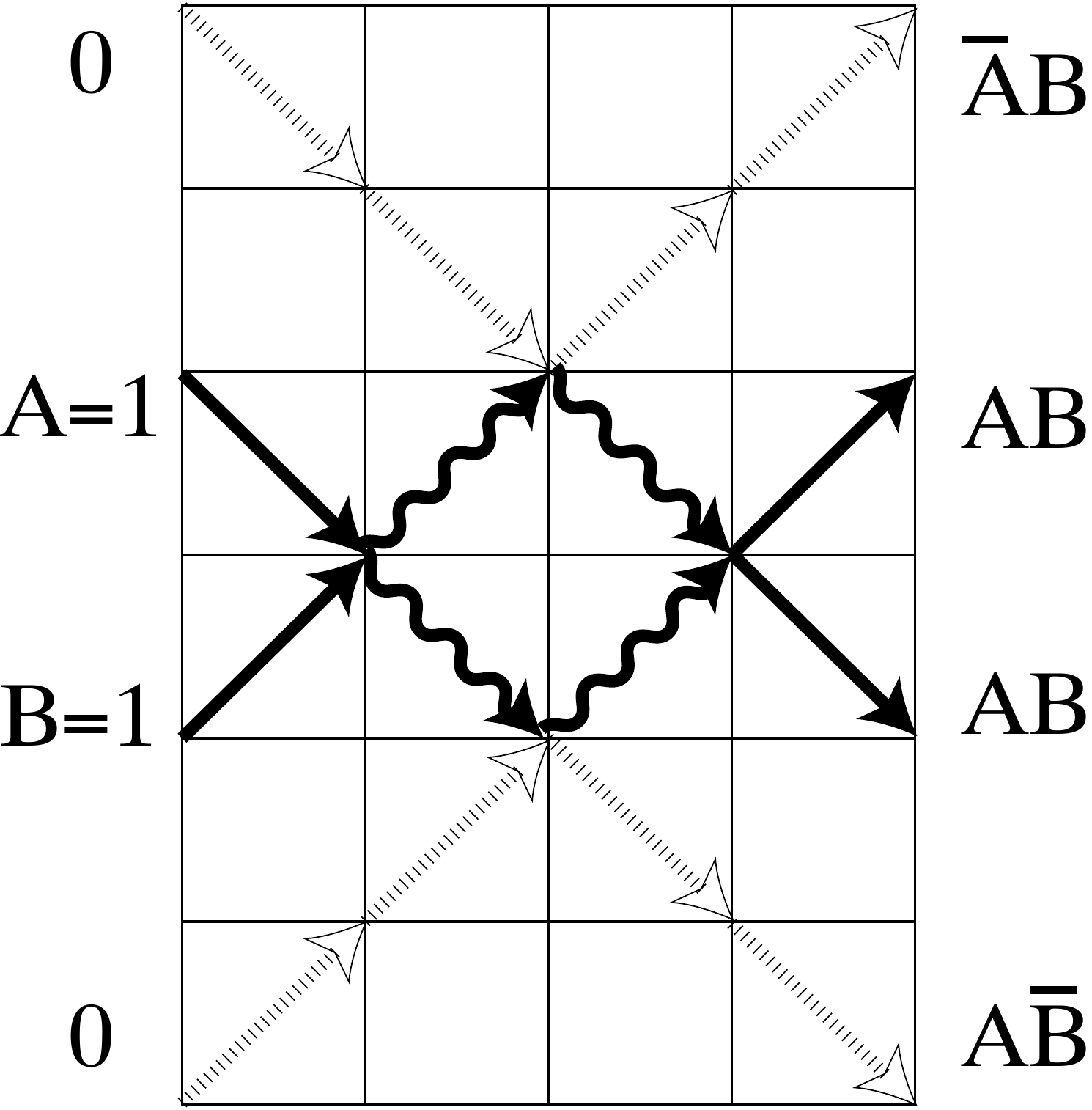} &
\includegraphics[height=1.4in]{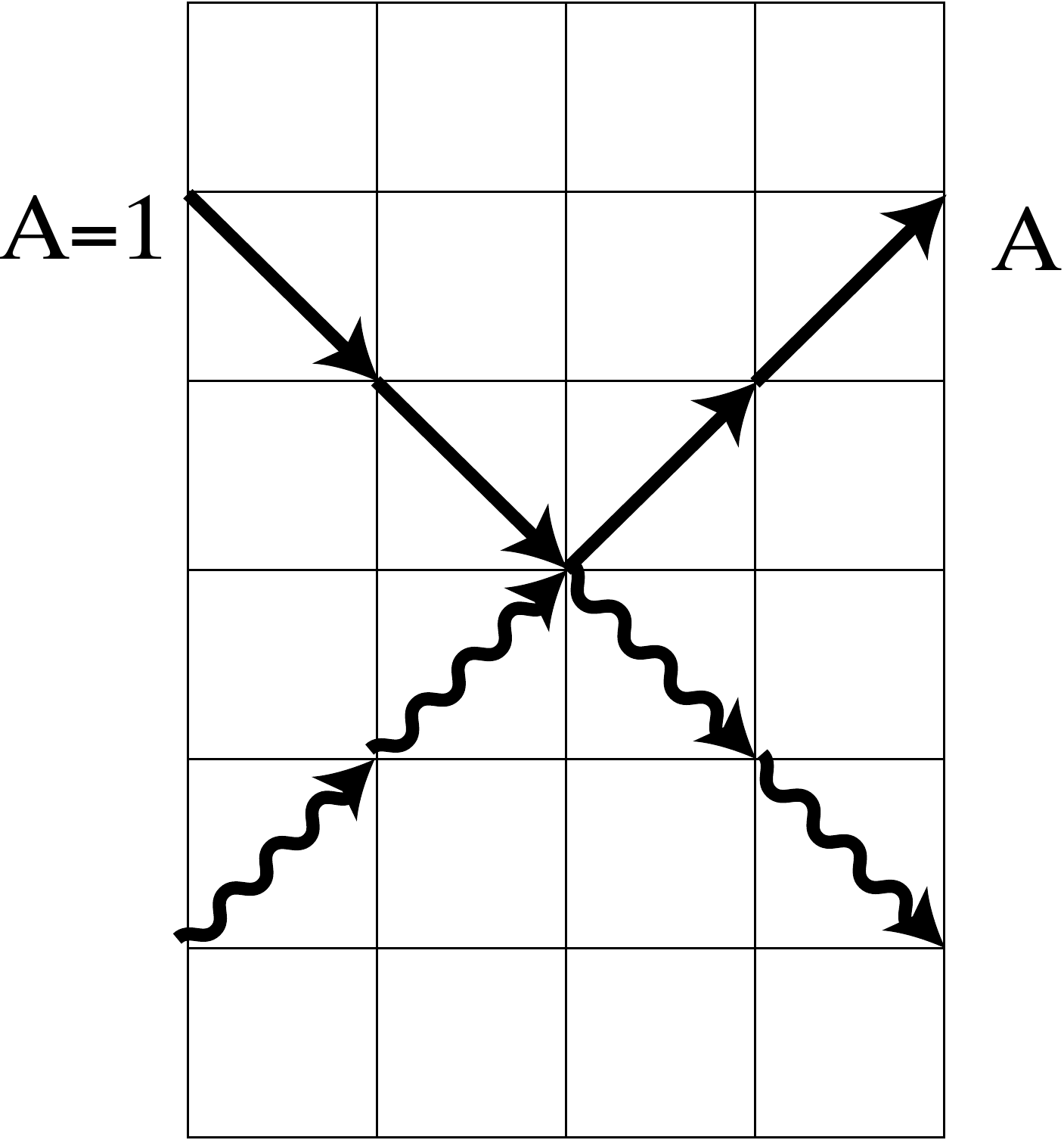} \\
\mbox{\bf (a)} & \mbox{\bf (b)} & \mbox{\bf (c)} & \mbox{\bf (d)} \\
\end{array}}
{A computation-universal LGA in which ones and zeros have the same
momentum. (a)~We use three kinds of particles that interact.  The 
three cases shown, plus rotations, inversions and the time-reversal of
these cases, are all of the interactions.  In all other cases,
particles don't interact.  (b)~We use the solid-black particles to
represent ``ones'' and the dotted particles to represent ``zeros''.
If only a single one comes into a ``collision,'' none of the interaction
cases applies, and so everything goes straight.  (c)~If two ones
collide, they turn into a third kind of particle (shown as a wavy
arrow), which is deflected by a zero (and deflects the zero).  The
inverse interaction recreates the two ones, displaced inwards from
their original paths (as in an SSM collision).  (d)~The wavy particle
also deflects (and is deflected by) a one.}

If we are trying to achieve macroscopic relativistic invariance along
with efficient macroscopic computational capability, we can see that
one potential problem in our ``bounce-back'' SSM gases
(Figures~\ref{fig.extra-collisions} and
\ref{fig.hex-version}) is a defect in their discrete rotational
symmetry.  Dual-rail pairs of signals aligned in one orientation can't
easily interact with dual-rail pairs that are aligned in a $60^\circ$
(triangular lattice) or $90^\circ$ (square lattice) rotated
orientation.  If this causes a problem macroscopically, we can always
try adding individual signal delays to the model, as in
Figure~\ref{fig.problem-delay}b.  This may have macroscopic problems
as well, however, since turning signals with the correct timing
requires several correlated interactions.  Of course the reason we
adopted dual-rail signalling to begin with was to avoid mixing
logic-value information with signal momentum---every dual-rail signal
has unit momentum and can be reflected without ``measuring'' the logic
value.  Perhaps we should simply decouple these two quantities at the
level of the individual particle, and use some other degree of freedom
(other than presence or absence of a particle) to encode the logic
state (eg., angular momentum).  An example of a model which decouples
logic values and momentum is given in Figure~\ref{fig.rel-example}.

In Figure~\ref{fig.rel-example}a, we define a rule which involves
three kinds of interacting particles.  Figures~\ref{fig.rel-example}b
and \ref{fig.rel-example}c show how an SSM style collision-gate can be
realized, using one kind of particle to represent an intermediate
state.  Single ones go straight, whereas pairs of ones are displaced
inwards.  Both ones and zeros are deflected by the wavy ``mirror''
particles, which can play the role of the mirror-streams in our
earlier constructions.  Deflecting a binary signal conserves momentum
without recourse to dual rail logic, and without contaminating the
mirror-stream.  Adding rest particles to this model allows signals
to cross (since the rule is, ``in all other cases particles don't
interact'').  Models similar to this ``proto-SSM'' would be
interesting to investigate on other lattices, in both 2D and 3D.

The use of rest particles to allow signals to cross in this and
earlier rules raises another issue connected with the macroscopic
limit.  If we want to support complicated macroscopic moving
structures that contain rest particles, we have to have the rest
particles move along with them!  (Or perhaps use moving signals to
indicate crossings.)  If we want to make rest particles ``move,'' they
can't be completely non-interacting.  Thus we might want to extend the
dynamics so that rest particles can both be created and destroyed.
This could be done by redefining some of the non-interacting collision
cases that have not been used in our constructions---we have actually
used very few of these cases.  These collisions would be different
from the springy collision of Figure~\ref{fig.ebm}a.  Even a
single-particle colliding with a rest particle can move it (as in
Figure~\ref{fig.problem-delay}b for example).

These are all issues that can be approached both theoretically, and by
studying large-scale simulations\cite{cc}.

\section{Semi-classical models of \\ dynamics}

The term {\em semi-classical} has been applied to analyses in which a
classical physics model can be used to reproduce properties of a
physical system that are fundamentally quantum mechanical.  Since the
finite and extensive character of entropy (information) in a finite
physical system is such a property\cite{atoms}, all CA models can in a
sense be considered semi-classical.  It is interesting to ask what
other aspects of quantum dynamics can be captured in classical CA
models.

One such aspect is the relationship in quantum systems between energy
and maximum rate of state change.  A quantum system takes a finite
amount of time to evolve from a given state to a different state
(i.e., a state that is quantum mechanically orthogonal).  There is a
simple relationship between the energy of a quantum system in the
classical limit and the maximum rate at which the system can pass
through a succession of distinct (mutually orthogonal) quantum states.
This rate depends only on how much energy the system has.  Suppose
that the quantum mechanical average energy $E$ (which is the energy
that appears in the classical equations of motion) is measured
relative to the system's ground-state energy, and in units where
Planck's constant $h$ is one.  Then the maximum number of distinct
changes that can occur in the system per unit of time is simply $2E$,
and this bound is always achieved by some state\cite{speed}.

Now suppose we have an energy-conserving LGA started in a state with
total energy $E$, where $E$ is much less than the maximum possible
energy that we can fit onto the lattice.  Suppose also that the
smallest quantity of energy that moves around in the LGA dynamics is a
particle with energy ``one.''  Then with the given energy $E$, the
maximum number of spots that can possibly change on the lattice in one
time-step is $2E$ (just as in the quantum case): $E$ smallest energy
particles can each leave one spot and move to another, each causing
two changes if none of them lands on a spot that was just vacated by
another particle.  Since the minimum value of $\Delta E$ is 1 in this
dynamics, and the minimum value of $\Delta t$ is 1 since this is our
integer unit of time, it is consistent to think of this as a system in
which the minimum value of $\Delta E \Delta t$ is 1 (which for a
quantum system would mean $h=1$).  Thus simple LGA's such as the SSM
gases reproduce the quantum limit in terms of their maximum rate of
dynamical change.

This kind of property is interesting in a physical model of
computation, since simple models that accurately reflect real physical
limits allow us to ask rather sharp questions about quantifying the
physical resources required by various algorithms (cf.~\cite{mpf}).

\section{Conclusion}

We have described soft sphere models of computation, a class of
reversible and computation-universal lattice gases which correspond to
a discrete-time sampling of continuous classical mechanical systems.
We have described models in both 2D and 3D that use immovable mirrors,
and provided a technique for making related models without immovable
mirrors that are exactly momentum-conserving while preserving their
universality and spatial efficiency.  In the context of the 2D
momentum-conserving models, we have shown that it is possible to avoid
entropy generation associated with routing signals. For all of the
momentum conserving models we have provided both a non-relativistic
and a relativistic interpretation of the microscopic dynamics.  The
same relativistic interpretation applies generally to mass and
momentum conserving lattice gases.  We have also provided a
semi-classical interpretation under which these models give the
correct physical bound on maximum computation rate.

It is easy to show that reversible LGA's can all be turned into
quantum dynamics which reproduce the LGA state at integer
times\cite{marg-qc}.  Thus SSM gases can be interpreted not only as
both relativistic and non-relativistic systems, but also as both
classical and as quantum systems.  In all cases, the models are
digital at integer times, and so provide a link between continuous
physics and the dynamics of digital information in all of these
domains, and perhaps also a bridge linking informational concepts
between these domains.

\section*{Acknowledgements}

This work was supported by DARPA under contract number
DABT63-95-C-0130.  This work was stimulated by the SFI Constructive CA
workshop.

{}

\end{document}